\documentclass[prx,aps,nofootinbib,notitlepage,longbibliography,twocolumn, superscriptaddress,preprintnumber]{revtex4-1}
\usepackage{amsmath,amssymb,amsfonts}
\usepackage{slashed}
\usepackage{graphicx}
\usepackage{bm}
\usepackage{float}
\usepackage[T1]{fontenc}
\usepackage{multirow}
\usepackage[utf8]{inputenc}
\usepackage{gauss}
\usepackage[normalem]{ulem}
\usepackage[top=1.0in,bottom=1.0in,left=1.0in,right=1.0in]{geometry}
\usepackage[colorlinks,linkcolor=blue,citecolor=blue,urlcolor=blue]{hyperref}
\usepackage{subcaption}
\usepackage{ragged2e}
\usepackage[compatibility=false]{caption}
\usepackage{tikz}
\usetikzlibrary{quantikz2}
\DeclareCaptionJustification{justified}{\justifying}
\usepackage{booktabs}
\newcommand{\be}{\begin{equation}}
\newcommand{\ee}{\end{equation}}
\newcommand{\bea}{\begin{eqnarray}}
\newcommand{\eea}{\end{eqnarray}}
\newcommand{\ba}{\begin{eqnarray}}
\newcommand{\ea}{\end{eqnarray}}

\DeclareMathOperator{\erf}{erf}
\DeclareMathOperator{\erfc}{erfc}
\usepackage{fancyhdr}

\newcommand{\des}{\hat a}
\newcommand{\cre}{\hat a^\dagger}
\newcommand{\cdx}{\text{CD}_X}
\newcommand{\nbar}{\bar{n}}
\newcommand{\definedas}{\mathrel{\mathop:}=}
\newcommand{\Yb}{\textsuperscript{171}Yb\textsuperscript{+}}
\newcommand{\tr}{\operatorname{Tr}}
\newcommand{\Hc}{\mathcal{H}_c}          
\newcommand{\Wvac}{W_{\rm vac}}          

\makeatletter
\newcommand{\appendixtocsectionsonly}{%
\let\l@subsection\@gobbletwo \let\l@subsubsection\@gobbletwo }
\makeatother

\makeatletter
\renewcommand*{\@fnsymbol}[1]{\ensuremath{\ifcase#1\or 1\or 2\or 3\or 4\or 5\or 6\or 7\or 8\or 9\or 10 \or 11\else\@ctrerr\fi}}
\makeatother
\begin{document}
\title{Trigonometric Continuous-Variable Quantum Gates: Realization with Trapped Ions and Nonperturbative Wigner Negativity}
\author{Tommaso Rainaldi}
\email{tommaso.rainaldi@stonybrook.edu}
\affiliation{Department of Physics and Astronomy, Stony Brook University, Stony Brook, NY 11794, USA}
\author{Jake Montgomery}
\email{jake.montgomery@stonybrook.edu}
\affiliation{Department of Physics and Astronomy, Stony Brook University, Stony Brook, NY 11794, USA}
\author{Christopher G. Yale}
\email{cgyale@sandia.gov}
\affiliation{Sandia National Laboratories, Albuquerque, NM, 87185, USA}
\author{Brian K. McFarland}
\email{bmcfarl@sandia.gov}
\affiliation{Sandia National Laboratories, Albuquerque, NM, 87185, USA}
\author{Melissa C. Revelle}
\email{mrevell@sandia.gov}
\affiliation{Sandia National Laboratories, Albuquerque, NM, 87185, USA}
\author{Daniel Lobser}
\email{dlobser@sandia.gov}
\affiliation{Sandia National Laboratories, Albuquerque, NM, 87185, USA}
\author{Edward C. Tortorici}
\email{ectorto@sandia.gov}
\affiliation{Sandia National Laboratories, Albuquerque, NM, 87185, USA}
\author{Susan M. Clark}
\email{sclark@sandia.gov}
\affiliation{Sandia National Laboratories, Albuquerque, NM, 87185, USA}
\author{George Siopsis}
\email{siopsis@tennessee.edu}
\affiliation{Department of Physics and Astronomy, The University of Tennessee, Knoxville, TN 37996, USA}
\author{Matt Grau}
\email{mgrau@odu.edu}
\affiliation{Department of Physics, Old Dominion University, Norfolk, VA 23529, USA}
\author{Felix Ringer}
\email{felix.ringer@stonybrook.edu}
\affiliation{Department of Physics and Astronomy, Stony Brook University, Stony Brook, NY 11794, USA}
\begin{abstract}
We experimentally realize trigonometric continuous-variable gates on a trapped-ion processor, for which a motional mode acquires a phase proportional to the cosine of its position quadrature, and, for the first time, implement the two-mode generalization, coupling two modes through a single nonlinear phase. Such gates provide an experimentally accessible, nonpolynomial primitive for periodic interactions acting on both compact and noncompact degrees of freedom, including rotor models, sine-Gordon-type systems, and lattice gauge theories. Scanning gate strength, spatial frequency, and circuit depth, we resolve via blue-sideband spectroscopy the parity selection rule that fingerprints the exact cosine evolution, and find that an open-system model incorporating residual thermal occupation and motional dephasing reproduces the data. We then derive the asymptotics of the Wigner negativity generated by these gates and find three scaling regimes. The negativity is beyond all algebraic orders in the gate strength while the negative regions sit in far phase-space tails, becomes linear once they reach the bulk, where it saturates a first-order bound we establish, and logarithmic at strong gate strength. These results expose a general mechanism, first identified here through the cosine gate, by which every finite-order perturbative estimate of a non-Gaussian resource can vanish even though the resource itself remains nonzero. Together, our results establish trigonometric gates as controllable, experimentally realizable building blocks for bosonic quantum simulation, expanding the class of nonlinear dynamics accessible to continuous-variable quantum processors.
\end{abstract}
\maketitle
\tableofcontents
\thispagestyle{fancy}
\fancyhead{}
\fancyfoot{}
\renewcommand{\headrulewidth}{0pt}
\pagestyle{plain}
\section{Introduction}
\label{sec:1}
Continuous-variable (CV) quantum systems provide a natural computational language when the physical degrees of freedom are intrinsically bosonic. Rather than encoding oscillator states into qubit registers, hybrid CV--discrete-variable (DV) architectures can manipulate bosonic modes directly while using ancillary qubits for control, nonlinear operations, and readout~\cite{Mohapatra:2026tyk,Heris:2026dmx}. Such architectures arise naturally in trapped ions, superconducting circuits, and cavity-QED platforms, where coherent qubit--oscillator interactions and conditional phase-space displacements are experimentally available~\cite{Leibfried:2003zz,Blais:2020wjs,Stavenger:2022wzz,Liu:2024mbr,Araz:2024dcy,Fluhmann:2018rvj,deNeeve:2020ugg,Heeres:2015cnj,Eickbusch:2021uod,Wang:2019xqo}. Such hardware natively supports entangling qubit-oscillator interactions, sideband couplings, spin-dependent forces~\cite{Haljan:2004nbn}, programmable phononic networks~\cite{Chen:2022auh}, and non-Gaussian resources~\cite{Meekhof:1996zz,Um:2015vwj}. This combination is particularly attractive for quantum simulation, where bosonic modes can represent oscillators, phonons, scalar fields, or gauge degrees of freedom without first introducing a finite-dimensional encoding, complementing qubit encodings of fermionic and spin degrees of freedom~\cite{Jordan:1928wi,Chiari:2025lwq} in lattice field theories and related lattice models~\cite{Kogut:1974ag,Jordan:2011ci,Zohar:2015hwa,Banerjee:2012pg,Banuls:2019bmf,Kreshchuk:2020aiq,Haase:2020kaj,Zohar:2021nyc,Klco:2018zqz,Bauer:2022hpo,Briceno:2023xcm,Davoudi:2021ney,Katz:2022gra,Gustafson:2024kym,Farrell:2023fgd,Ale:2024uxf,Abel:2025zxb,Crane:2024tlj,Ale:2025sxz,Saner:2025nrq,Than:2025gso,Zou:2026cfk,Athanasakos:2026upp,Cogburn:2026aqy}, in molecular and vibrational quantum dynamics~\cite{10.1021/jp992939g,Shen:2017txh,Wang:2019xqo,Valahu:2022kfg,Dutta:2024cso}, and in the quantum Rabi model~\cite{Lv:2017tmn}. Hybrid instruction sets and compilation strategies have further been developed for quantum simulation, error correction, and quantum algorithms~\cite{Liu:2024mbr,Sutherland:2021vsj,Eickbusch:2021uod,Chen:2021nbo,Kemper:2025ldr}, with bosonic error-correcting codes such as GKP, cat, and binomial encodings~\cite{Gottesman:2000di,Fluhmann:2018rvj,Hu:2018mmi,Campagne-Ibarcq:2019nmy} realized experimentally across several platforms~\cite{Brock:2024vkc,deNeeve:2020ugg,Matsos:2023iqx,Matsos:2024qer}.

A central question for these architectures is which nonlinear CV operations should be treated as elementary primitives. Conventional formulations of CV universality are usually expressed in terms of polynomial functions of the canonical quadratures $\hat x$ and $\hat p$. Gaussian operations, generated by Hamiltonians at most quadratic in the quadratures, together with a suitable non-Gaussian interaction provide a universal set \cite{Lloyd:1998jk,Braunstein:2005zz,Andersen:2014xqr}. Polynomial nonlinearities are natural for many applications, but they are not always well matched to the structure of the target Hamiltonian. Periodic interactions arise directly in rotor models, sine-Gordon-type systems, compact lattice gauge theories, Josephson physics, and other problems involving compact or angular degrees of freedom. Representing such globally periodic structure by a truncated polynomial can require increasingly high order as the relevant phase-space region grows. There is also a structural difference. The lowest-order polynomial nonlinearity, the cubic phase gate, is unbounded from below and must itself be stabilized by a higher-order term whose error accumulates under repeated application \cite{Moore:2025qzc}. A trigonometric generator is bounded by construction, and repeated application stays within the same one-parameter family. Trigonometric CV gates therefore provide an alternative primitive adapted directly to this periodic structure.

In this work we consider the elementary phase gate
\begin{equation}
    U(\theta,c)
    =
    e^{-i\theta\cos(c\hat x)}
\label{eq:intro_single_gate}
\end{equation}
and its multimode generalization
\begin{equation}
    U_{12}(\theta;c_1,c_2)
    =
    e^{-i\theta\cos(c_1\hat x_1+c_2\hat x_2)} .
\label{eq:intro_two_gate}
\end{equation}
The gate strength $\theta$ controls the accumulated nonlinear phase, while $c$ sets its spatial frequency in phase space. Through the Jacobi--Anger expansion, Eq.~\eqref{eq:intro_single_gate} can equivalently be viewed as a coherent infinite superposition of momentum displacements,
\begin{equation}
    e^{-i\theta\cos(c\hat x)}
    =
    \sum_{k\in\mathbb Z}
    (-i)^k J_k(\theta)e^{ikc\hat x},
\label{eq:intro_jacobi_anger}
\end{equation}
revealing a structure that is intrinsically nonpolynomial and particularly well suited to hybrid qubit--qumode control.

Trigonometric gates were introduced as theoretical and compilation primitives for hybrid CV--DV quantum processing in Refs.~\cite{Rainaldi:2025ymn,Chalermpusitarak:2025cod}. These works developed a Fourier-like alternative to polynomial CV synthesis and showed how periodic nonlinearities can be constructed from qubit-controlled phase-space operations. Such a gate has very recently been demonstrated on a trapped-ion platform and used to synthesize a programmable double-well potential \cite{McGarry:2026mkk}, and related work has realized non-Gaussian operations from engineered spin--motion couplings \cite{Bazavan:2024yya}. In Ref.~\cite{McGarry:2026mkk} the trigonometric gate appears as a compilation layer: a single spatial frequency and a single gate strength are used, and the free parameter is the Fourier phase that tilts the well. The present work is complementary and operates at the level of the primitive. We isolate the cosine gate itself, map its behavior across gate strength, spatial frequency, and circuit depth, extend it to two motional modes, and analyze the non-Gaussian resource that it generates.

Acting on the vacuum, the cosine gate produces a pure non-Gaussian state for every nonzero $\theta$, so by Hudson's theorem its Wigner function must acquire negative regions. How much negativity is generated, and whether it is accessible to perturbation theory, is the subject of Sec.~\ref{sec:mana}, where we quantify it by the mana, the Wigner logarithmic negativity of the state. We find three scaling regimes. Two occur at weak gate strength and are distinguished by the spatial frequency $c$. At small $c$ the negative regions appear far out in the phase-space tails, receding to infinity as $\theta\rightarrow0$, and the mana is suppressed beyond every algebraic order in $\theta$. Thus an arbitrarily weak nonlinear gate generates Wigner negativity that no finite-order expansion about the Gaussian limit can detect. At larger $c$ the negative regions instead reach the bulk of the state and the mana grows linearly with $\theta$, saturating a first-order bound that we establish. The third regime occurs at strong gate strength, where the mana grows logarithmically.

Closed-form phase-space results of this kind were previously available only for low-order polynomial gates. The cubic and quartic phase gates act on the Wigner function as an Airy transform \cite{Moore:2025qzc}, but this construction does not extend to the nonpolynomial generator considered here. In our analysis its role is played by the Bessel resummation developed in Sec.~\ref{sec:phase_space}. Ref.~\cite{Moore:2025qzc} also showed that a nonlinear phase gate produces Wigner negativity at any nonzero gate strength, while noting that the resulting negative volume is vanishingly small and difficult to resolve numerically. Here we make that observation quantitative and provide the rate. The mechanism is not specific to the cosine gate. It applies to any weak non-Gaussian phase acting on a Gaussian state, so perturbative estimates of non-Gaussian resource generation can vanish to all orders while the resource itself is nonzero.

We realize these gates using collective motional modes of trapped $^{171}{\rm Yb}^{+}$ ion chains on the Quantum Scientific Computing Open User Testbed (QSCOUT). Conditional phase-space displacements generated by spin-dependent forces, together with rotations of an ancillary gate qubit, produce finite-step approximations to Eq.~\eqref{eq:intro_single_gate}. Increasing the number of steps systematically approaches the ideal trigonometric evolution, while measurement of the gate qubit provides a natural herald for leading finite-step errors. We characterize the resulting motional states through blue-sideband spectroscopy and, for a representative operating point, through a phase-sensitive measurement of the characteristic function. We study the dependence on gate strength, spatial frequency, and circuit depth and compare the observations with finite-step simulations incorporating residual thermal occupation and motional dephasing. We also implement the two-mode gate in Eq.~\eqref{eq:intro_two_gate}, and the measured mode-resolved populations are consistent with the predicted coupled nonlinear evolution.

Our results combine an experimentally accessible nonpolynomial CV primitive with an analytically controlled description of how its non-Gaussian resources emerge. Two of these results are new in kind. First, the two-mode gate of Eq.~\eqref{eq:intro_two_gate} couples two motional modes through a single nonlinear phase and obeys a total-parity selection rule that distinguishes it from independent single-mode gates. Second, the weak-angle analysis establishes a sharp bound on the resource extractable at first order in the gate strength. These results make trigonometric phase gates a practical non-Gaussian primitive for continuous-variable quantum processing and for quantum simulation of systems with intrinsically periodic or compact degrees of freedom.

The remainder of this paper is organized as follows. Section~\ref{sec:4} develops the exact Fock-space and phase-space structure of one- and two-mode trigonometric gates. Section~\ref{sec:mana} analyzes their Wigner logarithmic negativity, covering tail nucleation, resolved-displacement linear response, and stationary-phase growth. Section~\ref{sec:2} describes the QSCOUT implementation, finite-step construction, postselection, and multimode extension. Section~\ref{sec:6} presents the experimental measurements and comparison with open-system simulations. We conclude in Section~\ref{sec:7} with implications for non-Gaussian CV processing and periodic bosonic quantum simulation. Detailed derivations of the weak-angle and stationary-phase asymptotics, finite-step convergence, ancilla constructions, and experimental analysis are provided in the Appendices.
\section{Trigonometric continuous-variable gates}\label{sec:4}
In this section, we present analytical results for the matrix elements of trigonometric gate primitives and the associated characteristic and Wigner functions.
\subsection{One-mode cosine gate and exact Fock-space structure}
While the argument of the trigonometric gates introduced in~\cite{Rainaldi:2025ymn} can, in principle, be any Hermitian operator $\hat{A}$, we limit ourselves in this work to the case where $\hat{A}$ is a linear function of the quadratures. When $\hat{A}$ is linear in the quadratures, i.e.\ $\hat{A}=\bm{c}\cdot\hat{\mathbf{Q}}$ with $\bm{c}\in \mathbb{R}^{2N}$ and $\mathbf{\hat{Q}} = (\hat{x}_1,\dots,\hat{x}_N,\hat{p}_1,\dots,\hat{p}_N)^T$, the exact cosine gate can be regarded as an infinite series of displacements, where each is weighted by a Bessel function. The relevant examples for this paper are the simplest cases involving the cosine of one or two position quadratures. In terms of operators, we have $\hat{A}_\text{1-mode} = c\hat{x}$ and $\hat{A}_\text{2-mode} = \bm{c}\cdot\hat{\bm{x}}$, with $\bm{c}=(c_1,c_2)$ and $\hat{\bm{x}} = (\hat{x}_1,\hat{x}_2)$. The generalization to more than two modes is straightforward. Throughout we use the displacement and conditional-displacement operators
\begin{equation}
    {\rm D}(\alpha) = e^{\alpha\cre-\alpha^*\des},
    \quad
    {\rm CD}_{\bm n\cdot\bm\sigma}(\alpha)
    = e^{(\alpha\cre-\alpha^*\des)\otimes\,\bm n\cdot\bm\sigma},
\label{eq:gates}
\end{equation}
where $\bm{n}$ denotes a Bloch vector and $\bm{\sigma}=( X,Y,Z)$. Starting with the single-qumode case, we consider the trigonometric gate unitary $e^{-i \theta \cos(c \hat{x})}$. After applying the Jacobi-Anger expansion, we find
\begin{align}
    e^{-i \theta \cos(c \hat{x})} & =
    \sum_{k\in\mathbb{Z}} (-i)^k J_k(\theta) e^{i c k \hat{x}} \nonumber \\
    &= \sum_{k\in\mathbb{Z}} (-i)^k J_k(\theta)\, {\rm D} \left( i k \frac{c}{\sqrt{2}} \right)\,.
\end{align}
Here, $J_k$ are Bessel functions and ${\rm D}$ are displacement gates, see Eq.~\eqref{eq:gates}. The Fock matrix elements of the displacement operators are well known, see Ref.~\cite{Cahill:1969it}. From these, we directly obtain the Fock-space transition matrix elements of the one-qumode cosine gate, which are given by
\begin{align}
    &\bra{m}e^{-i \theta\cos(c \hat{x})}\ket{n} = \sum_{k\in\mathbb{Z}}(-i)^kJ_k(\theta)\bra{m}e^{ic k\hat{x}}\ket{n}\nonumber\\
    & = \sum_{k\in\mathbb{Z}}(-i)^kJ_k(\theta)e^{-\frac{c^2k^2}{4}} \sqrt{\frac{\min(m,n)!}{\max(m,n)!}}\nonumber\\
    &\;\;\;\;\times\left(i\frac{c k}{\sqrt{2}}\right)^{|m-n|} L_{\min(m,n)}^{(|m-n|)}\left(\frac{c^2k^2}{2}\right)\,,
\label{eq:cos_gate_mat_element_mn}
\end{align}
where $L^{(\alpha)}_N(x)$ are the generalized Laguerre polynomials of degree $N$ with parameter $\alpha$.
\begin{figure*}[t]
\centering
\includegraphics[width=0.32\linewidth]{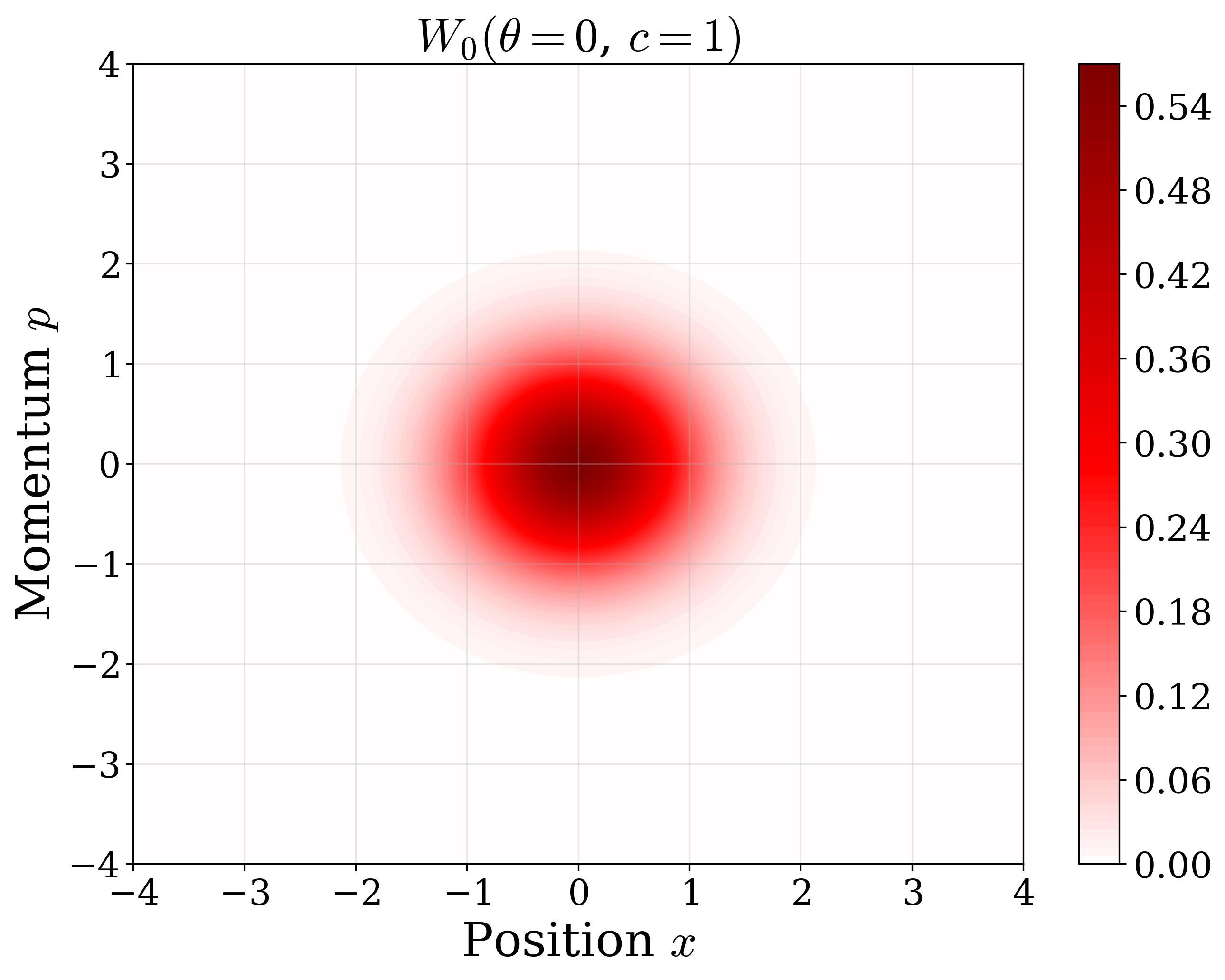}
\includegraphics[width=0.32\linewidth]{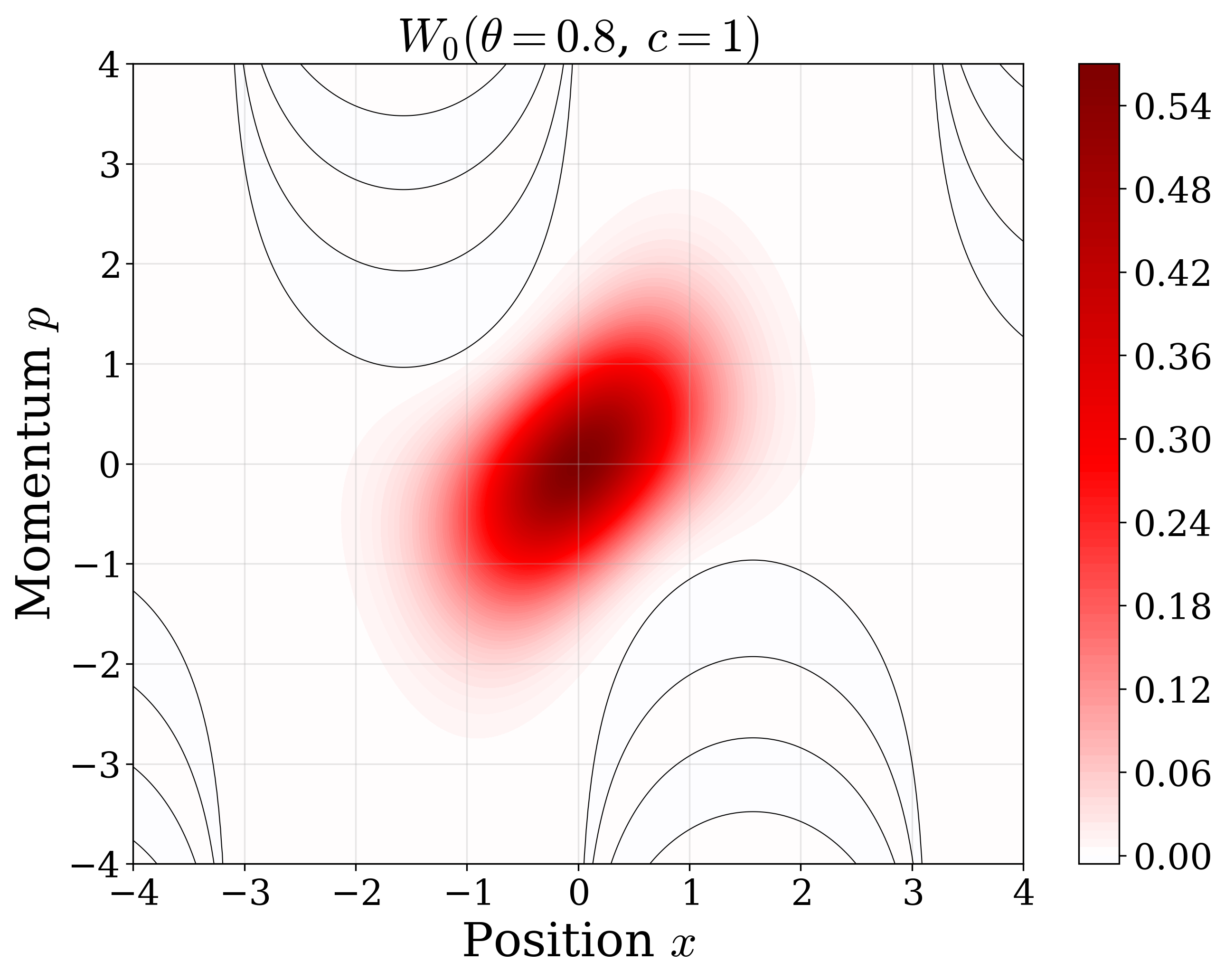}
\includegraphics[width=0.32\linewidth]{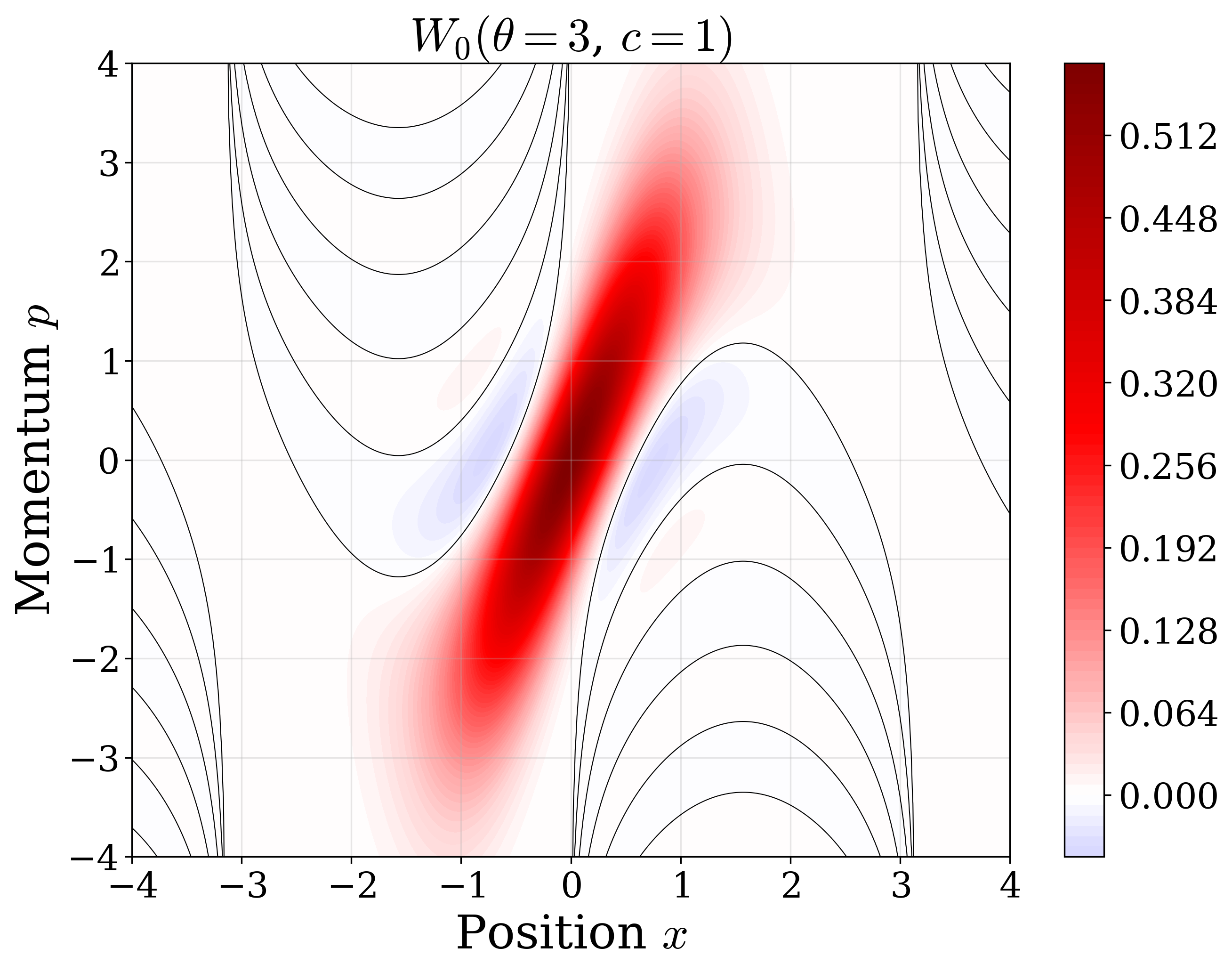}
\includegraphics[width=0.32\linewidth]{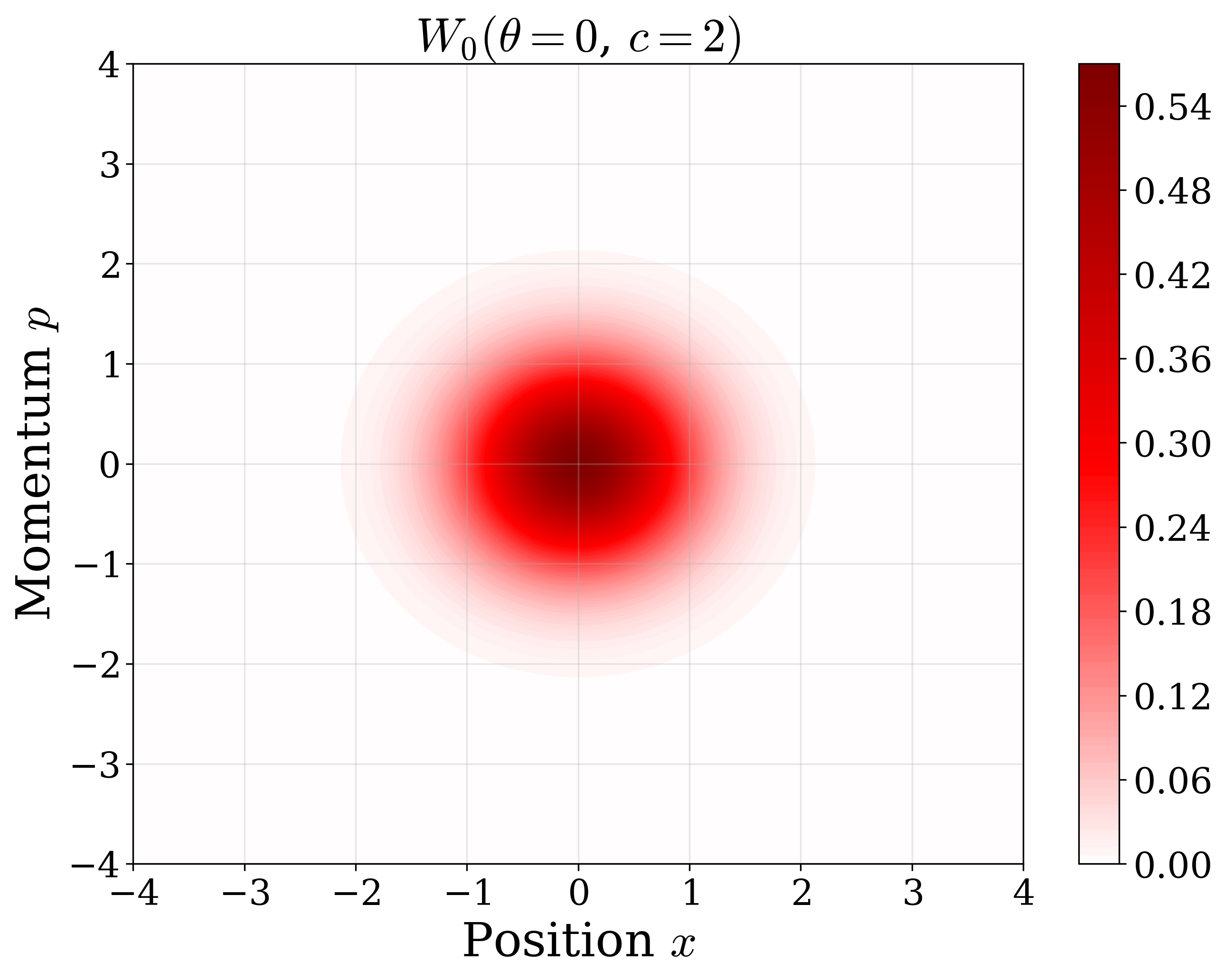}
\includegraphics[width=0.32\linewidth]{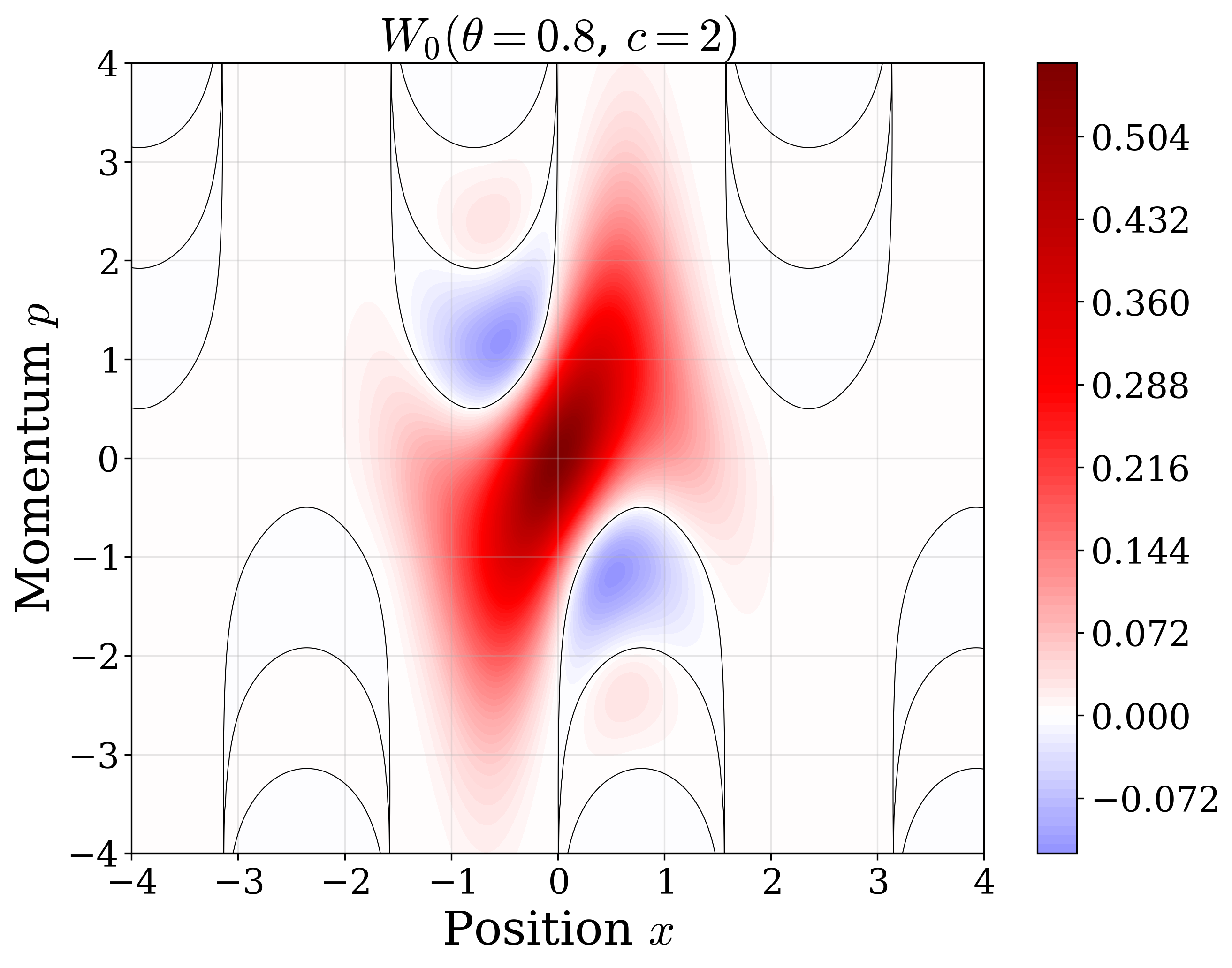}
\includegraphics[width=0.32\linewidth]{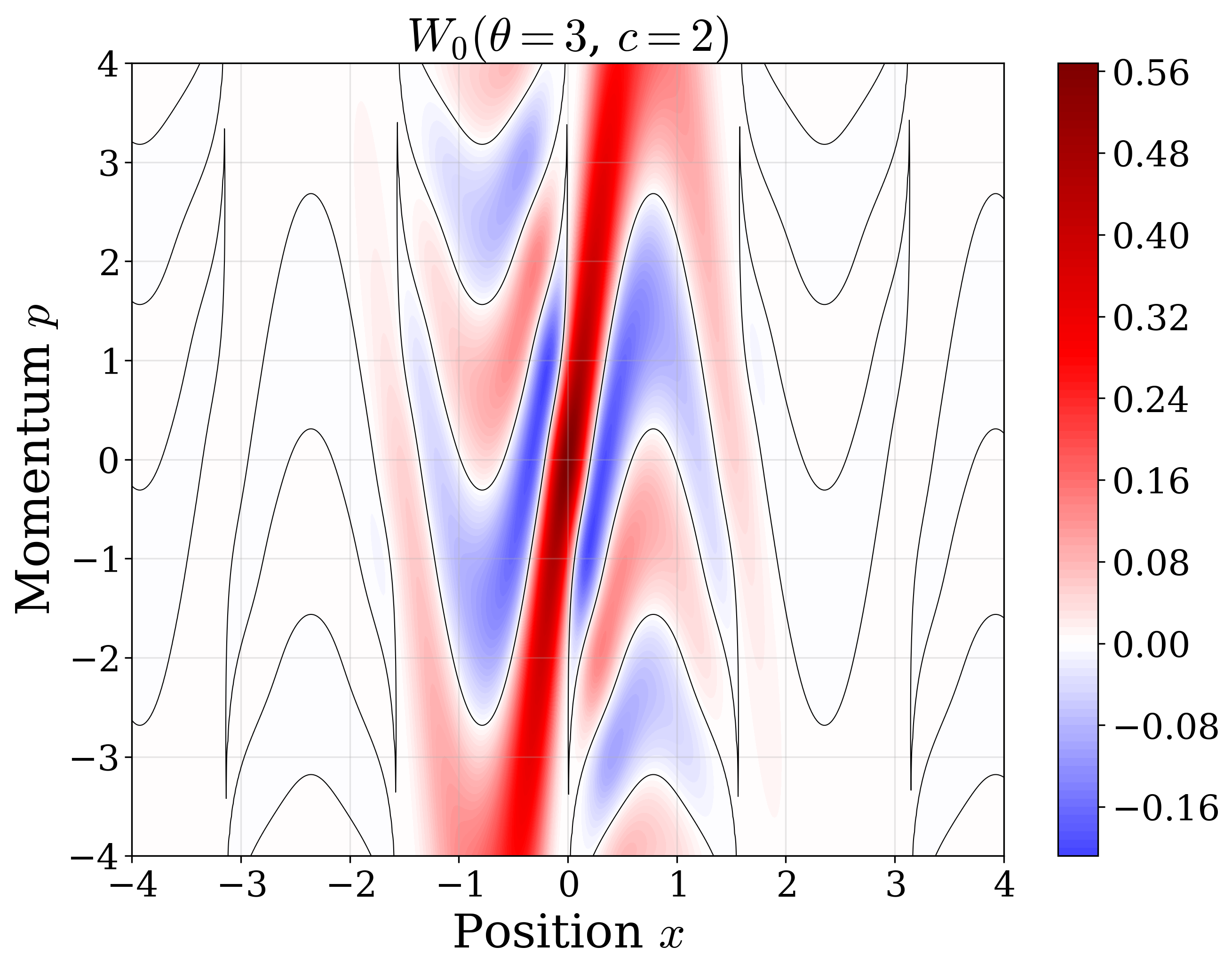}
        \caption{Wigner function of the single-qumode cosine gate acting on the Fock vacuum state $\ket{0}$ for different values of $\theta=0,0.8,3$ and fixed parameter $c = 1$ (upper row), $c=2$ (lower row).}
\label{fig:W0_c_2}
\end{figure*}
Non-vanishing matrix elements exist when $m$ and $n$ have equal parity, as can be seen directly from the coordinate space representation
\begin{equation}
    \bra{m}e^{-i \theta\cos(c \hat{x})}\ket{n}=\int_{-\infty}^{\infty}\text{d}x\,\psi_m(x)\psi_n(x)e^{-i\theta \cos(c x)},
\end{equation}
where
\begin{equation}
    \psi_n(x) = \frac{\pi^{-1/4}}{\sqrt{2^n n!}}e^{-\frac{x^2}{2}}H_n(x)\,,
\label{eq:fock_wavefunction}
\end{equation}
is the wave function of the $n$-th Fock state. From Eq.~\eqref{eq:cos_gate_mat_element_mn}, we also find the survival/transition probabilities
\begin{equation}
\begin{split}
    &P_{m\leftarrow n}(\theta,c)\equiv |\!\bra{m}e^{-i \theta\cos(c \hat{x})}\ket{n}\!|^2 =\\
    &  \sum_{k,k'\in\mathbb{Z}}(-i)^k(i)^{k'}J_k(\theta)J_{k'}(\theta)e^{-\frac{c^2(k^2+k'^2)}{4}} \frac{\min(m,n)!}{\max(m,n)!} \times\\
    &\left(\frac{c^2 kk'}{2}\right)^{|m-n|} L_{\min(m,n)}^{(|m-n|)}\left(\frac{c^2k^2}{2}\right)L_{\min(m,n)}^{(|m-n|)}\left(\frac{c^2k'^2}{2}\right).
\end{split}
\end{equation}

For the experimental measurements discussed below, we specifically focus on the case $n = 0$. The calculations simplify considerably because $L_{0}^{(m)}\left(\frac{c^2k^2}{2}\right) = 1$. Thus, we obtain
\begin{align}
    &P_{m\leftarrow 0}(\theta,c)
    =\nonumber\\
    &\begin{cases}
    0\,, & m \text{ odd},\\[6pt]
    \displaystyle
    \frac{c^{2m}}{2^m m!}
    \Big|\sum_{k\in\mathbb{Z}}(-1)^k i^{k+m}J_k(\theta)e^{-\frac{c^2k^2}{4}}k^m\Big|^2\,,
    & m \text{ even},
    \end{cases}
\label{eq:Probs_exact}
\end{align}
so that the only non-vanishing probabilities are obtained for even $m$.
\subsection{Multimode generalization and parity structure}
Next, we consider the two-qumode extension of the cosine gate. Mirroring the results of the single-mode case, we can once again regard the ideal gate as an infinite sum of displacements
\begin{equation}
\begin{split}
    &e^{-i \theta \cos(\bm{c}\cdot\hat{\bm{x}})} = \sum_{k\in\mathbb{Z}} (-i)^k J_k(\theta) e^{i c_1 k \hat{x}_1}e^{i c_2 k \hat{x}_2} \\
    &= \sum_{k\in\mathbb{Z}} (-i)^k J_k(\theta) \, {\rm D}_1 \left( i k \frac{c_1}{\sqrt{2}} \right) {\rm D}_2 \left( i k \frac{c_2}{\sqrt{2}} \right),
\end{split}
\end{equation}
with each mode getting displaced independently. The matrix elements in the two-mode Fock basis read
\begin{equation}
\begin{split}
    &\bra{m_1}\bra{m_2}e^{-i \theta \cos(c_1 \hat{x}_1 + c_2\hat{x}_2)}\ket{n_1}\ket{n_2} \\
    =& \sum_{k\in\mathbb{Z}}(-i)^kJ_k(\theta)\prod_{j=1}^2\bra{m_j}e^{ic_j k\hat{x}_j}\ket{n_j}\\
     =& \sum_{k\in\mathbb{Z}}(-i)^kJ_k(\theta)\prod_{j=1}^2e^{-\frac{c_j^2k^2}{4}} \sqrt{\frac{\min(m_j,n_j)!}{\max(m_j,n_j)!}}  \\
    &\times\left(i\frac{c_j k}{\sqrt{2}}\right)^{|m_j-n_j|}L_{\min(m_j,n_j)}^{(|m_j-n_j|)}\left(\frac{c_j^2k^2}{2}\right).
\end{split}
\label{eq:2mode_cos_gate_mat_element_m0n0_m1n1}
\end{equation}
Similar to Eq.~\eqref{eq:Probs_exact}, in this work, we are primarily interested in the Fock vacuum initialization where $\ket{n_1} = \ket{n_2} = \ket{0}$. In this case, the transition probabilities simplify to
\begin{equation}
\begin{split}
    &P_{m_{1,2} \leftarrow 0}(\theta,\bm{c})\\
    & = \left|\sum_{k\in\mathbb{Z}}(-i)^kJ_k(\theta)\prod_{j=1}^2 \frac{e^{-\frac{c_j^2k^2}{4}}}{\sqrt{m_j!}}
    \left(i\frac{c_j k}{\sqrt{2}}\right)^{m_j}\right|^2.
\end{split}
\label{eq:2mode_cos_gate_mat_element_m0_0_m1_0}
\end{equation}
The only non-vanishing probabilities are obtained for $(-1)^{m_1+m_2} = 1$. The total Fock parity must be the same as that of the initial state.
\subsection{Phase-space representation}\label{sec:phase_space}
\begin{figure*}[t]
        \centering
        \includegraphics[width=0.32\linewidth]{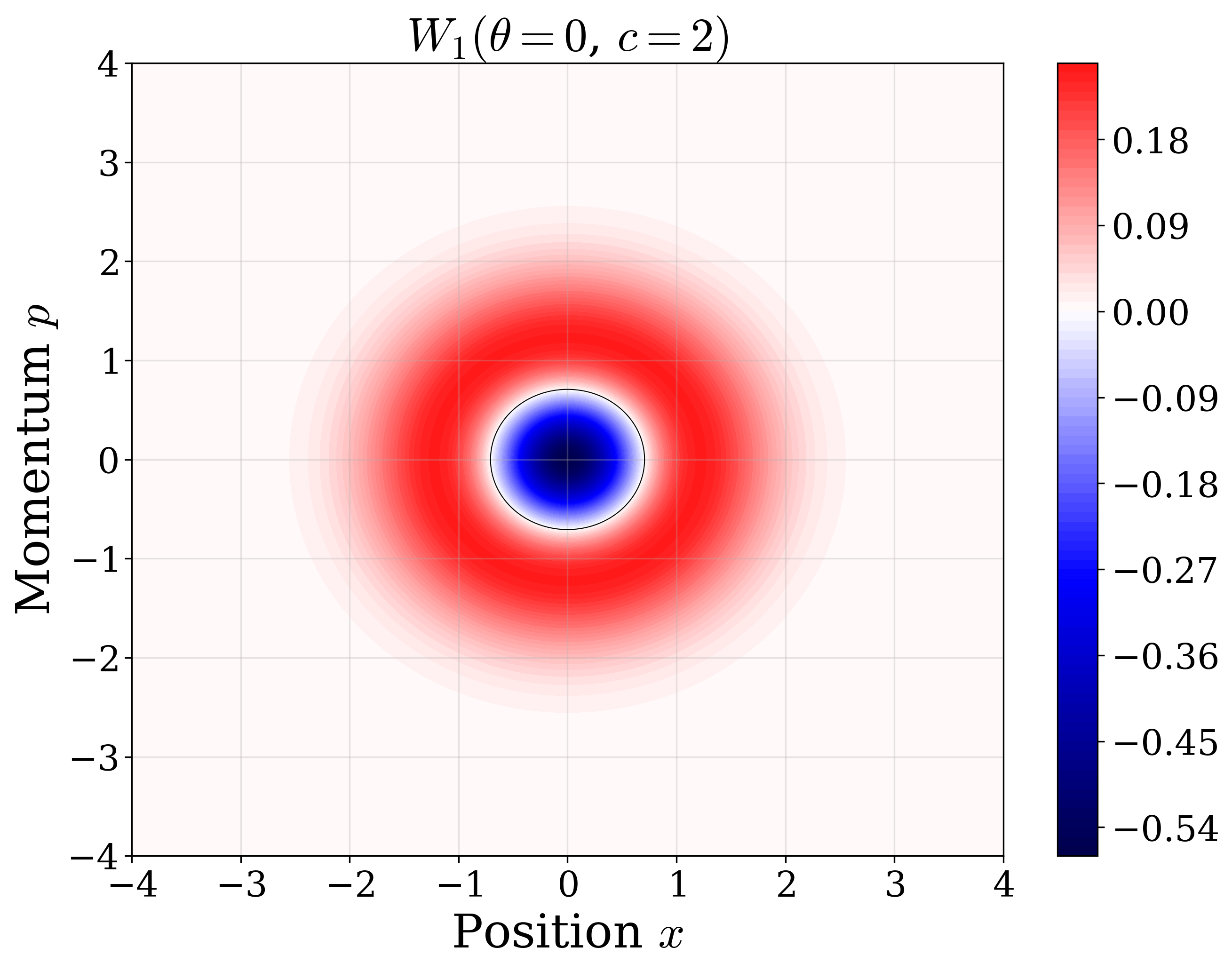}
        \includegraphics[width=0.32\linewidth]{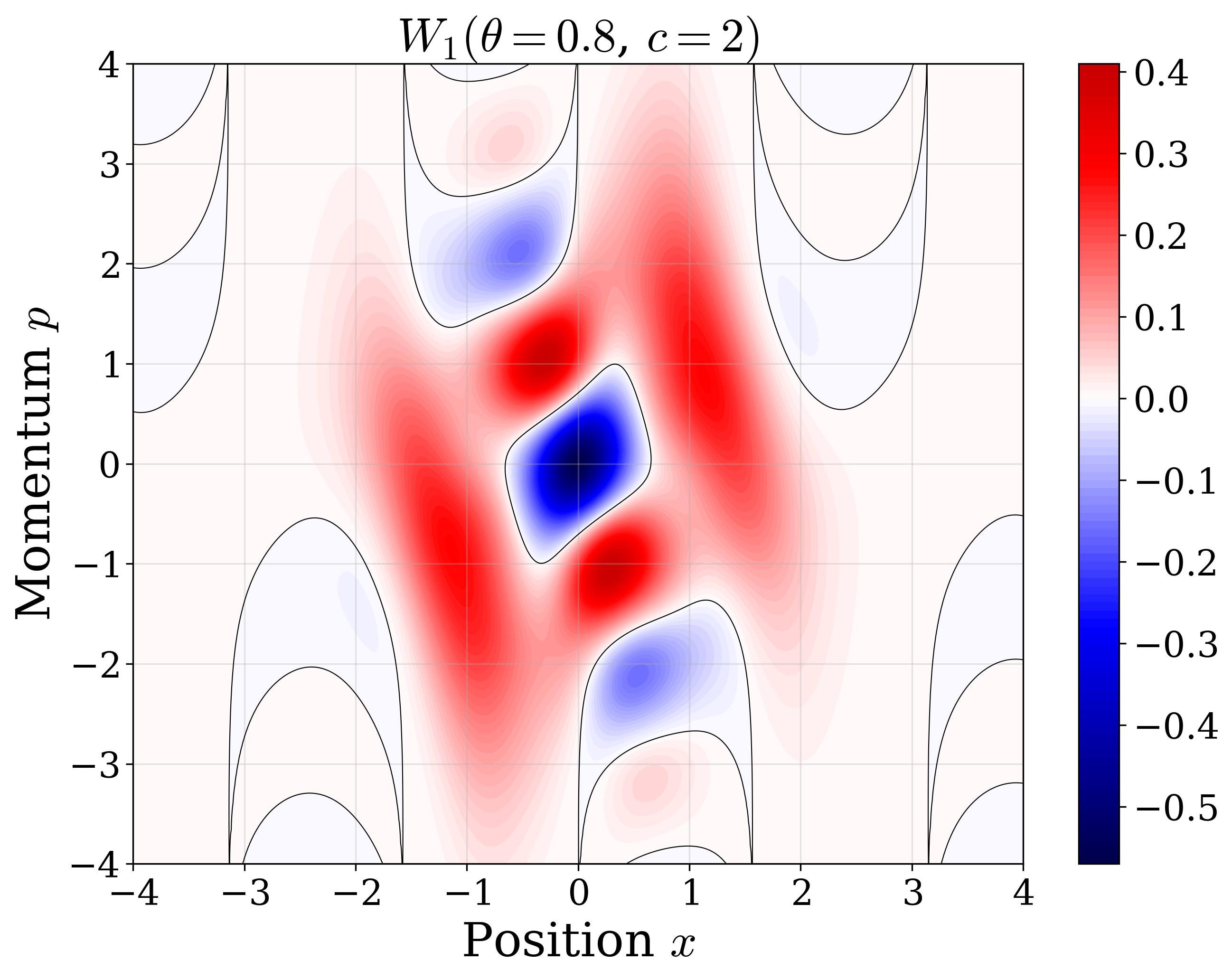}
        \includegraphics[width=0.32\linewidth]{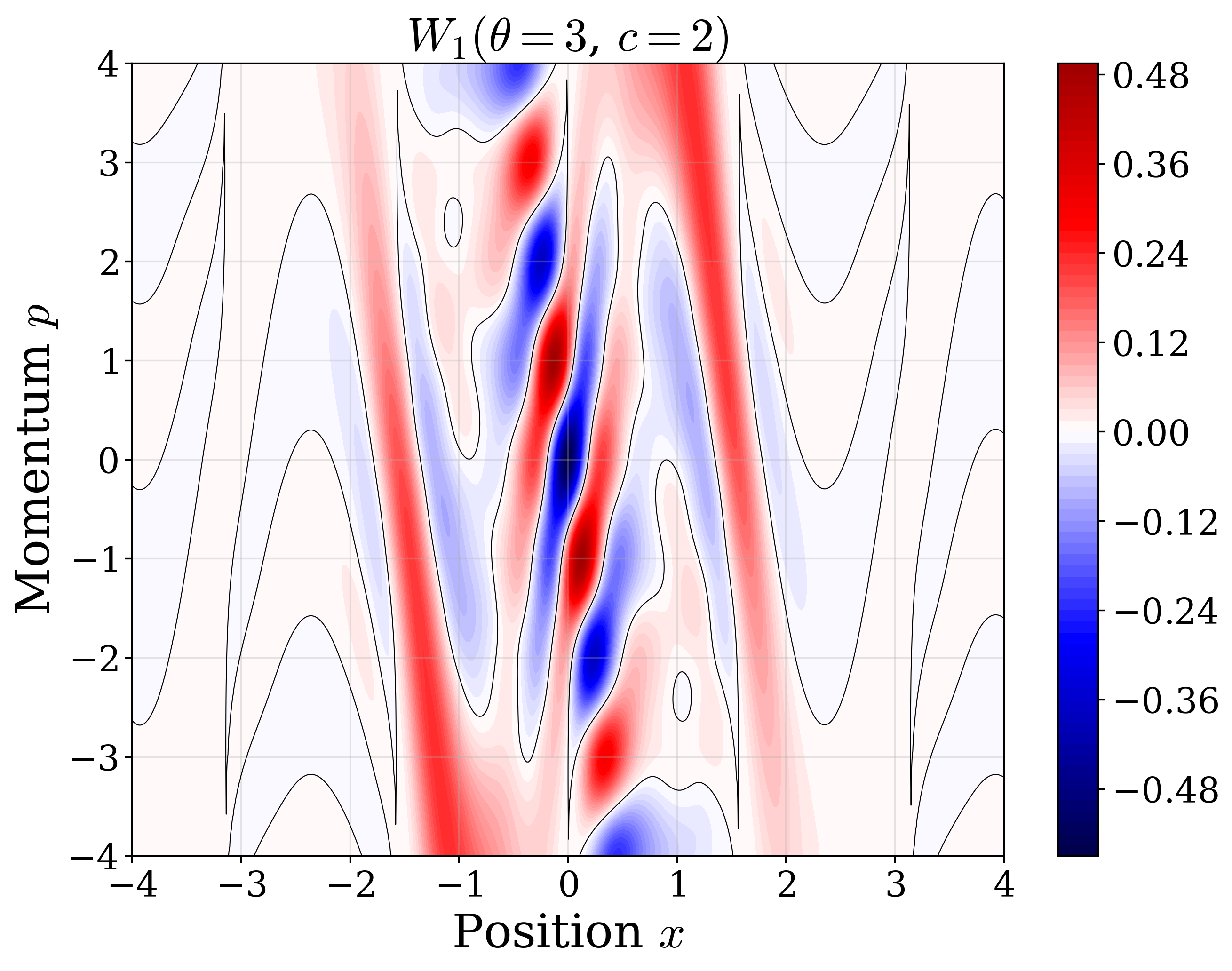}
        \caption{Wigner function of the single-qumode cosine gate acting on the Fock state $\ket{1}$ for different $\theta$ values and fixed parameter $c = 2$.}
        \label{fig:W1_c_2}
\end{figure*}
We can also obtain an analytical expression for the Wigner function of a given Fock state evolved under a single-quadrature cosine gate. We consider the $n$-th Fock state $\ket{n}$ as the initial state and evolve it with the cosine gate $\exp(-i \theta \cos(c \hat{x}))$. In position space, the evolved wavefunction is $\psi_n(x;\theta,c) = \psi_n(x) e^{-i \theta \cos(c x)}$, with $\psi_n$ given in Eq.~\eqref{eq:fock_wavefunction}. The Wigner function is defined as
\begin{equation}
\begin{split}
    &W_n(x, p;\theta, c) \\
    &= \frac{1}{2\pi} \int_{-\infty}^{\infty} {\rm d}y \, e^{i p y} \psi_n^*\left(x + \frac{y}{2};\theta,c\right) \psi_n\left(x - \frac{y}{2};\theta,c\right).
\end{split}
\label{eq:wigner_definition}
\end{equation}
After some algebra, we find
\begin{equation}\label{eq:Wignern}
    \begin{split}
        &W_n(x,p;\theta,c) = \frac{1}{2\pi}\frac{e^{-x^2} }{2^n n!\sqrt{\pi}} \sum_{k \in\mathbb{Z}}J_k(2\theta \sin(cx))\times\\
        &\int \text{d}y\,e^{i y (p - kc/2)}e^{-\frac{y^2}{4}}H_n\left(x-\frac{y}{2}\right)H_n\left(x+\frac{y}{2}\right)\,.
    \end{split}
\end{equation}
For later convenience, we also consider the characteristic function
\begin{equation}
\begin{split}
    &\chi_n(x, p; \theta, c) \\
    &= \int_{-\infty}^{\infty} {\rm d}u \, e^{-i u x} \psi_n^*\left(u + \frac{p}{2}; \theta, c\right) \psi_n\left(u - \frac{p}{2}; \theta, c\right)\,,
\end{split}
\end{equation}
which is related to the Wigner function by the Fourier transform
\begin{align} \label{eq:chi_W_relation}
    &\chi_n(x, p; \theta, c) \nonumber \\
    &= \int_{-\infty}^{\infty} \int_{-\infty}^{\infty} {\rm d}u {\rm d}v \, e^{-i(ux + vp)} W_n(u, v; \theta, c)\,.
\end{align}
\textbf{Zero Fock state $(n = 0)$}: Using the Fock ground state $\ket{0}$ as the initial state, we obtain the following result for the Wigner function
\begin{equation}
\begin{split}
        &W_0(x,p;\theta,c) \\
        &=\frac{e^{-x^2}}{\pi}\sum_{k \in\mathbb{Z}}e^{-(p-kc/2)^2}J_k(2\theta \sin(cx)).
\end{split}
\label{eq:W0_explicit}
\end{equation}

In Fig.~\ref{fig:W0_c_2}, we show numerical results for the Wigner function for different values of $\theta$ and representative choices of the parameter, $c = 1$ and $c = 2$. The non-polynomial evolution produces wave fronts that ``collide'' with the initial Gaussian peak centered at the origin. For $c = 1$, this collision resembles the evolution of a rotating squeezed state and produces only limited negativity. However, already for $c=2$, the number of wave fronts encountered by the Gaussian state increases, and the collision rapidly deforms the state, generating large peaks and troughs. This behavior is quantified systematically in Sec.~\ref{sec:mana}. For an analogous cosine gate involving the momentum quadrature, the wave fronts would approach the central Gaussian peak horizontally instead of vertically.

\textbf{First excited Fock state $(n = 1)$:} We obtain
\begin{equation}
    \begin{split}
        &W_1(x,p;\theta,c) \\
        &=\left(2x^2+2p^2-1\right)W_0(x,p;\theta,c)+ \\
        &\frac{e^{-x^2}}{\pi}\sum_{k \in\mathbb{Z}}e^{-(p-\frac{kc}{2})^2}J_k(2\theta \sin(cx))\left(\frac{k^2c^2}{2} - 2kcp \right).
    \end{split}
\end{equation}
In Fig.~\ref{fig:W1_c_2}, we show representative examples for $c = 2$ and several values of $\theta$. As in the previous case, the action of the cosine gate can be viewed in terms of incoming wave fronts. A key difference is that the initial $n=1$ Fock state already has a rotationally symmetric Wigner function with regions of negativity.

\textbf{General Fock state $n$}:
    \begin{align}
    W_n(x,p;\theta,c) &=\frac{e^{-x^2}}{\pi}\sum_{k \in\mathbb{Z}}e^{-(p-kc/2)^2}J_k(2\theta \sin(cx))\nonumber\\
        &\times (-1)^nL_n[2(x^2+(p-kc/2)^2)]\,,
    \label{eq:Wigner_cos_n}
    \end{align}
where $L_n(x)$ are the usual Laguerre polynomials. Its explicit form can be obtained from Eq.~\eqref{eq:Wignern}. Setting $\theta=0$ or $c=0$ naturally recovers the well-known Wigner function of the $n$-th Fock state. Similarly, the characteristic function is found to be
\begin{align}
    \chi_n(x, p; \theta, c) &=
     \sum_{k\in\mathbb{Z}} J_k\left( 2\theta \sin\left(\frac{cp}{2} \right)\right) e^{-\frac{(x + kc)^2 + p^2}{4}}\nonumber\\
     &\times L_n\left( \frac{(x + kc)^2 + p^2}{2} \right).
\end{align}
As a result, even for $c\neq 0$ and $\theta\neq 0$, the state has definite Fock parity and consequently satisfies the phase-space self-dual relation
\begin{equation}
    \chi_n(x, p; \theta, c)=\pi (-1)^n W_n\left(\frac{p}{2}, -\frac{x}{2}; \theta, c\right)\,.
\end{equation}
\section{Non-Gaussian resource dynamics}
\label{sec:mana}
\subsection{Wigner negativity and mana}
\label{sec:wigner_mana}
We now quantify the non-Gaussian resources generated by the trigonometric phase gate. We focus on the Fock vacuum $\ket{0}$ as the input state. Its Wigner function is Gaussian and everywhere nonnegative, so any Wigner negativity observed after the evolution
\begin{equation}
    \ket{\psi_\theta}
    =
    e^{-i\theta\cos(c\hat x)}\ket{0},
\label{eq:vacuum_evolved}
\end{equation}
is generated entirely by the nonlinear gate. For a normalized Wigner function $W(x,p)$, we define the Wigner $L_1$ norm as
\begin{equation}
    \|W\|_1
    =
    \int_{-\infty}^{\infty}dx
    \int_{-\infty}^{\infty}dp\,
    |W(x,p)|,
\label{eq:wigner_l1}
\end{equation}
and the corresponding Wigner logarithmic negativity \cite{Albarelli:2018ujl}, or mana, as
\begin{equation}
    \mathcal M(\theta,c)
    =
    \ln \left\|W(x,p;\theta,c)\right\|_1 .
\label{eq:mana_definition}
\end{equation}
Since $\int dx\,dp\,W=1$, an everywhere nonnegative Wigner function has $\|W\|_1=1$ and hence $\mathcal M=0$. Equivalently, introducing the total negative Wigner volume
\begin{equation}
    \mathcal N_W
    =
    -\int_{W<0}dx\,dp\;W(x,p;\theta,c),
\label{eq:negative_volume_main}
\end{equation}
one has
\begin{equation}
    \|W\|_1
    =
    1+2\mathcal N_W,
    \quad
    \mathcal M
    =
    \ln(1+2\mathcal N_W).
\label{eq:mana_negative_volume_main}
\end{equation}
Thus $\mathcal M$ provides a global measure of the negative phase-space volume generated by the gate. Throughout this section $W_0(x,p;\theta,c)$ denotes the evolved vacuum Wigner function of Eq.~\eqref{eq:W0_explicit}, and we write
\begin{equation}
    \Wvac(x,p)
    =
    W_0(x,p;0,c)
    =
    \frac{1}{\pi}e^{-x^2-p^2},
\label{eq:Wvac_def}
\end{equation}
for the unevolved Gaussian.

Equation~\eqref{eq:W0_explicit} makes clear that the action of the cosine gate is qualitatively different from that of a finite-order polynomial phase operation. Each Bessel order contributes a Gaussian component displaced to $p=kc/2$, and their coherent interference generates the oscillatory phase-space structures and negative regions visible in Fig.~\ref{fig:W0_c_2}. The dependence of the resulting negativity on the gate strength is nontrivial, and two distinct questions arise. First, how does negativity appear when an arbitrarily weak nonlinear gate is applied to an initially Gaussian state? Second, once the phase $\theta\cos(cx)$ becomes rapidly oscillatory, what determines the asymptotic growth of the integrated negativity?

The first question does not have a single answer, because the weak-angle limit is not uniform in the spatial frequency. The $k=\pm1$ terms in Eq.~\eqref{eq:W0_explicit} carry Bessel weight $J_{\pm1}(2\theta\sin cx)=O(\theta)$ but are displaced to $p=\pm c/2$, where the $k=0$ Gaussian has already decayed to $e^{-c^2/4}$. The weak-angle behavior is therefore organized by the size of $|\theta|$ relative to $\zeta_0(c)e^{-c^2/4}$, where $\zeta_0(c)$ is defined in Sec.~\ref{sec:weak_gate_negativity} and tends to unity at large $c$. Two regimes result. They share the weak-angle derivation of Sec.~\ref{sec:weak_gate_negativity}, but the mechanisms they describe are different and we keep them distinct throughout.

\textbf{Tail nucleation} ($|\theta|\ll\zeta_0(c)e^{-c^2/4}$): the displaced components are subdominant everywhere in the bulk, and the first sign change of the Wigner function is pushed out to $|p|\sim c^{-1}\ln(1/|\theta|)$. The resulting mana is nonperturbative in the gate strength and beyond all algebraic orders in $\theta$.

\textbf{Resolved-displacement linear response} ($\zeta_0(c)e^{-c^2/4}\ll|\theta|\ll1$): the displaced components dominate the undisplaced Gaussian on a region of nonzero measure, the negative lobes sit in the bulk, and $\mathcal M\simeq\kappa(c)|\theta|$ grows linearly.

The second regime is a window, not an expansion about $\theta=0$. At any fixed finite $c$ the ratio $|\theta|e^{c^2/4}/\zeta_0(c)$ tends to zero as $\theta\rightarrow0$, so the ultimate small-angle asymptotics is always the beyond-all-orders behavior, and the linear law is not the leading term of a Taylor series for $\mathcal M$. The width of that window grows with $c$. From Table~\ref{tab:mana_constants}, the crossover $\zeta_0(c)e^{-c^2/4}$ exceeds unity at $c=0.5$ and $c=1$, so the entire range $|\theta|<1$ lies in the tail-nucleation regime, whereas at $c=10$ the crossover is $1.4\times10^{-11}$ and every accessible gate strength lies in the linear-response window. The two are the noncommuting limits recorded in Eq.~\eqref{eq:noncommuting_main}.

In the opposite limit the Wigner integral becomes rapidly oscillatory and admits a semiclassical stationary-phase description. When regular nondegenerate stationary regions dominate, Sec.~\ref{sec:strong_gate_negativity} shows that
\begin{equation}
    \mathcal M
    =
    \frac{1}{2}\ln|\theta|+O(1).
\label{eq:mana_strong_overview}
\end{equation}

\begin{figure*}[t]
    \centering
    \includegraphics[width=0.98\linewidth]{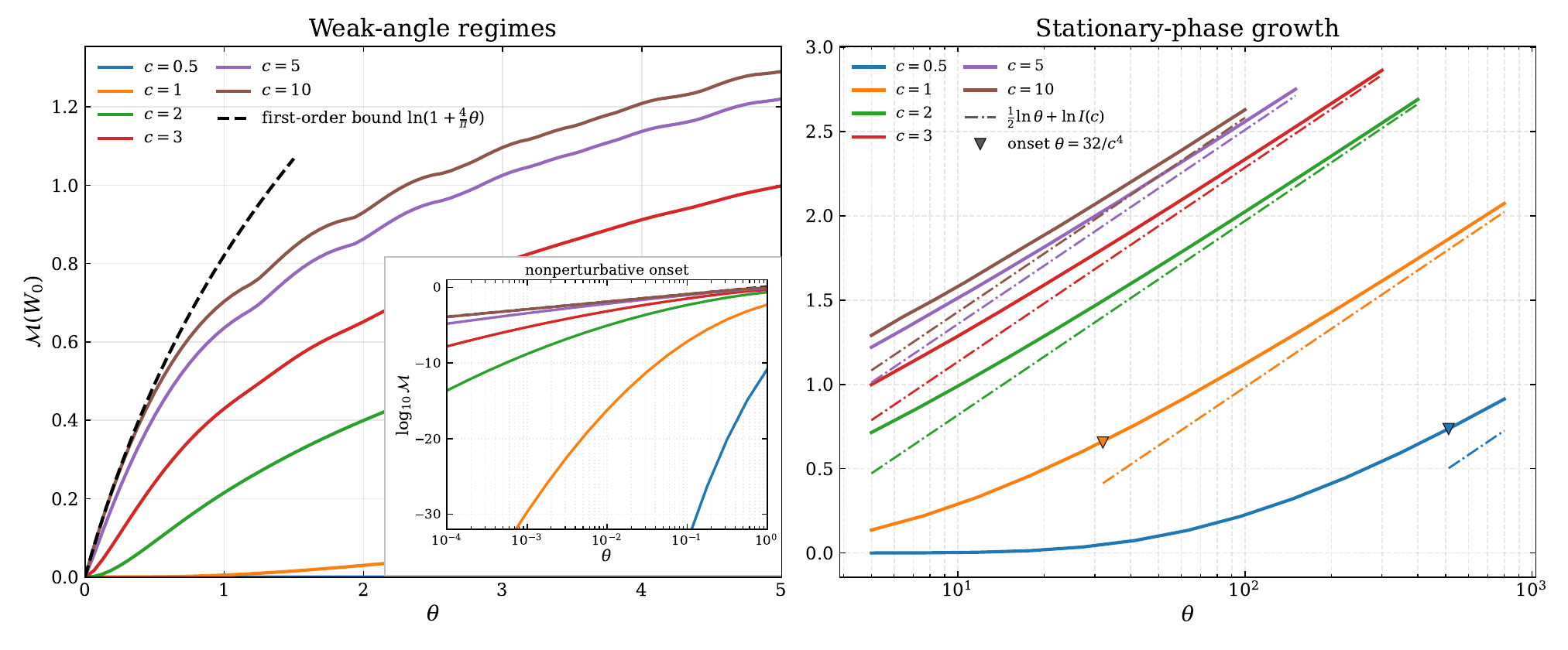}
        \caption{Mana $\mathcal M$ of the Fock vacuum evolved under the cosine gate, for $c=0.5,1,2,3,5,10$, obtained from Eq.~\eqref{eq:W0_explicit} as described in Appendix~\ref{app:weak_numerics}. Left: the two weak-angle regimes, with the first-order bound Eq.~\eqref{eq:mana_bound_main} at its maximal coefficient $\kappa(c)\rightarrow4/\pi$ dashed. The inset extends to $|\theta|=10^{-4}$ on logarithmic axes, where the crossover $|\theta|=\zeta_0(c)e^{-c^2/4}$ of Eq.~\eqref{eq:mana_weak_main} separates the two. Right: stationary-phase growth, with the asymptotes $\tfrac12\ln|\theta|+\ln I(c)$ of Eq.~\eqref{eq:cos_large_mana_main} dash-dotted where Eq.~\eqref{eq:strong_validity_main} holds, and triangles at the onset $|\theta|=32/c^4$.}
    \label{fig:W0_Mana}
\end{figure*}

The three regimes arise from physically different phase-space mechanisms. The spatial frequency $c$ plays an important role in all of them: it controls the separation of the momentum-displaced components in Eq.~\eqref{eq:W0_explicit}, the location of the first negative regions and the crossover between tail nucleation and linear response, and the onset of the semiclassical regime at strong coupling. Figure~\ref{fig:W0_Mana} summarizes the resulting behavior and compares exact numerical evaluation of Eq.~\eqref{eq:W0_explicit} with the asymptotic results derived below.
\begin{table*}[t]
\centering
        \caption{Constants entering the asymptotic results of this section, for the spatial frequencies of Fig.~\ref{fig:W0_Mana}. Here $\zeta_0$ is the smallest positive zero of $\Hc$, $A_c=-c\,\zeta_0\Hc'(\zeta_0)$, $C(c)$ is defined in Eq.~\eqref{eq:Cc_main} and $\kappa(c)$ in Eq.~\eqref{eq:mana_linear_main}. The last three columns bound the three regimes in $\theta$: the crossover $\zeta_0e^{-c^2/4}$ of Eq.~\eqref{eq:crossover_appendix}, the onset $32/c^4$ of Eq.~\eqref{eq:strong_validity_main}, and the offset $I(c)$ of Eq.~\eqref{eq:cos_large_mana_main}.}
\label{tab:mana_constants}
\begin{ruledtabular}
\begin{tabular}{cccccccc}
$c$ & $\zeta_0(c)$ & $A_c$ & $C(c)$ & $\kappa(c)$ & $\zeta_0e^{-c^2/4}$ & $32/c^4$ & $I(c)$\\
\colrule
$0.5$ & $4.3247$   & $0.0067$ & $2.8\times10^{-7}$ & $0.150$ & $4.1$  & $5.1\times10^{2}$ & $0.074$\\
$1$   & $1.7417$   & $0.4193$ & $2.0\times10^{-2}$ & $0.499$ & $1.4$  & $32$ & $0.268$\\
$2$   & $1.0781$   & $1.8448$ & $4.0\times10^{-1}$ & $1.060$ & $4.0\times10^{-1}$  & $2.0$ & $0.716$\\
$3$   & $1.0056$   & $2.9832$ & $8.2\times10^{-1}$ & $1.230$ & $1.1\times10^{-1}$  & $4.0\times10^{-1}$ & $0.980$\\
$5$   & $1.0$  & $5.0$ & $1.78$             & $1.273$ & $1.9\times10^{-3}$  & $5.1\times10^{-2}$ & $1.224$\\
$10$  & $1.0$  & $10.0$ & $5.03$             & $1.273$ & $1.4\times10^{-11}$ & $3.2\times10^{-3}$ & $1.313$\\
\end{tabular}
\end{ruledtabular}
\end{table*}
\subsection{Weak-angle asymptotics: tail nucleation and linear response}
\label{sec:weak_gate_negativity}
It is convenient to extract the overall Gaussian from Eq.~\eqref{eq:W0_explicit} using $e^{-(p-kc/2)^2}=e^{-p^2}e^{kcp-k^2c^2/4}$, giving
\begin{equation}
W_0(x,p;\theta,c)
=
\Wvac(x,p)\,G(x,p;\theta,c),
\label{eq:W_exact_weak}
\end{equation}
with the sign function
\begin{equation}
G(x,p;\theta,c)
=
\sum_{k\in\mathbb{Z}}
J_k\!\left(2\theta\sin(cx)\right)
e^{kcp-k^2c^2/4}.
\label{eq:Gdef_main}
\end{equation}
Since $\Wvac>0$, the negative regions of the Wigner function are exactly the regions where $G<0$. At every fixed point $(x,p)$ the expansion
\begin{equation}
W_0(x,p;\theta,c)
=
\Wvac(x,p)
+
\theta\,W^{(1)}(x,p)
+
O(\theta^2),
\label{eq:W_local_expansion}
\end{equation}
is an ordinary power series, with
\begin{equation}
W^{(1)}(x,p)
=
\frac{e^{-x^2}}{\pi}\sin(cx)
\left[
e^{-(p-c/2)^2}-e^{-(p+c/2)^2}
\right].
\label{eq:W1_main}
\end{equation}
On any fixed compact set the Wigner function is therefore positive for sufficiently small $\theta$, and since $\int dx\,dp\,W^{(1)}=0$ there is no ordinary $O(\theta)$ contribution to the $L_1$ norm from any fixed region of phase space. The relevant question is instead where $\theta W^{(1)}$ overwhelms $\Wvac$. From Eqs.~\eqref{eq:W_local_expansion}--\eqref{eq:W1_main} this happens once the natural tail variable
\begin{equation}
\zeta
=
|\theta\sin(cx)|\,
e^{c|p|-c^2/4},
\label{eq:tail_variable_main}
\end{equation}
is of order unity. Taking $|\theta|\rightarrow0$ and $|p|\rightarrow\infty$ with $\zeta$ fixed resums all relevant Bessel orders and gives
\begin{equation}
G
\sim
\Hc(\zeta),
\,\,\,
\Hc(\zeta)
\equiv
\sum_{k=0}^{\infty}
\frac{(-\zeta)^{k}}{k!}
e^{-\frac{c^2}{4}k(k-1)} .
\label{eq:Hc_main}
\end{equation}
The derivation is given in Appendix~\ref{app:weak_theta_negativity}. Equation~\eqref{eq:Hc_main} resums infinitely many Bessel orders even though $\theta$ is arbitrarily small. Keeping only $J_0$ and $J_{\pm1}$ suffices in the linear-response regime alone. Let $\zeta_0(c)$ denote the smallest positive zero of $\Hc$. The first sign change of the Wigner function then occurs at
\begin{align}
p_0(x)
&=
\frac{1}{c}
\left[
\ln\!\left(
\frac{\zeta_0(c)}{|\theta\sin(cx)|}
\right)
+
\frac{c^2}{4}
\right]
\nonumber\\
&=\mathcal P-\frac{1}{c}\ln|\sin(cx)| ,
\label{eq:p0_main}
\end{align}
with the $x$-independent term
\begin{equation}
\mathcal P(\theta,c)
=
\frac{1}{c}
\left[
\ln\!\left(\frac{\zeta_0(c)}{|\theta|}\right)
+\frac{c^2}{4}
\right] .
\label{eq:P_main}
\end{equation}

At fixed $c$ one has $|p_0|\sim c^{-1}\ln(1/|\theta|)$ as $\theta\rightarrow0$: the first negative regions do not develop perturbatively in the high-probability Gaussian core, but nucleate in increasingly remote momentum tails as the gate strength is reduced.

The integrated negativity is controlled by the competition between the growth of $-\Hc$ beyond its first zero and the Gaussian factor $e^{-p^2}$. Writing $|p|=p_0(x)+u$ and using $-\Hc(\zeta_0e^{cu})\simeq(A_c/c)(e^{cu}-1)$ with
\begin{equation}
A_c=-c\,\zeta_0(c)\,\Hc'\!\left[\zeta_0(c)\right]>0 ,
\label{eq:Ac_main}
\end{equation}
the momentum integral can be done in closed form, and the negative Wigner volume becomes
\begin{equation}
e^{\mathcal M}
=
1+
\frac{2A_c}{\pi c}
\int_{-\infty}^{\infty}dx\,
e^{-x^2}\,
\mathcal F\!\left[p_0(x),c\right]
\left[1+O(\mathcal P^{-1})\right],
\label{eq:mana_uniform_main}
\end{equation}
with $\mathcal F$ given in Eq.~\eqref{eq:Ffun_def}. This single quadrature is uniformly valid across the weak-angle region and reproduces both regimes introduced in Sec.~\ref{sec:wigner_mana}. Its structure is transparent: the Gaussian-weighted integral defining $\mathcal F$ has a saddle at
\begin{equation}
u_*=\frac{c}{2}-p_0 ,
\label{eq:saddle_main}
\end{equation}
which is pinned at the edge of the negative lobe when $p_0>c/2$ and detaches from it when $p_0<c/2$. Since $p_0(x)\ge\mathcal P$, with equality at the maxima of $|\sin(cx)|$, the crossover between the two regimes is at $c=2\mathcal P$, i.e.\ at $|\theta|=\zeta_0(c)e^{-c^2/4}$.

\textbf{Tail-nucleation regime ($c<2\mathcal P$)}: Here $\mathcal F\rightarrow(c/4)e^{-p_0^2}/p_0^2$, and Laplace evaluation of the $x$ integral about the maxima of $|\sin(cx)|$ gives
\begin{equation}
\mathcal M
\sim
C(c)\,
\mathcal P^{-5/2}e^{-\mathcal P^{2}},
\label{eq:mana_weak_main}
\end{equation}
with
\begin{equation}
C(c)
=
-\frac{\sqrt{c}}{2\sqrt{\pi}}\,
\zeta_0(c)\Hc'\!\left[\zeta_0(c)\right]\,
\vartheta_2\!\left(0,e^{-\pi^2/c^2}\right)
>0 ,
\label{eq:Cc_main}
\end{equation}
where $\vartheta_2$ is the second Jacobi theta function. The relative error is $O(\mathcal P^{-1})$, only logarithmically small, so Eq.~\eqref{eq:mana_weak_main} fixes the exponent but not the prefactor. Appendix~\ref{app:regime_I} quantifies the correction and resums its leading part in Eq.~\eqref{eq:mana_refined_app}. Equation~\eqref{eq:mana_weak_main} implies that at fixed $c$
\begin{equation}
\mathcal M
=
o(|\theta|^n)
\quad
\text{for every fixed } n>0 ,
\label{eq:beyond_all_orders_main}
\end{equation}
so that in this regime no finite-order perturbative expansion about $\theta=0$ can capture the onset of Wigner negativity. The suppression is log-Gaussian, $\ln\mathcal M=-\ln^2(1/|\theta|)/c^2+O[\ln(1/|\theta|)]$.

\textbf{Linear-response regime ($c>2\mathcal P$)}: Here the saddle Eq.~\eqref{eq:saddle_main} lies strictly inside the negative lobe, $\mathcal F\rightarrow\sqrt\pi|\theta\sin(cx)|/\zeta_0$, and Eq.~\eqref{eq:mana_uniform_main} collapses to ordinary linear response,
\begin{equation}
\mathcal M
\simeq
\kappa(c)\,|\theta| ,
\quad
\kappa(c)
=
\|W^{(1)}\|_1 .
\label{eq:mana_linear_main}
\end{equation}
The coefficient follows in closed form from Eq.~\eqref{eq:W1_main} and is given in Eq.~\eqref{eq:kappa_def}. It increases monotonically with $c$ towards $4/\pi$, see Table~\ref{tab:mana_constants}. Because $\Wvac>0$ everywhere and $\int(W_0-\Wvac)=0$, the negative part of $W_0-\Wvac$ carries half of its $L_1$ norm and dominates that of $W_0$, so $\mathcal N_W\le\tfrac12\|W_0-\Wvac\|_1$. Since $\mathcal M=\ln(1+2\mathcal N_W)$ increases with $\mathcal N_W$, this gives, for every $c$,
\begin{equation}
\mathcal M(\theta,c)
\le
\ln\!\left[1+\kappa(c)\,|\theta|\right]
=
\kappa(c)\,|\theta|+O(\theta^2).
\label{eq:mana_bound_main}
\end{equation}

Equation~\eqref{eq:mana_linear_main} shows that this bound is attained, so $\kappa(c)|\theta|$ is the maximal Wigner negativity obtainable at first order in the gate strength, and Eq.~\eqref{eq:mana_weak_main} measures how far below that maximum the cosine gate operates when $c$ is small. The two regimes give a precise, and non-monotonic, answer to how $c$ controls the accessible resource. At fixed $|\theta|$, increasing $c$ reduces the phase-space distance $\mathcal P$ to the first negative lobe and produces an extremely rapid, log-Gaussian growth of $\mathcal M$, but this growth terminates at $c\simeq c_\times=2\sqrt{\ln(\zeta_0(c)/|\theta|)}$, beyond which the mana saturates at the linear-response value $\kappa(c)|\theta|\rightarrow4|\theta|/\pi$. Eq.~\eqref{eq:mana_weak_main} is not valid past this point: taken literally it is non-monotonic in $c$, since $\mathcal P=c^{-1}\ln(\zeta_0/|\theta|)+c/4$ passes through a minimum, and at $c=10$ it underestimates the true mana by more than two orders of magnitude. Both behaviors are visible in the inset of Fig.~\ref{fig:W0_Mana}.

For any $c\neq0$ and $\theta\neq0$ the state $\ket{\psi_\theta}=e^{-i\theta\cos(c\hat x)}\ket{0}$ is pure and non-Gaussian and must therefore possess a Wigner-negative region. In the tail-nucleation regime this is reconciled with the pointwise positivity of Eq.~\eqref{eq:W_local_expansion} by the fact that the negative region escapes to $|p|\rightarrow\infty$ as $\theta\rightarrow0$: the limits $|\theta|\rightarrow0$ and $|p|\rightarrow\infty$ do not commute. In the linear-response regime no such subtlety arises and the negativity is visible already at first order. Correspondingly, the limits $|\theta|\rightarrow0$ and $c\rightarrow\infty$ do not commute either,
\begin{equation}
\lim_{c\rightarrow\infty}
\lim_{\theta\rightarrow0}
\frac{\mathcal M}{|\theta|}
=
0,
\quad
\lim_{\theta\rightarrow0}
\lim_{c\rightarrow\infty}
\frac{\mathcal M}{|\theta|}
=
\frac{4}{\pi}.
\label{eq:noncommuting_main}
\end{equation}
\subsection{Stationary-phase growth at strong gate strength}
\label{sec:strong_gate_negativity}
We next consider the opposite regime, where the phase imprinted by the trigonometric gate becomes rapidly oscillatory. In this limit the Wigner function admits a semiclassical description in terms of stationary-phase trajectories. The resulting behavior is qualitatively different from the two weak-angle mechanisms above. For a general single-coordinate phase gate
\begin{equation}
U(\theta)
=
e^{-i\theta f(\hat x)},
\label{eq:general_phase_gate_main}
\end{equation}
acting on a pure Gaussian input, rescaling the momentum as $p\sim\theta v$ writes the Wigner function as an oscillatory integral $K(x)\int dy\,A(y)e^{i\theta\Phi}$ over smooth Gaussian envelopes $K$ and $A$, with rapidly varying phase
\begin{equation}
\Phi(y;x,v)
=
vy+
f\left(x+\frac{y}{2}\right)
-
f\left(x-\frac{y}{2}\right),
\label{eq:phase_stationary_main}
\end{equation}
and stationary points at $f'(x+y/2)+f'(x-y/2)=-2v$. In a regular phase-space region where the relevant stationary points are isolated and nondegenerate, $\partial_y^2\Phi(y_*;x,v)\neq0$, standard stationary-phase theory gives $W=O(|\theta|^{-1/2})$. At the same time, the momentum rescaling implies $dx\,dp=|\theta|\,dx\,dv$, so that a regular stationary region contributes $\int dx\,dp\,|W|\sim|\theta|^{1/2}$ to the Wigner $L_1$ norm. Provided such regions occupy nonzero measure and their leading phase-averaged contribution does not vanish, one obtains
\begin{equation}
\mathcal M(\theta)
=
\frac{1}{2}\ln|\theta|
+
O(1).
\label{eq:mana_large_theta_main}
\end{equation}
The full stationary-phase derivation, including multiple stationary branches and the conditions required for Eq.~\eqref{eq:mana_large_theta_main}, is given in Appendix~\ref{app:large_theta}. Degenerate stationary points require separate treatment. In particular, the phase generically develops fold caustics at which $\partial_y\Phi=\partial_y^2\Phi=0$. Near such points the ordinary stationary-phase approximation is replaced by a uniform Airy approximation. Although the Wigner amplitude is then enhanced locally to $O(|\theta|^{-1/3})$, the corresponding caustic layer has width $O(|\theta|^{-2/3})$, so that its integrated contribution to the Wigner $L_1$ norm is only $O(1)$. Generic fold caustics are subleading compared with the $O(\sqrt{|\theta|})$ contribution from regular stationary regions. Higher degeneracies can lead to different local scaling and must be analyzed separately. For the cosine gate, $f(x)=\cos(cx)$, the phase becomes
\begin{equation}
\Phi(y;x,v)
=
vy
-
2\sin(cx)
\sin\left(\frac{cy}{2}\right),
\label{eq:cos_phase_main}
\end{equation}
with stationary points at $v=c\sin(cx)\cos(cy/2)$ and curvature $\partial_y^2\Phi=\tfrac{c^2}{2}\sin(cx)\sin(cy/2)$. The cosine phase possesses extended regions of regular nondegenerate stationary points for any $c\neq0$, and its caustics, at $y_n=2\pi n/c$, are generically of fold type except at the isolated values of $x$ for which $\sin(cx)=0$. The Gaussian envelope of the initial state exponentially suppresses distant stationary branches, so that the stationary-phase expansion remains controlled. The meaning of ``large gate strength'' must be made quantitative, since $|\theta|\gg1$ is not by itself the relevant condition. Writing
\begin{equation}
\theta\cos(cx)
=
\theta
\left[
1-\frac{c^2x^2}{2}+\frac{c^4x^4}{24}-\cdots
\right],
\end{equation}
the constant and quadratic terms generate a Gaussian unitary and contribute no Wigner negativity at all. The leading non-Gaussian phase is $\theta c^4x^4/24$, and the semiclassical regime requires it to be large over the support of the vacuum. With $\langle x^4\rangle=3/4$ for $\Wvac$, this gives
\begin{equation}
|\theta|\gg\frac{32}{c^4},
\label{eq:strong_validity_main}
\end{equation}
a far more stringent condition than $|\theta|\gg1$ at small spatial frequency: $|\theta|\gtrsim2$ suffices at $c=2$, but $|\theta|\gtrsim5\times10^2$ is required at $c=0.5$. This is confirmed by the numerical results in the right panel of Fig.~\ref{fig:W0_Mana}. Consequently, in this regime the cosine gate exhibits $e^{\mathcal M}\sim I(c)\sqrt{|\theta|}$, where $I(c)$ is a nonuniversal coefficient determined by the Gaussian input and by interference among the contributing stationary branches. It is evaluated in Appendix~\ref{app:large_theta} and listed in Table~\ref{tab:mana_constants}, saturating slowly as $c$ increases. Equivalently,
\begin{equation}
\mathcal M
=
\frac{1}{2}\ln|\theta|
+
\ln I(c)
+
o(1).
\label{eq:cos_large_mana_main}
\end{equation}
The logarithmic growth in Eq.~\eqref{eq:mana_large_theta_main} is not a consequence of unitarity alone, nor an upper bound on arbitrary single-mode non-Gaussian operations: a quadratic $f$ generates a Gaussian unitary, for which $\mathcal M=0$ identically. It characterizes the semiclassical regime in which regular stationary-phase regions dominate the Wigner integral for a nonlinear single-coordinate phase gate, and Appendix~\ref{app:large_theta} states the conditions precisely.
\section{Trapped-ion realization}\label{sec:2}
\subsection{QSCOUT bosonic modes and native interactions}\label{sec:qscout_modes}
In this section, we detail the experimental configuration and the specific parameters of the trapped-ion platform used to implement the trigonometric gates. The experiments were performed in a small register of trapped \Yb\ ions in the Quantum Scientific Computing Open User Testbed at Sandia National Laboratories~\cite{Clark:2021mxw}. An $N$-ion chain consists of up to $3N$ motional degrees of freedom, $N$ eigenmodes in the axial direction, and $N$ in each of the two orthogonal radial directions, or principal axes. We address these motional modes through motionally sensitive Raman transitions driven in a counter-propagating configuration. In this device, those beams are orthogonal to the axial direction, allowing access only to the two radial manifolds of motional modes. Additionally, the surface electrode trap provides the ability to rotate the radial principal axes~\cite{Revelle:2020xpg}, and so we rotate the axes so that the lower-frequency manifold of radial modes is more strongly coupled to the laser drives. All of the experiments are thus performed on motional eigenmodes within this lower-frequency manifold with either $N=3$ or $N=4$ ions.

The qubit states of \Yb\ correspond to the hyperfine $^{2}\text{S}_{1/2}$ ``clock'' transition with a splitting of 12.642~GHz and are driven by a pulsed 355-nm laser via a Raman transition detuned off the excited $^{2}\text{P}_{1/2}$. We choose the qubits as the following hyperfine states $\ket{\uparrow}\definedas \ket{0}=\ket{{}^{2}\mathrm{S}_{1/2},F=0,m_F=0}$ and $\ket{\downarrow}\definedas\ket{1}=\ket{{}^{2}\mathrm{S}_{1/2},F=1,m_F=0}$. The radial motional modes decorate the qubit transition with first-order red- and blue-sidebands $\sim2\pi \times 2.2$~MHz detuned from the transition. Driving these red- and blue-sideband transitions independently corresponds to the Jaynes--Cummings (JC) and anti-Jaynes--Cummings (AJC) interactions,
\begin{align}
    H_{\rm JC} &= \frac{\eta\Omega}{2}\left(\des\sigma_+ + \cre\sigma_-\right),
    \nonumber\\
    H_{\rm AJC} &= \frac{\eta\Omega}{2}\left(\des\sigma_- + \cre\sigma_+\right),
\label{eq:JCandAJC}
\end{align}
respectively. In the Lamb-Dicke regime, simultaneous red- and blue-sideband drives addressing ion $j$ and motional mode $k$ realize a spin-dependent force Hamiltonian of the form~\cite{Wineland:1997mg,Fluhmann:2019thesis}
\begin{align}
    &\hat H_\text{SDF}^{(j,k)}(t) = \nonumber\\
    &\frac{1}{2} \eta_{j,k}\Omega
    \sigma_{\phi_s}^{(j)}
    \left(
    \hat{a}_k e^{-i(\delta t + \phi_m)} +
    \hat{a}_k^\dagger e^{i(\delta t + \phi_m)}
    \right),
    \label{eq:SDF_ham}
\end{align}
where $\Omega$ is the carrier transition Rabi frequency, and $\eta_{j,k}$ is the ion $j$'s coupling strength to the mode $k$, i.e.\ the Lamb-Dicke factor, and for these experiments we set the sideband Rabi frequency for any given ion $j$ and participating mode $k$ to be $\eta_{j,k}\Omega = 2\pi\times16.66$~kHz. $\delta$ is then the detuning from the motional sideband, and $\phi_s$ and $\phi_m$ are the spin and motional phases. Note that we use natural units where $\hbar = 1$, a convention we maintain throughout the rest of the paper. For near resonant pulses, this interaction results in the conditional displacement $\text{CD}_{{\bm n}\cdot{\bm \sigma}}(\alpha)$, see Eq.~\eqref{eq:gates}, with $\alpha$ determined by the pulse area, detuning, and optical phase. These $\text{CD}_{{\bm n}\cdot{\bm \sigma}}(\alpha)$ gates can then be performed on any selected motional mode, with the magnitude of the displacement proportional to the pulse area and the Lamb-Dicke factor, $\eta_{j,k}$. The phase of the displacement is set by the $\phi_m$ of the interaction, which in practice is set by the difference in phase between the red- and blue-sideband drives.

In these experiments, we use ancillary qubits both for the conditional displacements and for several indirect probes of the motional state, since there are no direct readouts of the motional state available in trapped-ion systems. For clarity, we will refer to the qubit involved in the conditional displacement operations as the ``gate qubit,'' and the ones that probe the motional characteristics as the ``probe qubits.''

One method of readout is to perform a full characteristic function readout through mapping a series of real and imaginary conditional displacements, measuring the qubit state, and extracting $\langle Z\rangle = P_{\ket{\uparrow}} - P_{\ket{\downarrow}}$ for each displacement~\cite{Fluhmann:2019slh,Leibfried:1996zz}. Based on the probe qubit's initial state, $\ket{\uparrow}$ or $\ket{+}$, this will either reveal the real or the imaginary part, respectively, of the characteristic function, $\chi$. This method is resource intensive, but it provides the ability to extract both Fock state occupancies and phase relationships. Alternatively, applying a BSB gate of varying duration to the probe qubit can provide information on the Fock state occupancies~\cite{Meekhof:1996zz}, see Fig.~\ref{fig:qumode_readout_circuit}. Because the coupling of the Fock states increases as $\sqrt{n+1}$ with increasing Fock number $n$, we can extract occupancies from a fit to a series of sinusoids with frequencies scaling as $\sqrt{n+1}$~\cite{Lv:2016obh}, see Section~\ref{sec:5.B} for more details. Both of these methodologies are amenable to postselection, whereby we remove instances in which the gate qubit's state was flipped during the conditional displacements. For all the data presented here, postselection success was typically $80$--$90\%$.

\begin{figure*}[t]
        \centering
\includegraphics[width=0.95\columnwidth]{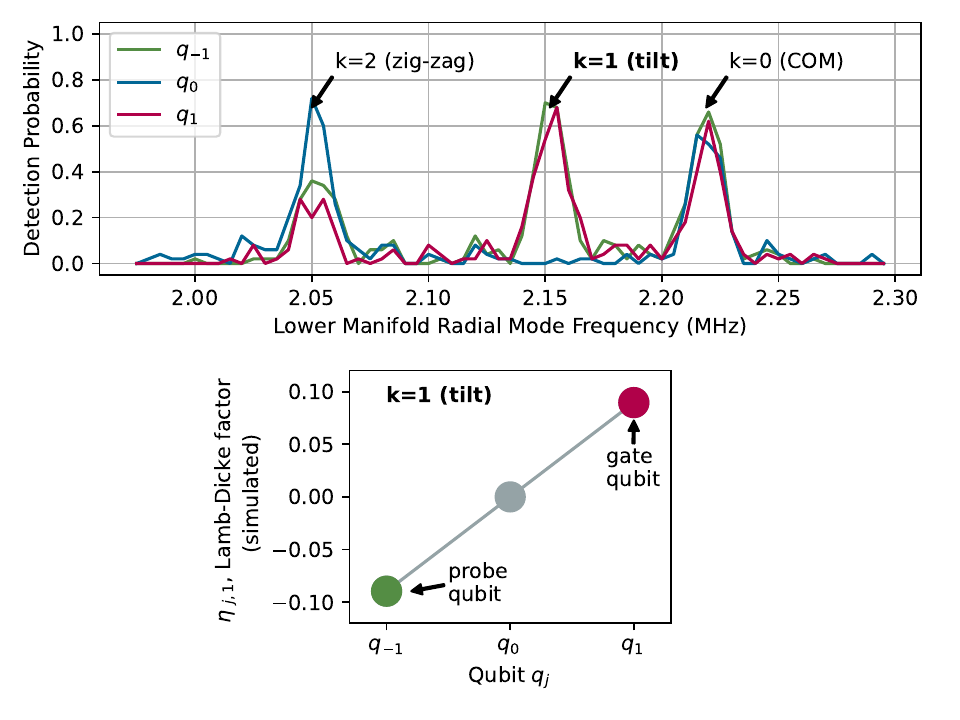}
\hspace*{.5cm}
\includegraphics[width=0.95\columnwidth]{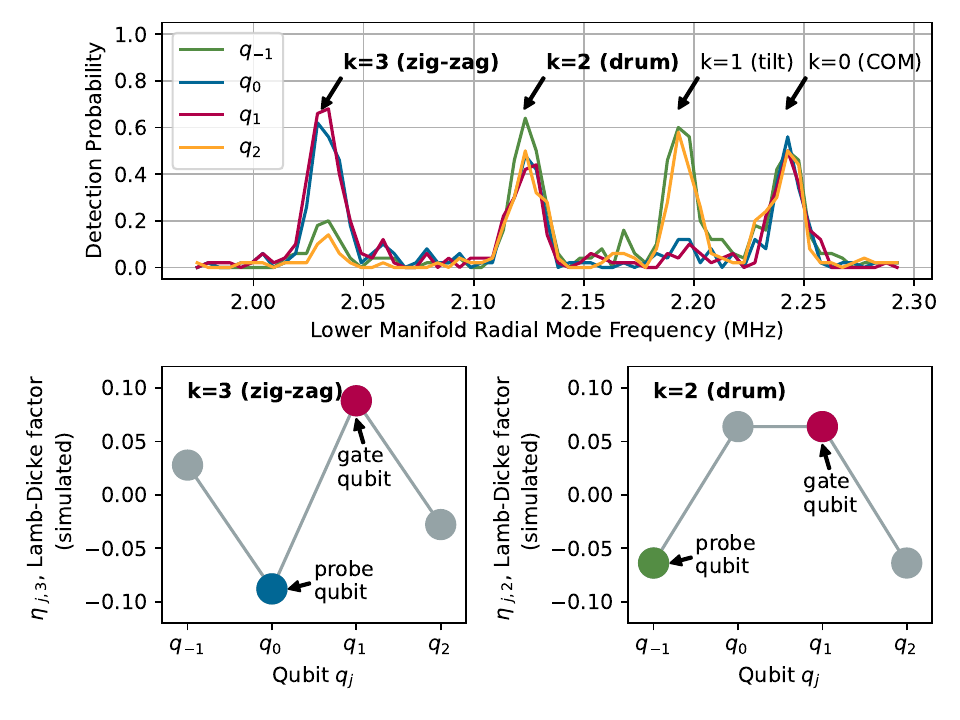}
        \caption{Trapped ion radial-mode spectroscopy and mode participation. Top row: measured detection probabilities for each qubit as a function of the lower-manifold radial-mode frequency for three-ion (left) and four-ion (right) chains. The resonances identify the collective radial motional modes, labeled by mode index and character. Bottom row: simulated Lamb-Dicke factors $\eta_{j,k}$ for the modes used in the one-qumode (three-ion) experiment, $k=1$ (tilt), and the two-qumode experiment (four-ion), $k=3$ (zig-zag) and $k=2$ (drum). Colored markers indicate the probe and gate qubits used in each implementation, and the gray markers denote the remaining ions.}
        \label{fig:34-ion_modes}
\end{figure*}

For the experiments described below, we use two different ion registers. For the single-mode case of the trigonometric gates, we use a three-ion chain: the gate qubit is coupled to a particular mode, while the probe qubit coupled to the same mode measures the Fock occupancy of that mode, see Fig.~\ref{fig:34-ion_modes} (left panels). If we assign the mode indices $k$ with the highest-energy mode $k\definedas0$ (center-of-mass), then we operate this particular gate on the $k=1$ (tilt) mode using one of the two outer ions as the gate qubit and the other as the probe qubit. In the case of the two-mode trigonometric gate, we use a four-ion chain, see Fig.~\ref{fig:34-ion_modes} (right panels). Here, we use the $k=2$ (drum) and $k=3$ (zig-zag) modes as the modes of choice. In particular, $k=2$ has equal magnitude participation across all ions, while $k=3$ is more strongly coupled to the two center ions. The gate qubit for the two-mode gate is thus one of the two center ions. For readout, the other center ion acts as the probe qubit for the $k=3$ mode, while one of the outer ions is the probe qubit for the $k=2$ mode.

In both register lengths, all non-center-of-mass modes are sideband cooled to near the vacuum state, with a residual phonon number, $\bar{n}\sim0.1$. The center-of-mass mode is unused due to less efficient sideband cooling, leading to $\bar{n}\sim0.5$. Motional dephasing is the predominant source of error, with measured coherence times ranging from $500$--$700\,\mu{\rm s}$ for various modes. While motional heating rates vary from mode to mode, non-center-of-mass modes have rates $\leq 100~\text{quanta}/s$. For instance, in the four-ion register, the tilt ($k=1$) and zig-zag ($k=3$) modes have rates $\sim10~\text{quanta}/s$, while the drum ($k=2$) mode heats more rapidly, at $\sim100~\text{quanta}/s$.
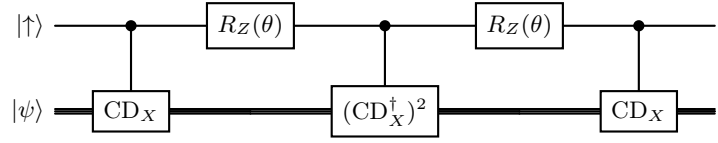
\begin{figure*}[t]
    \centering
\begin{quantikz}
    \lstick{$\ket{\uparrow}$} & \ctrl{1} & \gate{R_Z(\theta)} & \ctrl{1} & \gate{R_Z(\theta)} & \ctrl{1}  & \\
    \lstick{$\ket{\psi}$} \setwiretype{b} & \gate{\text{CD}_X} & \qw & \gate{(\text{CD}_X^\dagger)^2} & \qw & \gate{\text{CD}_X} & \qw
\end{quantikz}
        \caption{Circuit that approximates the action of the trigonometric operator $e^{-i\theta\cos(c\hat{x})}$ on a qumode state $\ket{\psi}$ up to order $\mathcal{O}(\theta^2)$ using one ancillary gate qubit. The argument of the conditional displacement is $\text{CD}_X = \text{CD}_X(ic/(2\sqrt{2}))$. Throughout this work, qubits (qumodes) are shown as single (triple) wires.}
    \label{fig:Trotter1_cos_gate_circuit}
\end{figure*}
\subsection{Finite-step implementation of the cosine gate}
\label{sec:finite_step_implementation}
We now describe how the ideal trigonometric evolution
\begin{equation}
    U_c(\theta)
    =
    e^{-i\theta\cos(c\hat x)}
\label{eq:ideal_cosine_gate_impl}
\end{equation}
is synthesized from the native hybrid qubit--qumode interactions available on the trapped-ion platform. The construction uses conditional displacements of a motional mode together with single-qubit rotations of an ancillary gate qubit. For a bosonic mode with annihilation operator $\hat a$, we use the conditional displacement of Eq.~\eqref{eq:gates} with $\bm n\cdot\bm\sigma=X$ \cite{Sinanan-Singh:2023xzs,Liu:2024ncw,Hong:2025dct,Tortorici:2026dhj},
\begin{equation}
    {\rm CD}_{X}(\alpha)
    =
    \exp\!\left[
        \left(
            \alpha \hat a^\dagger
            -
            \alpha^*\hat a
        \right)\otimes X
    \right],
\label{eq:conditional_displacement_impl}
\end{equation}
where $X$ acts on the gate qubit. For a purely imaginary displacement, $\alpha=i\lambda/\sqrt{2}$, this operation is equivalently
\begin{equation}
    {\rm CD}_{X}
    \!\left(
        \frac{i\lambda}{\sqrt{2}}
    \right)
    =
    e^{i\lambda\hat x\otimes X},
\label{eq:cd_position_impl}
\end{equation}
using
\begin{equation}
    \hat x
    =
    \frac{\hat a+\hat a^\dagger}{\sqrt{2}}.
\end{equation}
Such conditional displacements are generated directly by the spin-dependent-force interaction described in Sec.~\ref{sec:qscout_modes}. We also use rotations of the gate qubit,
\begin{equation}
    R_Z(\theta)
    =
    e^{-i\theta Z/2}.
\label{eq:Rz_impl}
\end{equation}
\begin{figure*}[t]
        \centering
        \includegraphics[width=0.32\linewidth]{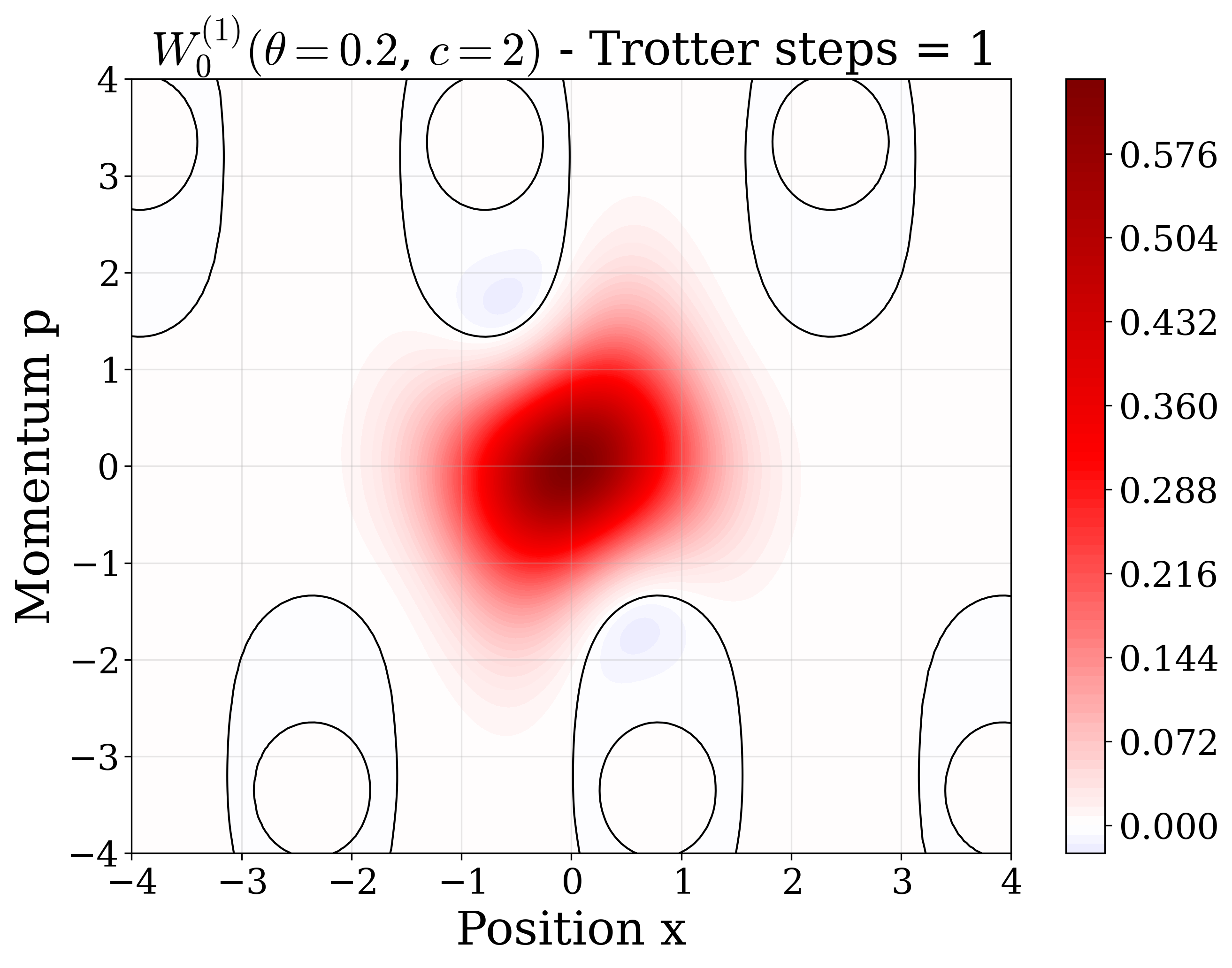}
        \includegraphics[width=0.32\linewidth]{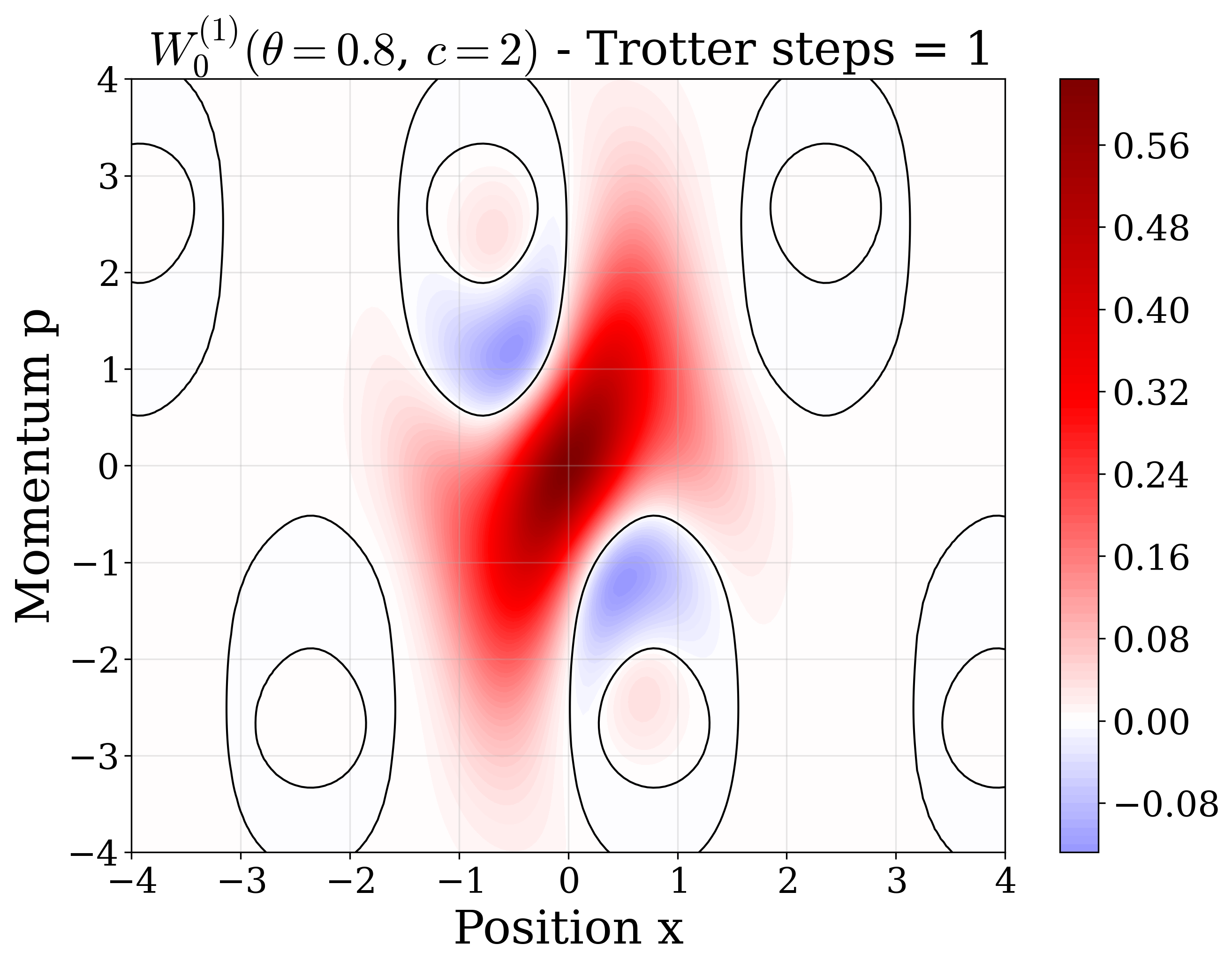}
        \includegraphics[width=0.32\linewidth]{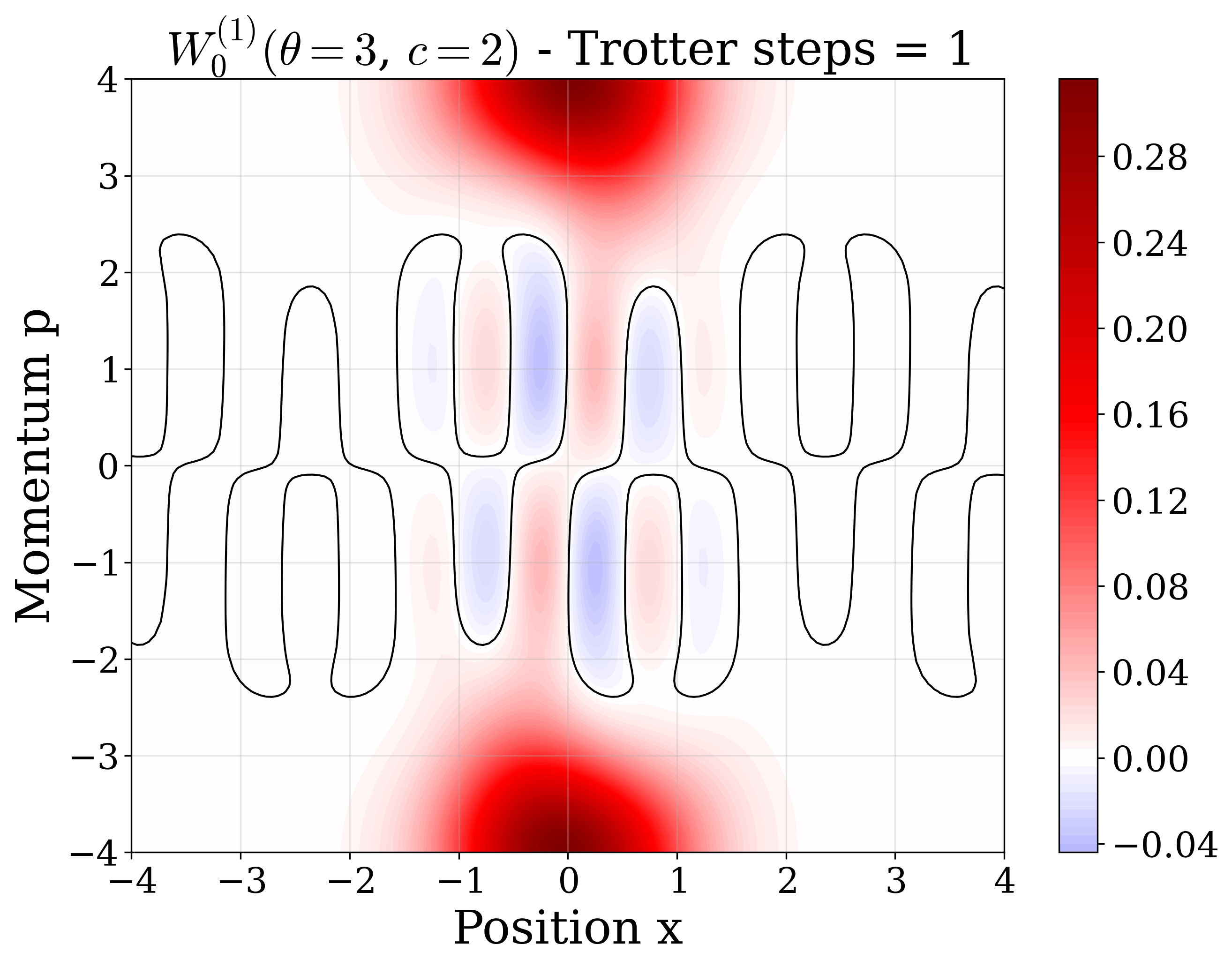}
        \caption{Wigner function of the Fock vacuum $\ket{0}$ evolved for one Trotter step using the circuit in Fig.~\ref{fig:Trotter1_cos_gate_circuit}, after tracing out the qubit. The results are shown for different values of $\theta$ at fixed $c=2$.}
        \label{fig:W0_1_c_2_steps_1}
\end{figure*}
\begin{figure*}[t]
        \centering
        \includegraphics[width=0.32\linewidth]{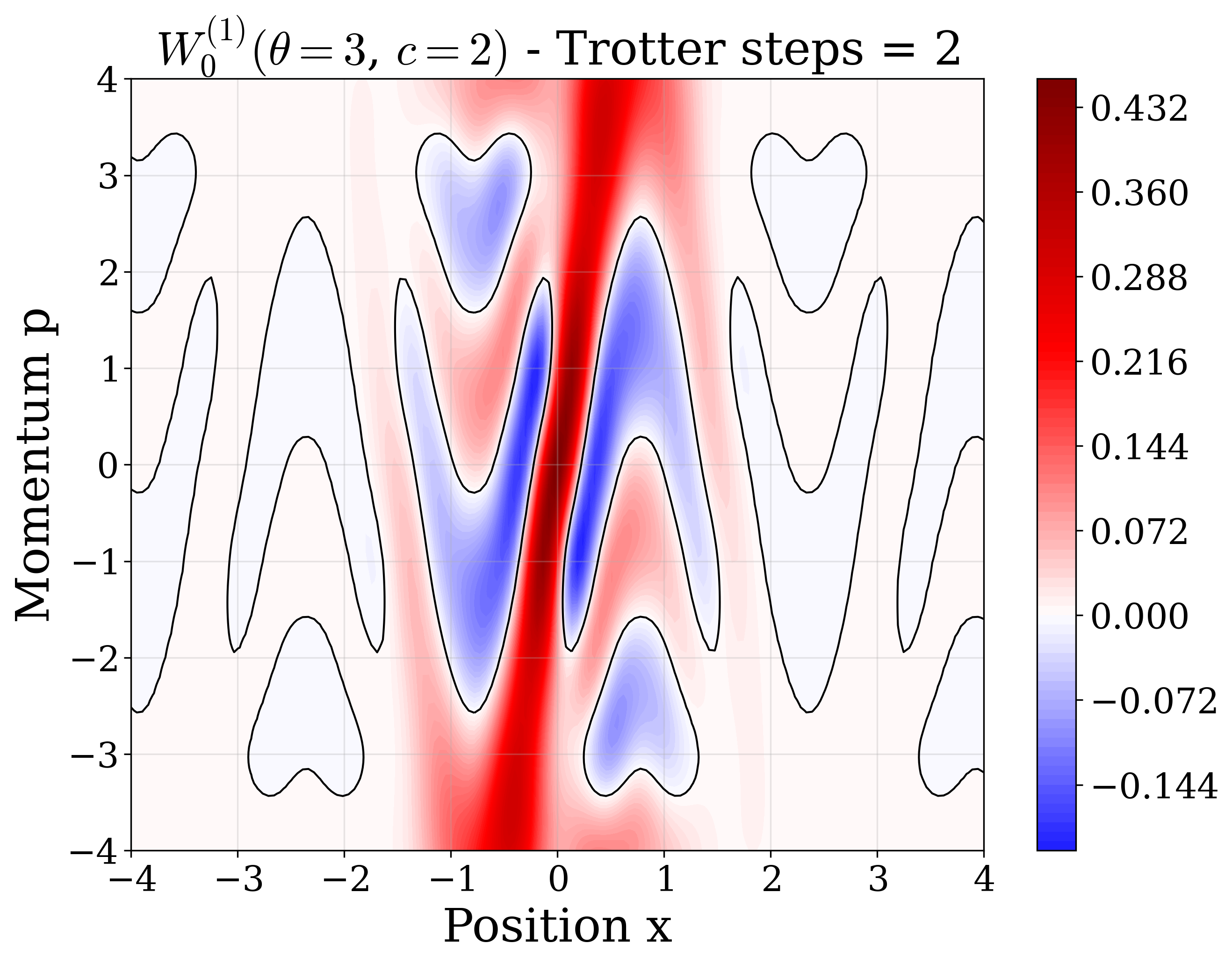}
        \includegraphics[width=0.32\linewidth]{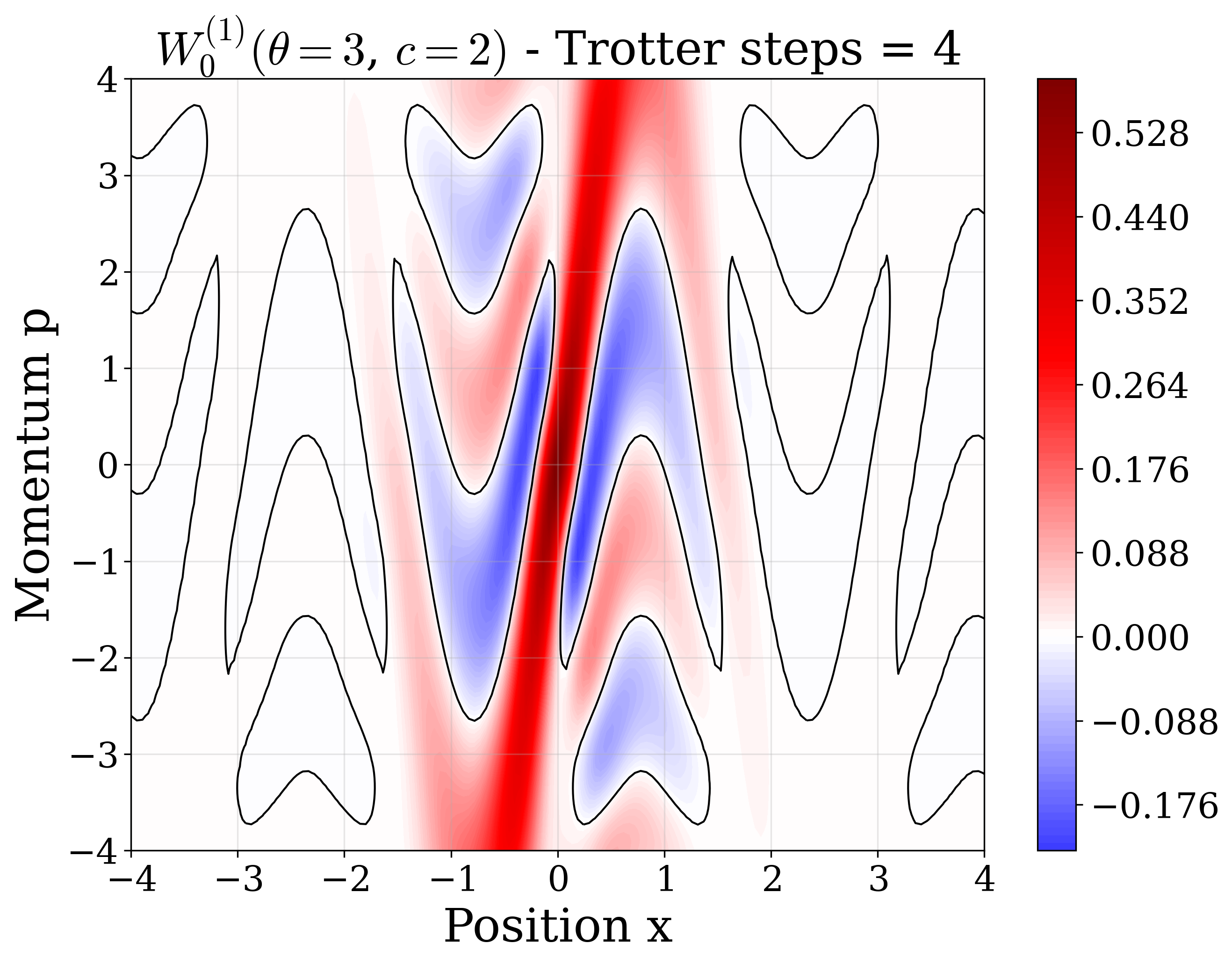}
        \includegraphics[width=0.32\linewidth]{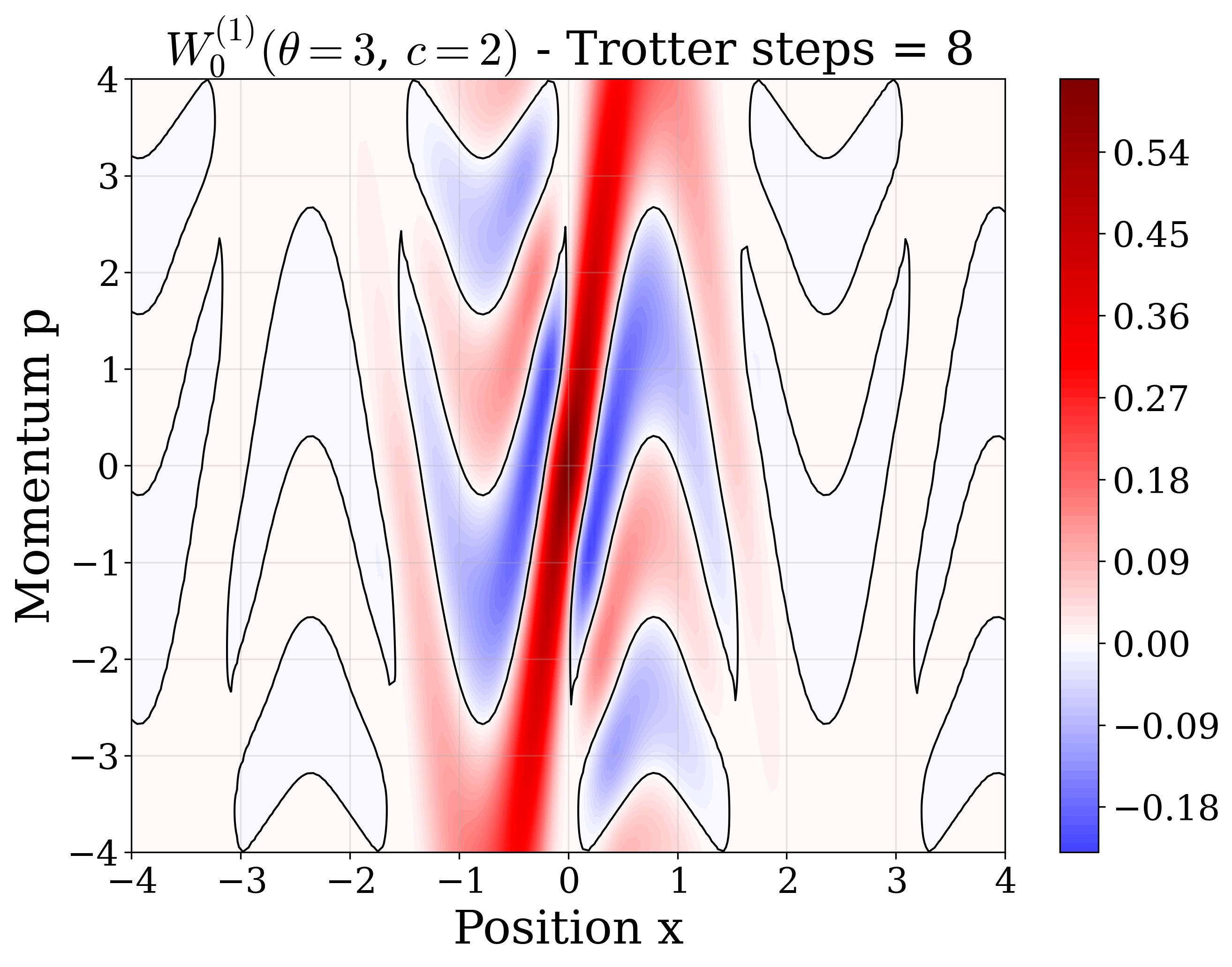}
        \caption{Wigner function of the Fock vacuum $\ket{0}$ evolved for $N\in\{2,4,8\}$ Trotter steps using the circuit in Fig.~\ref{fig:Trotter1_cos_gate_circuit}, after tracing out the qubit. The results are shown for fixed $\theta=3$ and $c=2$.}
        \label{fig:W0_1_c_2_t_3_steps_2_4_8}
\end{figure*}

Using Trotterization, the trigonometric gates can be approximated in terms of elementary CV-DV operations~\cite{Chalermpusitarak:2025cod,Rainaldi:2025ymn}. Within its regime of validity, the resulting circuit reproduces the action of the exact cosine gate. The circuit that approximates the cosine gate up to corrections of order $\mathcal{O}(\theta^2)$ is shown in Fig.~\ref{fig:Trotter1_cos_gate_circuit} using a single ancillary gate qubit. Here, we follow the convention of Ref.~\cite{Chalermpusitarak:2025cod}, which is unitarily equivalent to the construction of Ref.~\cite{Rainaldi:2025ymn}. See Appendix~\ref{app:3} for details. The circuit in Fig.~\ref{fig:Trotter1_cos_gate_circuit} corresponds to the following CV-DV hybrid unitary
\begin{align}
    \hat{\mathcal{C}}_1(\theta,c)\equiv \,&\text{CD}_X\left(i\frac{c}{2\sqrt{2}}\right)\cdot R_Z(\theta)\cdot \text{CD}^\dagger_X\left(i\frac{c}{\sqrt{2}}\right)\nonumber\\
    &\cdot R_Z(\theta)\cdot \text{CD}_X\left(i\frac{c}{2\sqrt{2}}\right)\,,
    \label{eq:op_Trotter1_1mode}
\end{align}
where the subscript denotes a first-order Trotter decomposition. Separating the qubit and qumode degrees of freedom, we can write
\begin{equation}
    \begin{split}
        \hat{\mathcal{C}}_1(\theta,c) = \hat{G}_1(\theta,c)\otimes \mathbb{I} + \hat{G}_2(\theta,c)\otimes Z + \hat{B}(\theta,c)\otimes X,
    \end{split}
    \label{eq:op_Trotter1_1mode_new}
\end{equation}
with
\begin{equation}
    \begin{split}
        \hat{G}_1(\theta,c) &= \cos^2\left(\frac{\theta}{2}\right)-\sin^2\left(\frac{\theta}{2}\right)\cos(2c\hat{x}) \\
        &= 1 + \mathcal{O}(\theta^2),\\
        \hat{G}_2(\theta,c) &= -i\sin(\theta)\cos(c\hat{x}) \\
        &= -i\theta\cos(c\hat{x}) + \mathcal{O}(\theta^2),\\
        \hat{B}(\theta,c) & = -i\sin^2\left(\frac{\theta}{2}\right)\sin(2c\hat{x}) = \mathcal{O}(\theta^2).
    \end{split}
    \label{eq:good_bad_defs}
\end{equation}
The operators $\hat{G}_{1,2}$ contain the desired leading-order contribution, while $\hat{B}$ represents the unwanted component that first appears at order $\mathcal{O}(\theta^2)$. In the following, we refer to $\hat{G}_{1,2}$ and $\hat{B}$ as the ``good'' and ``bad'' components of the circuit, respectively. The approximation can be systematically improved by increasing the number of Trotter steps~\cite{Suzuki:2002zuq,Wiebe:2008cbb}. For example, for two Trotter steps, we obtain
\begin{equation}
\begin{split}
    &\hat{\mathcal{C}}_2(\theta,c)\equiv\hat{\mathcal{C}}_1^2(\theta/2,c) \\
    & = (\hat{G}_1^2 + \hat{G}_2^2 + \hat{B}^2) \otimes \mathbb{I} + 2\hat{G}_1\hat{G}_2 \otimes Z + 2\hat{G}_1\hat{B} \otimes X,
\end{split}
\label{eq:op_Trotter2_1mode}
\end{equation}
where the argument $\theta$ of each bosonic operator is replaced by $\theta\mapsto\theta/2$. After a large number of Trotter steps, the circuit will asymptote to
\begin{align}
    &\hat{\mathcal{C}}_\infty(\theta,c)\equiv \lim_{N\to\infty} \prod_{n=1}^N \hat{\mathcal{C}}_1(\theta/N,c) \nonumber\\
    &= \cos(\theta\cos(c\hat{x}))\otimes\mathbb{I} -i \sin(\theta\cos(c\hat{x}))\otimes Z + 0\otimes X\nonumber \\
    &= e^{-i\theta\cos(c\hat{x})\otimes Z}.
\end{align}
The number of conditional displacement gates $\cdx$ after $N$ Trotter steps is $\#\cdx = 2N+1$. More details about the systematic improvement can be found in Appendix~\ref{app:2}.

The circuit performance can be improved by postselecting on the measurement outcome of the gate qubit: if a qubit flip is observed, the corresponding run is discarded.
\begin{figure*}[t]
    \centering
\begin{quantikz}
    \lstick{$\ket{\uparrow}$} & \ctrl{1}& \ctrl{2} & \gate{R_Z(\theta)} & \ctrl{1} & \ctrl{2} & \gate{R_Z(\theta)} & \ctrl{1} & \ctrl{2} & \\
    \lstick{$\ket{\psi_1}$} \setwiretype{b} & \gate{\text{CD}_X} & \qw & \qw & \gate{\left(\text{CD}^\dagger_X\right)^2} & \qw &\qw & \gate{\text{CD}_X} & \qw & \qw\\
    \lstick{$\ket{\psi_2}$} \setwiretype{b} & \qw & \gate{\text{CD}_X} & \qw & \qw & \gate{\left(\text{CD}^\dagger_X\right)^2} &  \qw & \qw &  \gate{\text{CD}_X} & \qw
\end{quantikz}
        \caption{Circuit that approximates the action of the trigonometric operator $e^{-i\theta\cos(c_1\hat{x}_1 + c_2\hat{x}_2)}$ acting on the two-qumode state $\ket{\psi_1}\otimes\ket{\psi_2}$ up to corrections of order $\mathcal{O}(\theta^2)$. The argument of the sequences conditional displacement is $\text{CD}_X = \text{CD}_X(ic_j/(2\sqrt{2}))$, with $c_j = c_1, c_2$.}
    \label{fig:Trotter1_cos_gate_2modes_circuit}
\end{figure*}
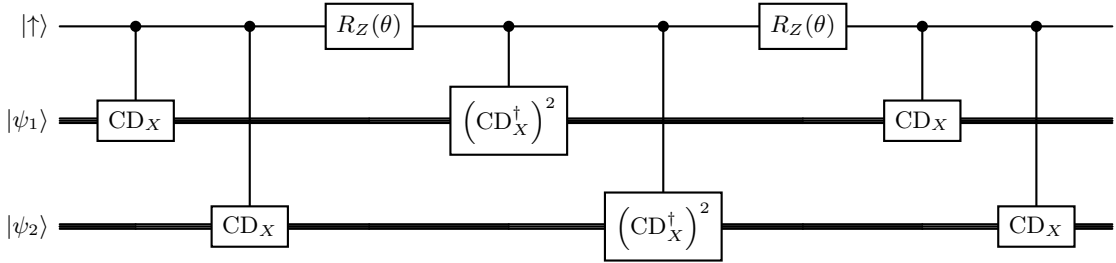

If we focus on the single first-order Trotter step and initialize the qubit to the $\ket{\uparrow}$ state, the resulting state can be written as
\begin{equation}
    \begin{split}
        \hat{\mathcal{C}}_1(\theta,c)\ket{\psi}\ket{\uparrow} & = \hat{G}\ket{\psi}\ket{\uparrow}  + \hat{B}\ket{\psi}\ket{\downarrow}\,,
    \end{split}
    \label{eq:Trotter1_circ_1mode}
\end{equation}
where $\ket{\psi}$ is the initial state of the qumode and $\hat{G} \equiv \hat{G}_1 + \hat{G}_2$. Looking back at the results in Eq.~\eqref{eq:good_bad_defs}, we can write the good and bad components in terms of displacement operators
\begin{equation}
    \begin{split}
        &\hat{G}_1(\theta,c)= \\
        &   \cos^2\left(\frac{\theta}{2}\right)\mathbb{I}-\frac{1}{2}\sin^2\left(\frac{\theta}{2}\right)\left[{\rm D}(i\sqrt{2}c)+{\rm D}^\dagger(i\sqrt{2}c)\right],\\
        &\hat{G}_2(\theta,c) = -\frac{i}{2}\sin(\theta)\left[{\rm D}\left(i\frac{c}{\sqrt{2}}\right)+{\rm D}^\dagger\left(i\frac{c}{\sqrt{2}}\right)\right],\\
        &\hat{B}(\theta,c)  = -\frac{1}{2}\sin^2\left(\frac{\theta}{2}\right)\left[{\rm D}(i\sqrt{2}c)-{\rm D}^\dagger(i\sqrt{2}c)\right].
    \end{split}
    \label{eq:Trotter1_circ_1mode_operatos}
\end{equation}
The density matrix of the oscillator state $\ket{\psi}$ and qubit $q$ system after one Trotter step is
\begin{equation}
\begin{split}
    \hat{\rho}^{[1]}_{\psi,q}(\theta,c)&\equiv \hat{\mathcal{C}}_1(\theta,c)\ket{\psi}\ket{\uparrow}\bra{\uparrow}\bra{\psi}\hat{\mathcal{C}}^\dagger_1(\theta,c) \\
    &= \begin{pmatrix}
        \hat{G}\ket{\psi}\bra{\psi}\hat{G}^\dagger & \hat{G}\ket{\psi}\bra{\psi}\hat{B}^\dagger\\
        \hat{B}\ket{\psi}\bra{\psi}\hat{G}^\dagger & \hat{B}\ket{\psi}\bra{\psi}\hat{B}^\dagger
    \end{pmatrix}.
    \label{eq:den_mat_GB}
\end{split}
\end{equation}
Tracing out the qubit gives the mixed qumode state described by the following density matrix
\begin{equation}
    \hat{\rho}^{[1]}_{\psi}(\theta,c)\equiv \tr_q\left[\hat{\rho}^{[1]}_{\psi,q}\right] = \hat{G}\ket{\psi}\bra{\psi}\hat{G}^\dagger +\hat{B}\ket{\psi}\bra{\psi}\hat{B}^\dagger.
    \label{eq:reduced_rho}
\end{equation}
The Wigner function of this reduced state can be computed as
\begin{equation}\label{eq:W0_Trotter_11}
    W_{\psi}^{[1]}(\alpha;\theta,c) \equiv \frac{2}{\pi} \tr\left[\hat{\rho}^{[1]}_{\psi} (\theta,c) {\rm D}(\alpha) \hat{\Pi} \,{\rm D}^\dagger(\alpha)\right] ,
\end{equation}
with $\alpha = (x+ip)/\sqrt{2}$ and $\hat{\Pi} \equiv (-1)^{\hat{a}^\dagger\hat{a}}$ the parity operator. Note that the superscript $[N]$ labels the number of Trotter steps and should not be confused with the perturbative order $W^{(1)}$ of Eq.~\eqref{eq:W1_main}.

In Fig.~\ref{fig:W0_1_c_2_steps_1}, we illustrate the Wigner function for the case where the state is initialized in the Fock vacuum, $\ket{\psi}=\ket{0}$. The mana of the finite-step approximation is obtained by integrating the absolute value of Eq.~\eqref{eq:W0_Trotter_11}, and approaches the exact result of Sec.~\ref{sec:mana} as the number of steps is increased. In Fig.~\ref{fig:W0_1_c_2_t_3_steps_2_4_8}, we show the Wigner function of the evolved vacuum for $N\in\{2,4,8\}$ Trotter steps at fixed $\theta=3$ and $c=2$. As expected, increasing the number of Trotter steps drives the state toward the exact result shown in the rightmost panel of the lower row of Fig.~\ref{fig:W0_c_2}.

If we initialize the mode to the zero Fock state $\ket{\psi}=\ket{n=0}$, the transition probabilities to other Fock states are
\begin{align}
&P_{0 \to n} = \bra{n} \hat{\rho}^{[1]}_{0}(\theta,c) \ket{n} = |\bra{n} \hat{G} \ket{0}|^2 + |\bra{n} \hat{B} \ket{0}|^2\nonumber\\
&= \begin{cases}
\displaystyle \left| f_0(\theta)\delta_{n,0} - f_1(\theta,n) \right|^2 & n \text{ even}, \\
\displaystyle \sin^4\left(\frac{\theta}{2}\right) \frac{(2c^2)^n}{n!} e^{-2c^2} & n \text{ odd},
\end{cases}
\label{eq:Probs_Trotter1}
\end{align}
with $f_0(\theta) = \cos^2\left(\frac{\theta}{2}\right)$ and
\begin{equation*}
    f_1(\theta,n)= \frac{(ic)^n}{\sqrt{n!}} \left( \sin^2\left(\frac{\theta}{2}\right)  \frac{e^{-c^2}}{2^{-\frac{n}{2}}} + i\sin(\theta) \frac{e^{-\frac{c^2}{4}}}{2^{\frac{n}{2}}} \right).
\end{equation*}

The structure of Eq.~\eqref{eq:Probs_Trotter1} reflects a useful parity property of the first-order circuit: $\hat{G}$ connects the vacuum only to even Fock states, while $\hat{B}$ connects it only to odd Fock states. As a result, postselecting on the qubit state $\ket{\uparrow}$ leaves the even transition probabilities unchanged, while removing the odd transitions. The same parity-based interpretation extends to multiple Trotter steps, although the explicit form of the transition probabilities becomes more involved. Equivalently, the bosonic component supported on the $\ket{\uparrow}$ qubit state contains the parity-preserving contribution associated with the exact trigonometric gate, whereas the component supported on the $\ket{\downarrow}$ qubit state contains the parity-violating contribution generated by the finite-step approximation. \subsection{Two-mode implementation}
We now extend the construction to trigonometric gates whose argument is a weighted sum of commuting Hermitian operators. We focus on the minimal two-mode case, which is the relevant setting for the experiments presented below. Specifically, we consider the gate
\begin{equation}
    e^{-i \theta \cos(c_1\hat{x}_1+c_2\hat{x}_2)}=e^{-i \theta \cos(\bm{c}\cdot \hat{\bm{x}})},
\end{equation}
with $\bm{c} = (c_1,c_2)\in\mathbb{R}^2$ and $\hat{\bm{x}} = (\hat{x}_1,\hat{x}_2)$ denoting two independent position quadratures. From Refs.~\cite{Chalermpusitarak:2025cod,Rainaldi:2025ymn}, we obtain the circuit shown in Fig.~\ref{fig:Trotter1_cos_gate_2modes_circuit}. The extension to additional commuting quadratures is obtained by applying the corresponding conditional displacements to each mode using the same control qubit.
\begin{figure}[t]
    \centering
    \begin{quantikz}[column sep=0.4cm, row sep=0.35cm]
        \lstick{$\ket{\uparrow}$}
        & \qw
        & \gate[
            wires=2,
            style={minimum width=2.8cm}
          ]{e^{-i\theta\cos(c\hat{x})}}
        & \qw
        & \qw
        & \qw
        & \octrl{2}
        & \qw
        \\
        \lstick{$\ket{\psi}$}\setwiretype{b}
        & \qw
        &
        & \qw
        & \qw
        & \gate[
            wires=2,
            style={minimum width=1.7cm}
          ]{\text{BSB}\left(\frac{\eta \Omega}{2} t\right)}
        & \qw
        & \qw
        \\
        \lstick{$\ket{\uparrow}$}
        & \qw
        & \qw
        & \qw
        & \qw
        &
        & \meter{}
    \end{quantikz}
        \caption{Qumode-readout circuit. After applying the hybrid trigonometric gate acting on the qumode and gate qubit, the qumode is coupled to the probe qubit through a blue sideband gate, see Eq.~\eqref{eq:JCandAJC}, and the probe qubit's probability $P_{\downarrow}(t)$ is measured when the gate qubit remains in $\ket{\uparrow}$. }
    \label{fig:qumode_readout_circuit}
\end{figure}
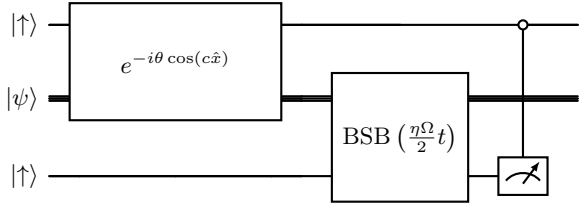
\begin{figure*}[t]
    \centering
    \includegraphics[width = 0.98\linewidth]{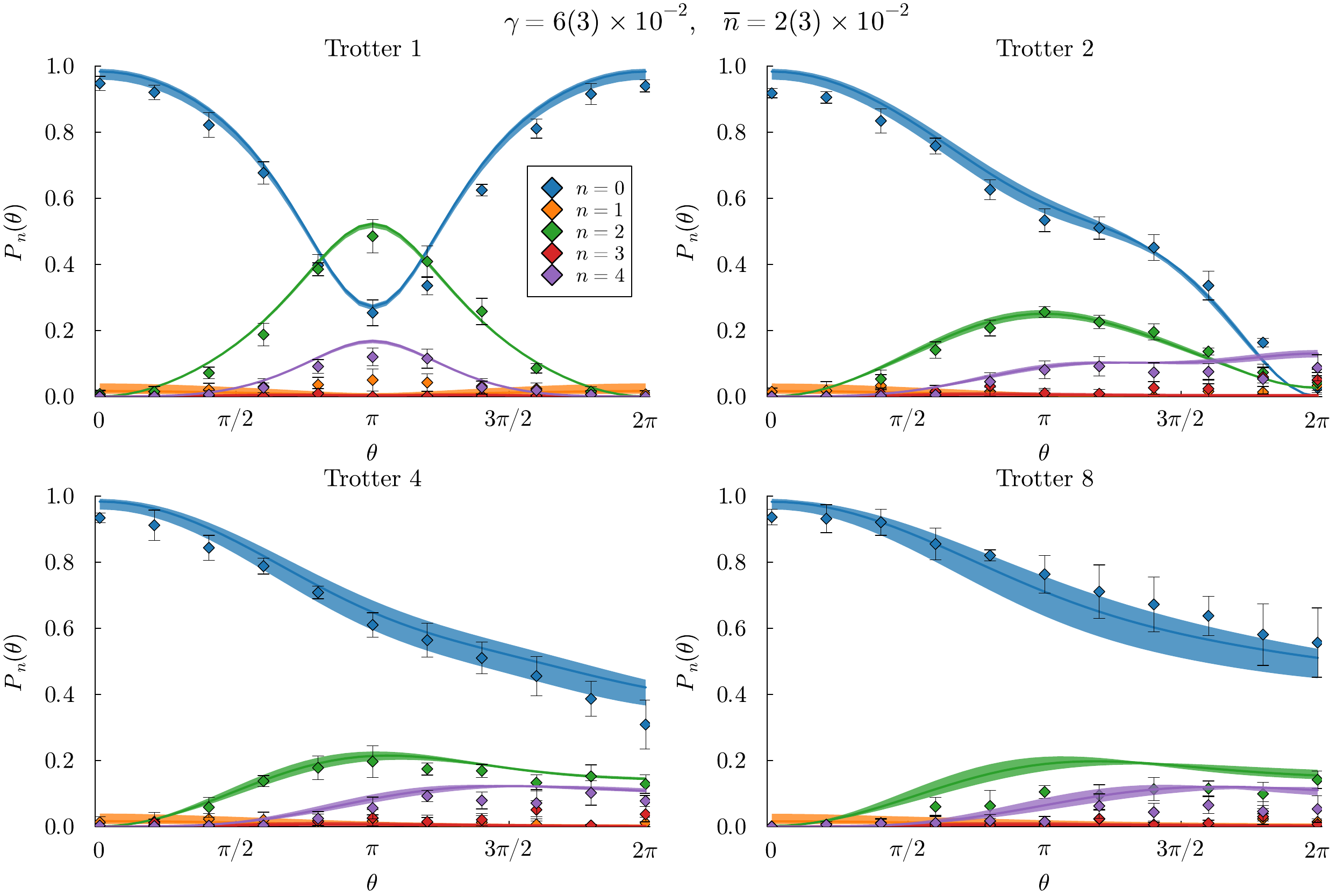}
        \caption{Postselected transition probabilities as a function of $\theta$ for the $c=1$ circuit Fig.~\ref{fig:Trotter1_cos_gate_circuit} (and its higher Trotter step variants) using Trotter steps of 1 (top left), 2 (top right), 4 (bottom left), and 8 (bottom right). The data (diamonds) is plotted alongside the optimized simulations (lines with uncertainty bands). The parameters that optimize the simulation are displayed at the top of each plot.}
    \label{fig:sim_trot_circuits}
\end{figure*}
The circuit in Fig.~\ref{fig:Trotter1_cos_gate_2modes_circuit}, which approximates the cosine gate of a sum of two quadratures, can be analyzed in terms of the operators that compose it. In analogy with Eq.~\eqref{eq:op_Trotter1_1mode}, we define the hybrid unitary
\begin{align}
    &\hat{\mathcal{C}}_{1,(2)}(\theta,\bm{c})\equiv \nonumber \\
    &\prod_{j=1}^2\text{CD}_{X,j}\left(i\frac{c_j}{2\sqrt{2}}\right)\cdot R_Z(\theta)\cdot \prod_{j=1}^2\text{CD}^\dagger_{X,j}\left(i\frac{c_j}{\sqrt{2}}\right)\nonumber \\
    &\cdot R_Z(\theta)
    \cdot \prod_{j=1}^2\text{CD}_{X,j}\left(i\frac{c_j}{2\sqrt{2}}\right)\,.
\label{eq:op_Trotter1_2modes}
\end{align}
This unitary can be decomposed as
\begin{equation}
    \begin{split}
        &\hat{\mathcal{C}}_{1,(2)}(\theta,\bm{c}) =\\
        &\hat{G}_{1,(2)}(\theta,\bm{c})\otimes \mathbb{I} + \hat{G}_{2,(2)}(\theta,\bm{c})\otimes Z + \hat{B}_{(2)}(\theta,\bm{c})\otimes X,
    \end{split}
\end{equation}
with
\begin{equation}
    \begin{split}
        &\hat{G}_{1,(2)}(\theta,\bm{c}) \\
        &= \cos^2\left(\frac{\theta}{2}\right)-\sin^2\left(\frac{\theta}{2}\right)\cos(2\bm{c}\cdot \hat{\bm{x}}) = 1 + \mathcal{O}(\theta^2),\\
        &\hat{G}_{2,(2)}(\theta,\bm{c}) \\
        &= -i\sin(\theta)\cos(\bm{c}\cdot \hat{\bm{x}}) = -i\theta\cos(\bm{c}\cdot \hat{\bm{x}}) + \mathcal{O}(\theta^2),\\
        &\hat{B}_{(2)}(\theta,\bm{c})  = -i\sin^2\left(\frac{\theta}{2}\right)\sin(2\bm{c}\cdot \hat{\bm{x}}) = \mathcal{O}(\theta^2).
    \end{split}
    \label{eq:good_bad_defs_2modes}
\end{equation}

The interpretation is analogous to the one-mode case. For transitions out of the two-mode vacuum $\ket{0}_1\otimes\ket{0}_2$, the ideal gate conserves the total parity rather than the parity of each individual mode. For example, the transition $\ket{0}_1\otimes\ket{0}_2\to\ket{1}_1\otimes\ket{1}_2$ is allowed because the total parity is preserved, unlike in the single-mode case where the transition from the vacuum to the first Fock state is forbidden by parity. Similarly to the one-mode case, the total-parity-violating bosonic component $\hat{B}_{(2)}\ket{0}_1\ket{0}_2$ has support only on the $\ket{\downarrow}$ qubit state. Postselecting on the $\ket{\uparrow}$ qubit state then removes contributions such as $\ket{0}_1\otimes\ket{0}_2\to\ket{1}_1\otimes\ket{0}_2$, which violate total parity conservation.
\section{Experimental characterization}\label{sec:6}
In this section, we discuss the qumode readout and the modeling of trapped ion device errors. We then present experimental results for the trigonometric gates introduced above.
\begin{figure*}[t]
    \centering
    \includegraphics[width=0.48\linewidth]{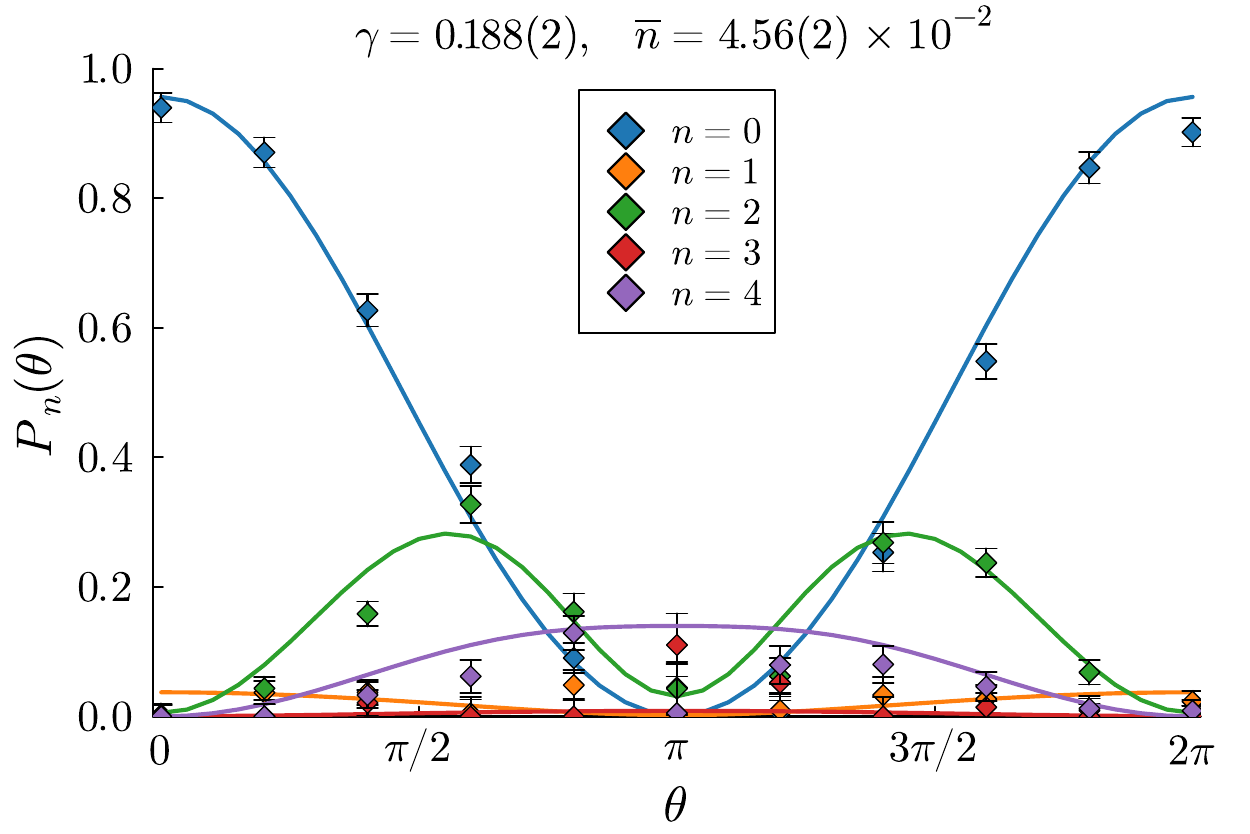}
    \includegraphics[width=0.48\linewidth]{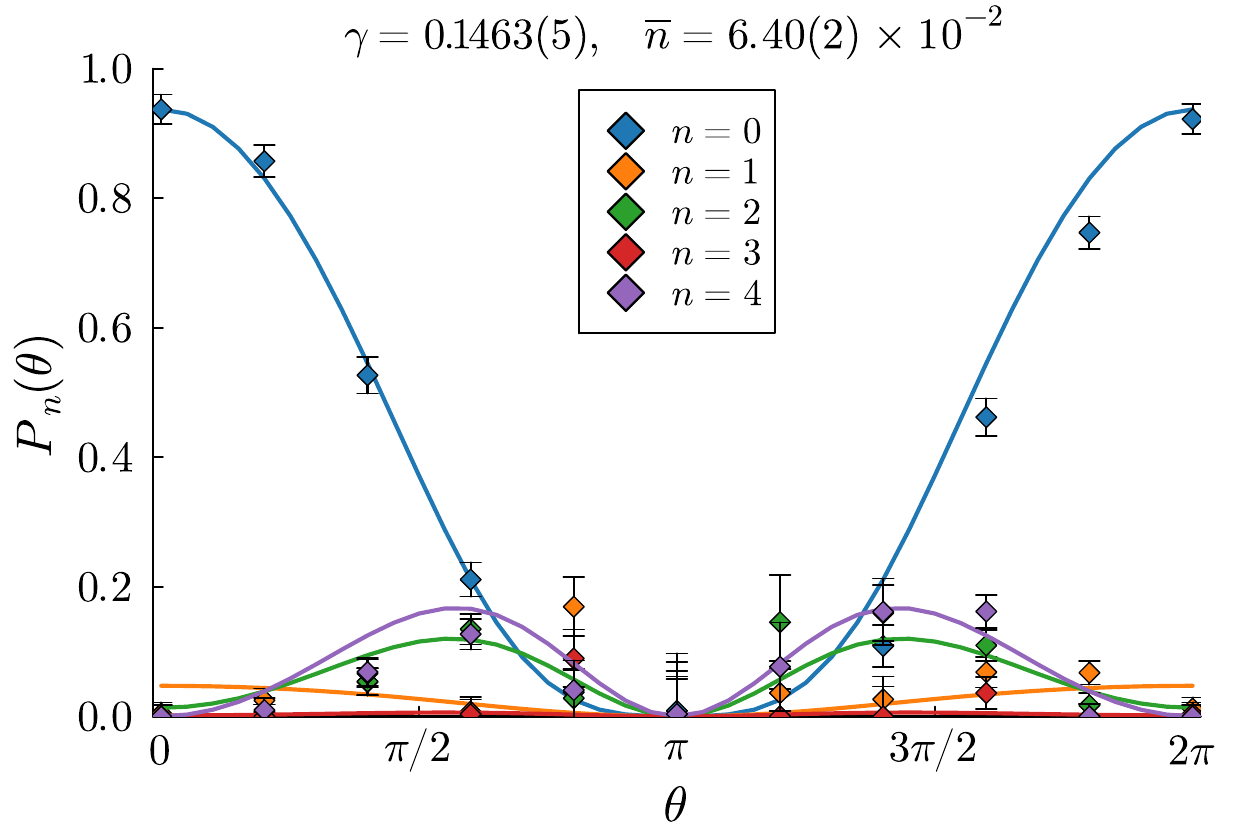}
        \caption{Postselected transition probabilities as a function of $\theta$ for the $c = 2$ (left) and $c= 3$ (right), Trotter one circuit Fig.~\ref{fig:Trotter1_cos_gate_circuit}. The width of the uncertainties of the simulations (lines) are too small to resolve.}
    \label{fig:sim_c_circuits}
\end{figure*}
\subsection{Motional-state readout and noise model}
\label{sec:5.B}
We characterize the motional output states using two complementary measurements. Fock-state populations are reconstructed from blue-sideband (BSB) spectroscopy, while a phase-sensitive measurement of the characteristic function is used for one representative single-mode output state. Because the motional degrees of freedom are not measured directly, both protocols map properties of the oscillator state onto an ancillary probe qubit. Details of the reconstruction procedures, statistical analysis, Hilbert-space truncation, and numerical simulations are given in Appendix~\ref{app:experimental_methods}.

For BSB spectroscopy, the probe qubit is initialized in $\ket{\uparrow}$ and coupled to the selected motional mode. For a state with Fock populations $P_n$, the resulting qubit population contains oscillations at the number-dependent sideband frequencies $\propto\sqrt{n+1}$. Including an effective dephasing rate $\Gamma$ during the readout, we model the measured signal as
\begin{equation}
P_{\downarrow}(t)
\simeq
\frac{1}{2}
\left[
1-
\sum_{n=0}^{\Lambda}
P_n
e^{-\Gamma t}
\cos\!\left(
\eta\Omega\sqrt{n+1}\,t
\right)
\right],
\label{eq:bsb_readout_main}
\end{equation}
where $\eta\Omega$ is the effective BSB coupling and $\Lambda$ is the Fock-space cutoff used in the reconstruction. Fitting the measured BSB trace to Eq.~\eqref{eq:bsb_readout_main} yields the populations $\{P_n\}$ shown below. We distinguish the readout-dephasing parameter $\Gamma$ from the motional dephasing rate $\gamma$ accumulated during the trigonometric-gate circuit itself.

The finite-step gate construction naturally supports heralding through the ancillary gate qubit. After the gate sequence, shots in which the gate qubit is measured in $\ket{\downarrow}$ are discarded, and the motional observables are reconstructed from the conditional $\ket{\uparrow}$ ensemble. The corresponding postselection probability is
\begin{equation}
p_{\rm succ}
=
\tr\!
\left[
\left(
\mathbb I_{\rm m}
\otimes
\ket{\uparrow}\bra{\uparrow}
\right)
\rho_{\rm out}
\right],
\label{eq:psucc_main}
\end{equation}
and is typically $50$--$90\%$ depending on $\theta/N$. All experimental Fock-state distributions reported below are normalized within this postselected ensemble. The success probability is therefore a separate performance metric and is not contained in the normalized population distributions.

To compare the measured populations with the finite-step circuit predictions, we include two leading imperfections of the motional register: residual thermal occupation after sideband cooling and motional phase noise accumulated during the circuit. The initial oscillator state is modeled as
\begin{equation}
\rho_{\rm th}(\bar n)
=
\sum_{n=0}^{\infty}
\frac{\bar n^n}{(1+\bar n)^{n+1}}
\ket{n} \bra{n} ,
\label{eq:thermal_main}
\end{equation}
with the experimentally used non-center-of-mass modes cooled close to the vacuum. Motional dephasing during the gate sequence is described by the effective Lindblad equation
\begin{equation}
\frac{d\rho}{dt}
=
-i[H(t),\rho]
+
\gamma
\left[
\hat n\rho\hat n
-
\frac{1}{2}
\left\{
\hat n^2,\rho
\right\}
\right],
\label{eq:lindblad_main}
\end{equation}
where $H(t)$ follows the experimentally implemented sequence of conditional displacements and qubit rotations. For two motional modes, the corresponding model contains independent terms of the form Eq.~\eqref{eq:lindblad_main} with parameters $(\bar n_1,\bar n_2,\gamma_1,\gamma_2)$. After propagating the full qubit--qumode density matrix, the same postselection used experimentally is applied to the simulated state. The resulting conditional population is
\begin{equation}
P_n^{[N]}
=
\frac{
\bra{ n,\uparrow}
\rho_{\rm out}^{[N]}
\ket{n,\uparrow}
}{
p_{\rm succ}^{[N]}
},
\label{eq:conditional_population_main}
\end{equation}
where $N$ denotes the number of finite steps. The thermal occupations and effective dephasing rates are inferred by comparison with the reconstructed population data as described in Appendix~\ref{app:experimental_methods}. Accordingly, agreement with this model should be interpreted as showing that thermal initialization and motional dephasing provide a compact description of the dominant observed deviations from ideal evolution. This does not amount to a parameter-free identification of a unique microscopic noise mechanism.

In addition to the population measurements, we reconstruct the characteristic function
\begin{equation}
\chi(\beta)
=
\tr
\left[
\rho_{\rm m}{\rm D}(\beta)
\right]
\label{eq:chi_measurement_main}
\end{equation}
for a representative output state using conditional-displacement interferometry. Unlike the Fock populations, $\chi(\beta)$ retains phase information and is therefore sensitive to small rotations of the implemented displacement axis. This measurement provides an independent phase-sensitive characterization of the non-Gaussian structure generated by the gate and is discussed in Sec.~\ref{sec:phase_space_characterization}.
\begin{figure*}[t]
    \centering
    \includegraphics[width=0.95\linewidth]{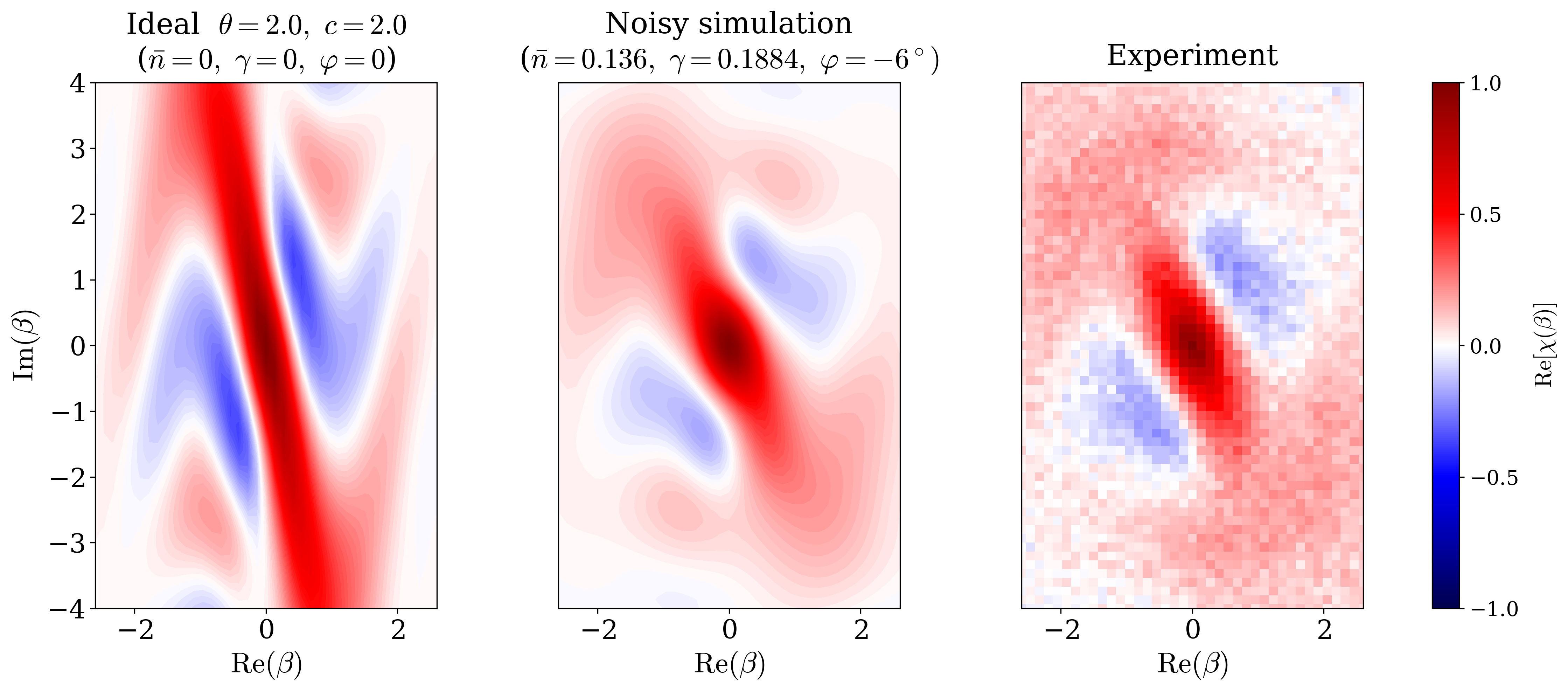}
        \caption{Real part of the characteristic function $\chi(\beta) = \tr[\hat{\rho}_\psi {\rm D}(\beta)]$ of the postselected qumode state obtained by applying the two-step Trotter approximation of the cosine gate, see Eq.~\eqref{eq:op_Trotter2_1mode}, with $\theta = 2$, $c = 2$. Left: ideal simulation. Middle: simulation including qumode dephasing ($\gamma = 0.1884$), displacement miscalibration angle $\varphi = -6^{\circ}$ and thermal initialization ($\bar{n} = 0.136$). Right: Experimental reconstruction.}
    \label{fig:chi_comparison_dephasing}
\end{figure*}

\begin{figure*}[t]
    \centering
    \includegraphics[width=0.98\linewidth]{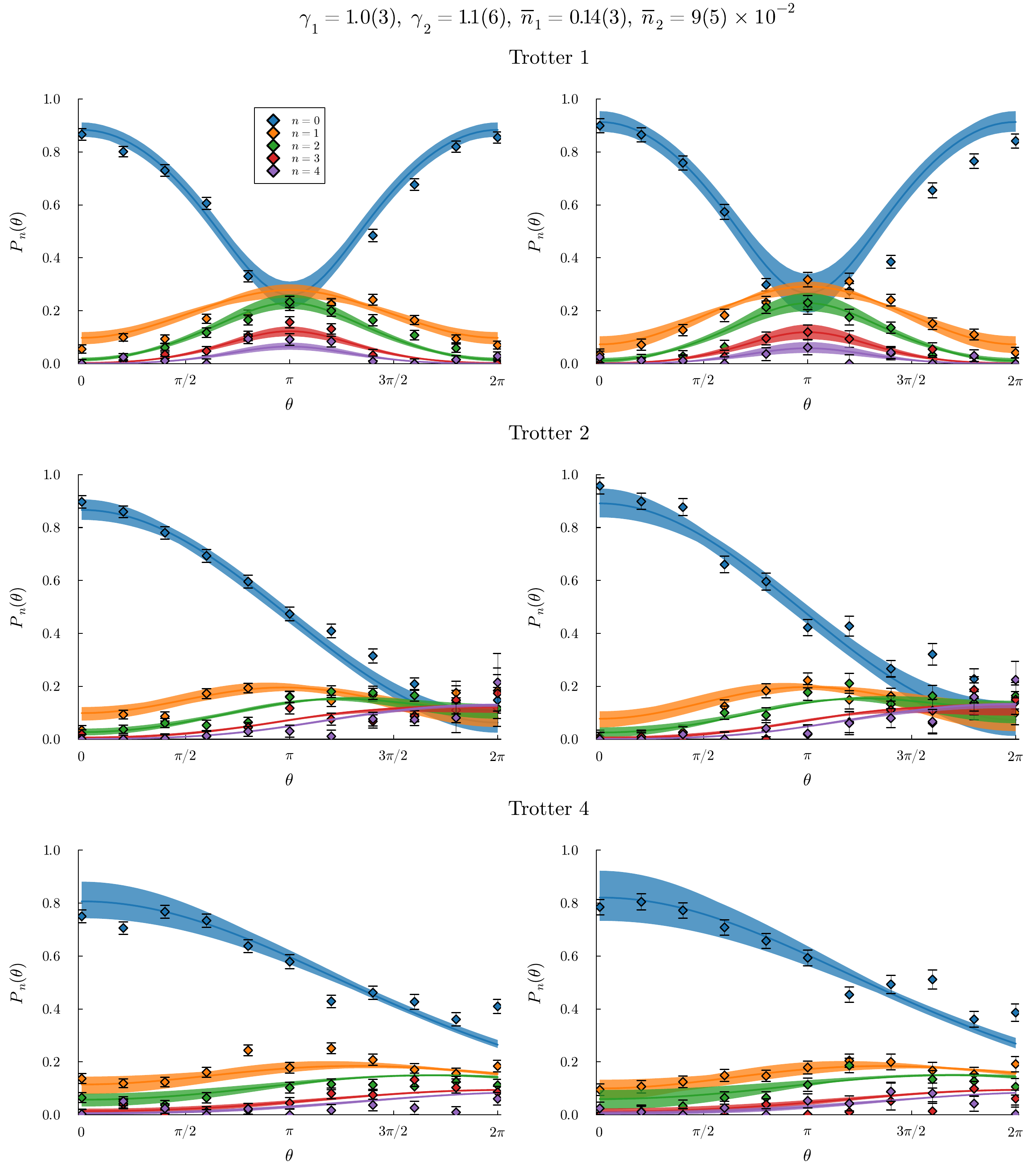}
        \caption{Data for the transition probabilities $P_{n}(\theta) $ of the circuit in Fig.~\ref{fig:Trotter1_cos_gate_2modes_circuit} (and its higher Trotter order variants) for qumode 2 (left column) and qumode 3 (right column) on the QSCOUT platform, using one (top row), two (middle row) and four (bottom row) Trotter steps. The optimal dephasing and the initial phonon occupation of the corresponding mode are labeled atop the plots.}
    \label{fig:sim_X1X2}
\end{figure*}

\subsection{Single-mode gate characterization and Trotter convergence}

Using the BSB readout of the qumodes, we reconstruct transition probabilities for the one- and two-qumode circuits at several Fock levels $n$ and Trotter steps $N$, and compare them to classical simulations with fitted dephasing rates and initial thermal phonon occupations. The results for the one-qumode circuit with parameter $c=1$ and Trotter steps $N=1,2,4,8$ are shown in Fig.~\ref{fig:sim_trot_circuits}. Overall, we find good agreement between the experimental data and the simulations. The agreement is strongest for the transition to the lowest Fock level, corresponding to $\rho(\nbar)\to \ket{0}\!\bra{0}$, which has the strongest signal overall, while transitions to higher Fock levels can show somewhat larger deviations. The optimized values of the dephasing rate $\gamma$ and thermal occupation $\nbar$ are obtained via a global fit to the four circuits, and are displayed at the top, with $\gamma$ given in units of the reciprocal of the physical $\cdx$ gate time. Each Trotter step comprises three $\cdx$ gates, the second of which is twice the duration of the other two, so that a single step takes the equivalent of four $\cdx$ gate times. The $R_Z$ gate in each step contributes negligibly to the circuit duration.

Data and simulations for the one-Trotter-step circuits with parameters $c=2,3$ are shown in Fig.~\ref{fig:sim_c_circuits}. The optimized parameters are again consistent with the experimental conditions. In this case, the circuit structure is fixed while the gate strength is varied, so that the fitted dephasing rates are expected to be more directly comparable. The fitted phonon occupations are also consistent with the one-qumode results in Fig.~\ref{fig:sim_trot_circuits}. The fits are done for each individual circuit and achieve much stronger agreement. This gives small uncertainties in the parameters and in turn small uncertainty bands on the simulation curves.

The marginal Fock-state distributions for the two-qumode circuit with $c=1$ and $N=1,2,4$ Trotter steps are shown in Fig.~\ref{fig:sim_X1X2}, together with the corresponding optimized simulations. The simulations generally reproduce the measured distributions, including the nontrivial population transfer to higher Fock levels, indicating that the error model captures the dominant features of the trapped-ion implementation.
\subsection{Phase-space characterization}\label{sec:phase_space_characterization}
Finally, in Fig.~\ref{fig:chi_comparison_dephasing} we show the characteristic function $\chi(\beta) = \tr[\hat{\rho}_\psi {\rm D}(\beta)]$ (related to the Wigner function via Eq.~\eqref{eq:chi_W_relation}) of the qumode state obtained after postselecting the gate qubit in $\ket{\uparrow}$, for two Trotter steps as in Eq.~\eqref{eq:op_Trotter2_1mode} at $\theta = 2$, $c = 2$. The left panel displays the ideal noiseless result ($\bar{n} = 0$, $\gamma = 0$, $\varphi = 0$), establishing the ideal target state structure. In the middle panel we consider the three dominant experimental imperfections:
\begin{itemize}
    \item Qumode dephasing at rate $\gamma = 0.1884$, chosen as the one obtained from the fit in the left panel of Fig.~\ref{fig:sim_c_circuits}.
    \item Miscalibration of the conditional displacements $\text{CD}_X(\alpha)\to\text{CD}_X(\alpha e^{i\varphi})$. For a gate duration of $t \approx 15~\mu\text{s}$ and an expected detuning of $\delta/2\pi \approx 750\text{--}1500~\text{Hz}$, the accumulated phase error $-\varphi = \delta t$ is roughly $0.07\text{--}0.14~\text{rad}$, which matches the $4^\circ\text{--}8^\circ$ expectation represented in Fig.~\ref{fig:chi_comparison_dephasing} with the middle value $\varphi = -6^\circ$.
    \item A non-vanishing mean phonon occupation number measured as $\bar{n} = 0.136$ due to an imperfect initialization.
\end{itemize}
The resulting characteristic function is in good qualitative agreement with the experimentally reconstructed one measured via the method described in~\cite{Fluhmann:2019slh} (right panel), confirming that qumode dephasing is the main source of deviation from the ideal case, followed by smaller contributions from the imperfect initialization and from the miscalibration of the conditional displacement's argument. We quantify their consistency with the measured characteristic function in Sec.~\ref{sec:nongaussian_resource}.

\subsection{Non-Gaussian resource of the implemented gate}
\label{sec:nongaussian_resource}

The comparison of Fig.~\ref{fig:chi_comparison_dephasing} is qualitative, and the three parameters entering it are fixed independently of the characteristic-function data. We now use those data to test the noise model quantitatively and to quantify the non-Gaussian resource of the implemented gate, using two independent and complementary methods: a chi-squared fit of the open-system model to the data followed by a full-phase-space evaluation of the fitted state (Sec.~\ref{sec:nongaussian_resource_fit}), and a direct Fourier reconstruction of the Wigner function from the measured characteristic function itself (Sec.~\ref{sec:nongaussian_resource_direct}). The two methods agree within their combined uncertainties on every quantity they share, which we take as an independent cross-validation of both the noise model and the reconstruction procedure.

Two properties of the measurement simplify both analyses. First, the evolved state has definite total Fock parity, $\hat\Pi\hat\rho\hat\Pi=\hat\rho$ with $\hat\Pi$ the parity operator of Eq.~\eqref{eq:W0_Trotter_11}, so that $\chi(-\beta)=\chi(\beta)=\chi^*(\beta)$ and the characteristic function is real throughout. The single-quadrature measurement of Sec.~\ref{sec:phase_space_characterization} is therefore complete. 
Second, the $39\times61$ grid on $\operatorname{Re}(\beta)\in[-2.6,2.6]$, $\operatorname{Im}(\beta)\in[-4,4]$ provides $2379$ measured values, each an average over $18$ accepted scans of $100$ shots, postselected on the gate qubit remaining in $\ket{\uparrow}$. The postselection success rate for each grid point was $0.775$ with a standard deviation of $0.010$. One further scan was discarded because acquisition was aborted early after a lengthy ion loss and reload process, an exclusion made on provenance, not as a numerical outlier cut.

\subsubsection{Model fit and full-phase-space resource}
\label{sec:nongaussian_resource_fit}

Since each measured value is a binomial proportion estimated from the $N_\ell$ shots surviving postselection, we fit the open-system model of Sec.~\ref{sec:6} to the data by minimizing
\begin{align}
\chi^2(\bar n,\gamma,\varphi)
&=
\sum_\ell
\frac{
\left[\chi^{\rm exp}_\ell-\chi^{\rm sim}_\ell(\bar n,\gamma,\varphi)\right]^2
}{
\sigma_\ell^2
},
\nonumber\\
&\sigma_\ell^2=\frac{1-\chi_\ell^2}{N_\ell},
\label{eq:chi_fit_objective}
\end{align}
with $N_\ell$ the measured $\beta$-dependent number of postselected shots. This gives
\begin{align}
\bar n&=0.147\pm0.007 ,
\nonumber\\
\gamma&=0.189\pm0.004 ,
\nonumber\\
\varphi&=-4.7^\circ\pm0.2^\circ .
\label{eq:chi_fit_params}
\end{align}
The uncertainties are marginal $\Delta\chi^2=1$ intervals, obtained from the diagonal of $2\mathbf H^{-1}$ with $\mathbf H$ the Hessian of Eq.~\eqref{eq:chi_fit_objective} at the minimum, and scaled by $\sqrt{\chi^2/{\rm dof}}$ to account for the residual systematic discussed below. The parameters are only mildly correlated, the largest correlation being $-0.40$ between $\bar n$ and $\gamma$, and the displacement-axis offset is essentially uncorrelated with the two damping parameters. The characteristic function therefore determines $\varphi$ independently, which the Fock-population measurements cannot do because they are invariant under a rigid phase-space rotation, see Appendix~\ref{app:phase_calibration}. The fitted $\bar n$ and $\gamma$ lie within $8\%$ and $1\%$ of their independently determined values, and $\varphi$ falls inside the $4^\circ$ to $8^\circ$ range expected from the residual sideband detuning, so that the measured characteristic function provides an independent, phase-sensitive confirmation of the open-system model used throughout Sec.~\ref{sec:6}. The remaining discrepancy is small in absolute terms but exceeds the statistical resolution of the measurement, the weighted residual at the optimum being $\chi^2/{\rm dof}=3.0$, so a systematic not captured by the three-parameter error model is present. The intervals in Eq.~\eqref{eq:chi_fit_params} should thus be considered lower bounds on the model uncertainty, and the few-percent offsets from the independently determined parameters as consistency at the level of the model, not of the statistical errors.

The negative Wigner volume and mana below are evaluated directly from the density matrix over the full phase space: the damped, thermally mixed fitted/external states already converge by cutoff $30$, but the undamped reference states ($\bar n=\gamma=0$) need cutoff $\geq60$.

\begin{table}[t]
\centering
        \caption{Negative Wigner volume $\mathcal N_W$ of Eq.~\eqref{eq:negative_volume_main} and mana $\mathcal M$ of Eq.~\eqref{eq:mana_definition}, evaluated over the full phase space, for the simulated state at the externally fixed parameters of Fig.~\ref{fig:chi_comparison_dephasing} and at the values fitted to the measured characteristic function.}
\label{tab:chi_fit}
\begin{ruledtabular}
\begin{tabular}{lccccc}
 & $\bar n$ & $\gamma$ & $\varphi$ & $\mathcal N_W$ & $\mathcal M$\\
\colrule
external            & $0.136$ & $0.188$ & $-6.0^\circ$ & $0.0367$ & $0.0707$\\
fit: $\bar n$ fixed & $0.136$ & $0.192$ & $-4.6^\circ$ & $0.0357$ & $0.0689$\\
fit: all free       & $0.147$ & $0.189$ & $-4.7^\circ$ & $0.0348$ & $0.0673$\\
\end{tabular}
\end{ruledtabular}
\end{table}

We take the midpoint of the three parameter prescriptions of Table~\ref{tab:chi_fit} as the central value and half their spread as the model uncertainty:
\begin{align}
\mathcal N_W^{\rm model}&=0.0357\pm0.0009 ,
\nonumber\\
\mathcal M^{\rm model}&=0.0690\pm0.0017 ,
\label{eq:mana_experiment_model}
\end{align}
to be compared with $\mathcal N_W=0.2524$ and $\mathcal M=0.4087$ for the ideal two-step circuit and $\mathcal N_W=0.2456$, $\mathcal M=0.3996$ for the exact cosine gate. Motional dephasing and residual thermal occupation suppress the mana of the implemented gate by roughly a factor of six relative to the ideal finite-step circuit, while leaving it clearly nonzero. This conclusion is insensitive to the parameter uncertainties, the mana varying by only $2\%$ across all fit variants. We note also that the finite-step circuit slightly exceeds the exact gate, consistent with the additional phase-space structure visible in Fig.~\ref{fig:W0_1_c_2_steps_1}.

\subsubsection{Direct Fourier reconstruction from the data}
\label{sec:nongaussian_resource_direct}

A direct determination of $\mathcal N_W$ from the measured characteristic function, without passing through the open-system model, is a valuable independent check of Eq.~\eqref{eq:mana_experiment_model}. The characteristic function is available only over the finite rectangle $\operatorname{Re}(\beta)\in[-2.6,2.6]$, $\operatorname{Im}(\beta)\in[-4,4]$, where $|\chi|$ has fallen only to $\simeq0.1$ on the boundary. A hard truncation of the inverse Fourier transform at this window produces sinc-like ringing that inflates the apparent negative volume by $50$--$90\%$ on simulated states whose answer is known, with the error set by the window extent, not by the sampling, so refining the grid does not help. We therefore multiply the measured characteristic function by the Gaussian apodization
\begin{equation}
 G_\lambda(\beta)=\exp\!\left[-\frac{\lambda^2|\beta|^2}{2}\right]
 \end{equation}
before applying the inverse transform.

\textbf{Apodization calibration:} We choose $\lambda=0.20$ as the nominal
reconstruction width by a closure test against the ideal, undamped two-step circuit: the windowed and full-phase-space $\mathcal N_W$ agree to $<0.1\%$ ($0.25233$ vs.\ $0.25239$), a calibration that does not depend on the noise model being correct. As a robustness check we additionally calibrate $\lambda$ directly against the noisy/physical-error-model state, solving for the width at which the windowed reconstruction exactly reproduces that state's known full-phase-space $\mathcal N_W$, both with the model's externally fixed parameters ($\lambda^\ast\simeq0.186$) and with the values fitted in Sec.~\ref{sec:nongaussian_resource_fit} ($\lambda^\ast\simeq0.185$), which agree with each other to within $0.001$.

Applying $\lambda=0.20$ to the experimental data gives the finite-window point estimate $\mathcal N_W=0.0422$. Statistical uncertainty is obtained by resampling the 18 accepted experimental scans as complete units, propagating finite-shot fluctuations and scan-to-scan drift through calibration, Fourier reconstruction, and the nonlinear negative-volume integral. Because noise can itself produce a small positive bias in the extracted negative volume, empirical residual maps are also added to the noisy theoretical characteristic function and propagated through the same estimator at the same $\lambda$. This gives a model-dependent bias of $0.0009$, leading to
\begin{equation}
\mathcal N_W^{\rm exp}
 =
 0.0414
 \pm0.0038_{\rm stat}
 \pm0.0084_{\rm win},
 \label{eq:nw_exp_direct}
 \end{equation}
where the window uncertainty is the largest deviation of this same bias-corrected estimator, recomputed at each endpoint, over $\lambda=0.20\pm0.05$. The corresponding mana is $\mathcal M^{\rm exp}=0.079\pm0.007_{\rm stat}\pm0.016_{\rm win}$. Repeating the full procedure at the model-calibrated $\lambda^\ast$ gives $\mathcal N_W^{\rm exp}=0.0436 \pm0.0039_{\rm stat}\pm0.0084_{\rm win}$, reported as a consistency check, not the primary result, since calibrating the reconstruction against the same model it is then compared to would make that comparison less independent.

The open-system model predicts $\mathcal N_W=0.0367$ over the full phase space at the externally fixed parameters (Table~\ref{tab:chi_fit}) and $\mathcal N_W=0.0347$ after the same $\lambda=0.20$ finite-window reconstruction. The extracted experimental value, Eq.~\eqref{eq:nw_exp_direct}, is comparable to the combined statistical and finite-window sensitivity of this model-windowed prediction and does not constitute significant tension. We regard the model-based full-phase-space evaluation of Sec.~\ref{sec:nongaussian_resource_fit} and the direct data-driven reconstruction as complementary analyses of the same characteristic-function data, not independent measurements: the direct reconstruction's calibration and bias correction both rely on the same open-system model being tested. Their agreement is nonetheless a useful internal consistency check, and the two together give a fuller picture than either alone. The model fit supplies the controlled full-phase-space extrapolation, while the Fourier reconstruction shows what can be extracted more directly from the measured characteristic function.

\subsection{Two-mode gate characterization}

For the two-qumode data, the BSB readout yields mode-resolved marginal Fock-state distributions, not the full joint distribution. The measurement verifies population transfer into each addressed mode, but does not directly measure correlations between the two modes. Nevertheless, the enhancement of odd Fock-level populations in the two-mode case, compared with their suppression in the one-mode results of Fig.~\ref{fig:sim_trot_circuits}, provides an indirect signature of the nontrivial two-mode dynamics shown in Fig.~\ref{fig:sim_X1X2}.
\section{Conclusions and Outlook}
\label{sec:7}
We have experimentally realized and characterized a family of trigonometric continuous-variable gates of the form
\begin{equation}
    U(\theta,c)
    =
    e^{-i\theta\cos(c\hat x)}
\end{equation}
on trapped-ion motional modes, together with a multimode extension generated by
\begin{equation}
    U_{12}(\theta)
    =
    e^{-i\theta\cos(c_1\hat x_1+c_2\hat x_2)} .
\end{equation}
Using conditional displacements and an ancillary qubit, we implemented finite-step approximations to these nonpolynomial interactions and observed their systematic evolution toward the ideal trigonometric dynamics as the circuit depth was increased. Fock-state measurements resolve the characteristic parity structure and population transfer of the one-mode gate, while the two-mode measurements are consistent with the predicted coupled nonlinear evolution. Phase-sensitive characteristic-function measurements provide a complementary characterization of the non-Gaussian motional state. Fitting the open-system model to those data and evaluating the resource over the full phase space gives $\mathcal M=0.0690$, roughly a factor of six below the ideal finite-step circuit yet still nonzero, and a direct Fourier reconstruction of the measured characteristic function is consistent with that value.

Beyond the gate implementation, the trigonometric evolution reveals several qualitatively distinct mechanisms for generating Wigner negativity. Starting from the Gaussian vacuum, an arbitrarily weak cosine phase produces a pure non-Gaussian state and necessarily generates Wigner negativity. Whether that negativity is perturbatively accessible depends on the spatial frequency, through the combination $|\theta|e^{c^2/4}$. For $|\theta|\ll\zeta_0(c)e^{-c^2/4}$ the first negative regions nucleate in the momentum tails at
\begin{equation}
    |p|
    \sim
    \frac{1}{c}
    \ln\frac{1}{|\theta|},
    \quad
    |\theta|\rightarrow0,
\end{equation}
leading to
\begin{equation}
    \mathcal M
    \sim
    C(c)\,\mathcal P^{-5/2}e^{-\mathcal P^{2}},
    \quad
    \mathcal P=
    \frac{1}{c}
    \left[
        \ln\!\left(\frac{\zeta_0(c)}{|\theta|}\right)
        +\frac{c^2}{4}
    \right],
\end{equation}
so that at fixed $c$ the onset of Wigner negativity is beyond all algebraic orders in the gate strength. This provides a concrete example in which the non-Gaussian character of the state is immediate while the associated Wigner resource is invisible to any finite-order expansion about the Gaussian limit. In the resolved-displacement linear-response regime $\zeta_0(c)e^{-c^2/4}\ll|\theta|\ll1$, by contrast, the same lobes lie in the bulk of the momentum distribution and the mana grows linearly, $\mathcal M\simeq\kappa(c)|\theta|$, saturating the first-order bound $\mathcal M\le\kappa(c)|\theta|+O(\theta^2)$ that we establish for all $c$. The beyond-all-orders statement is therefore a fixed-$c$ statement, and the limits $|\theta|\rightarrow0$ and $c\rightarrow\infty$ do not commute. The third regime is governed by a different mechanism. When the imprinted phase becomes rapidly oscillatory, the Wigner integral is controlled by semiclassical stationary-phase structure. For nonlinear single-coordinate phase gates whose dominant stationary regions are regular and nondegenerate,
\begin{equation}
    \mathcal M
    =
    \frac{1}{2}\ln|\theta|+O(1).
\end{equation}
For the cosine gate, regular stationary regions coexist with fold caustics, whose integrated contribution is subleading in this limit. The onset of the logarithmic behavior requires $|\theta|\gg32/c^4$, since the quadratic part of the cosine generates a Gaussian unitary and contributes no negativity. The three regimes correspond to distinct phase-space mechanisms: non-Gaussian resources evolve from tail-dominated nucleation, through linear response saturating the first-order bound, to semiclassical stationary-phase growth. The dependence on $c$ is strong but not monotonic. Increasing $c$ reduces the phase-space distance to the first Wigner-negative regions and produces an extremely rapid, log-Gaussian increase of the accessible negativity, but only until $c\simeq2\sqrt{\ln(\zeta_0(c)/|\theta|)}$, beyond which the mana saturates at $\kappa(c)|\theta|\rightarrow4|\theta|/\pi$ and larger $c$ yields no further gain at weak gate strength. At strong gate strength $c$ instead controls the onset of the semiclassical regime.

The experimental deviations from ideal evolution are well described by a compact open-system model including residual thermal occupation and motional dephasing. The agreement should be interpreted as evidence that these effects account for the principal imperfections observed in the present implementation, not as a unique microscopic identification of the underlying noise. Similarly, the present two-mode measurements yield mode-resolved populations but not the full joint Fock distribution. Measurements of joint parity, cross-mode correlations, or the two-mode characteristic function would provide a more complete characterization of the coupled multimode operation.

Trigonometric gates are especially natural for quantum problems whose Hamiltonians contain periodic or compact degrees of freedom, including rotor models, sine-Gordon-type interactions, and compact lattice gauge fields. In such settings, directly synthesizing periodic nonlinearities can avoid representing intrinsically global phase-space structure through high-order polynomial approximations. Extending the present approach to larger multimode registers, correlated phase-space measurements, and longer coherent gate sequences would enable direct studies of nonlinear bosonic dynamics in regimes that are difficult to access with Gaussian operations and polynomial nonlinearities alone.
\vspace{2.5ex}
\begin{acknowledgments}
This material is supported in part by the US Department of Energy, Office of Science, Office of Advanced Scientific Computing Research under its Quantum Testbed Program. Sandia National Laboratories is a multimission laboratory managed and operated by National Technology \& Engineering Solutions of Sandia, LLC, a wholly owned subsidiary of Honeywell International Inc., for the US Department of Energy's National Nuclear Security Administration under contract DE-NA0003525. This paper describes objective technical results and analysis. Any subjective views or opinions that might be expressed in the paper do not necessarily represent the views of the US Department of Energy or the United States Government. TR, JM, and FR are supported by the DOE, Office of Science, Office of Nuclear Physics, Early Career Program under contract No. DE-SC0025881. GS acknowledges support by NSF award DGE-2152168 and DOE awards DE-SC0024325 and DE-SC0024328. This research used resources of the National Energy Research Scientific Computing Center (NERSC), a DOE Office of Science User Facility using NERSC award NP-ERCAP0037787. The authors would like to thank Stony Brook Research Computing and Cyberinfrastructure, and the Institute for Advanced Computational Science at Stony Brook University, for access to the SeaWulf computing system, made possible by grants from the National Science Foundation (\#1531492 and Major Research Instrumentation award \#2215987), with matching funds from Empire State Development's Division of Science, Technology and Innovation (NYSTAR) program (contract C210148). SAND2026-25868O.
\end{acknowledgments}
\vfill
\bibliographystyle{utphys}
\bibliography{bibliography}

@article{Tortorici:2026dhj,
    author = "Tortorici, Edward C. and McGarrigle, Ethan C. and McFarland, Brian K. and Johnson, Wes L. and Lobser, Daniel S. and Revelle, Melissa C. and Ruzic, Brandon P. and Clark, Susan M. and Yale, Christopher G.",
    title = "{QSCOUT's Qubit-Boson Gate Set}",
    eprint = "2607.08560",
    archivePrefix = "arXiv",
    primaryClass = "quant-ph",
    month = "7",
    year = "2026"
}

@article{HUDSON1974249,
title = {When is the wigner quasi-probability density non-negative?},
journal = {Reports on Mathematical Physics},
volume = {6},
number = {2},
pages = {249-252},
year = {1974},
issn = {0034-4877},
doi = {https://doi.org/10.1016/0034-4877(74)90007-X},
url = {https://www.sciencedirect.com/science/article/pii/003448777490007X},
author = {R.L. Hudson}
}

@article{KatoMcLeod1971,
  author  = {Kato, Tosio and McLeod, J. B.},
  title   = {The functional-differential equation
             $y'(x)=ay(\lambda x)+by(x)$},
  journal = {Bull. Amer. Math. Soc.},
  volume  = {77},
  number  = {6},
  pages   = {891--937},
  year    = {1971},
  doi     = {10.1090/S0002-9904-1971-12805-7}
}

@article{Moore:2025qzc,
    author = "Moore, Darren W. and Filip, Radim",
    title = "{Nonlinear phase gates as Airy transforms of the Wigner function}",
    eprint = "2504.04851",
    archivePrefix = "arXiv",
    primaryClass = "quant-ph",
    doi = "10.1038/s41534-025-01006-z",
    journal = "npj Quantum Inf.",
    volume = "11",
    number = "1",
    pages = "159",
    year = "2025"
}

@article{Rainaldi:2025ymn,
    author = "Rainaldi, Tommaso and Ale, Victor and Grau, Matt and Kharzeev, Dmitri and Rico, Enrique and Ringer, Felix and Shome, Pubasha and Siopsis, George",
    title = "{Trigonometric continuous-variable gates and hybrid quantum simulations of the sine-Gordon model}",
    eprint = "2512.19582",
    archivePrefix = "arXiv",
    primaryClass = "quant-ph",
    doi = "10.1007/JHEP03(2026)125",
    journal = "JHEP",
    volume = "03",
    pages = "125",
    year = "2026"
}

@article{Heeres:2015cnj,
    author = "Heeres, Reinier W. and Vlastakis, Brian and Holland, Eric and Krastanov, Stefan and Albert, Victor V. and Frunzio, Luigi and Jiang, Liang and Schoelkopf, Robert J.",
    title = "{Cavity State Manipulation Using Photon-Number Selective Phase Gates}",
    eprint = "1503.01496",
    archivePrefix = "arXiv",
    primaryClass = "quant-ph",
    doi = "10.1103/PhysRevLett.115.137002",
    journal = "Phys. Rev. Lett.",
    volume = "115",
    number = "13",
    pages = "137002",
    year = "2015"
}

@article{Eickbusch:2021uod,
    author = "Eickbusch, Alec and Sivak, Volodymyr and Ding, Andy Z. and Elder, Salvatore S. and Jha, Shantanu R. and Venkatraman, Jayameenakshi and Royer, Baptiste and Girvin, S. M. and Schoelkopf, Robert J. and Devoret, Michel H.",
    title = "{Fast universal control of an oscillator with weak dispersive coupling to a qubit}",
    eprint = "2111.06414",
    archivePrefix = "arXiv",
    primaryClass = "quant-ph",
    doi = "10.1038/s41567-022-01776-9",
    journal = "Nature Phys.",
    volume = "18",
    pages = "1464--1469",
    year = "2022"
}

@article{Sutherland:2021vsj,
    author = "Sutherland, R. T. and Srinivas, R.",
    title = "{Universal hybrid quantum computing in trapped ions}",
    eprint = "2105.05768",
    archivePrefix = "arXiv",
    primaryClass = "quant-ph",
    doi = "10.1103/PhysRevA.104.032609",
    journal = "Phys. Rev. A",
    volume = "104",
    number = "3",
    pages = "032609",
    year = "2021"
}

@article{Wang:2019xqo,
    author = "Wang, Christopher S. and Curtis, Jacob C. and Lester, Brian J. and Zhang, Yaxing and Gao, Yvonne Y. and Freeze, Jessica and Batista, Victor S. and Vaccaro, Patrick H. and Chuang, Isaac L. and Frunzio, Luigi and Jiang, Liang and Girvin, S. M. and Schoelkopf, Robert J.",
    title = "{Efficient multiphoton sampling of molecular vibronic spectra on a superconducting bosonic processor}",
    eprint = "1908.03598",
    archivePrefix = "arXiv",
    primaryClass = "quant-ph",
    doi = "10.1103/PhysRevX.10.021060",
    journal = "Phys. Rev. X",
    volume = "10",
    number = "2",
    pages = "021060",
    year = "2020"
}

@article{Farrell:2023fgd,
    author = "Farrell, Roland C. and Illa, Marc and Ciavarella, Anthony N. and Savage, Martin J.",
    title = "{Scalable Circuits for Preparing Ground States on Digital Quantum Computers: The Schwinger Model Vacuum on 100 Qubits}",
    eprint = "2308.04481",
    archivePrefix = "arXiv",
    primaryClass = "quant-ph",
    reportNumber = "IQuS@UW-21-060, NT@UW-23-13",
    doi = "10.1103/PRXQuantum.5.020315",
    journal = "PRX Quantum",
    volume = "5",
    number = "2",
    pages = "020315",
    year = "2024"
}

@article{Kogut:1974ag,
    author = "Kogut, John B. and Susskind, Leonard",
    title = "{Hamiltonian Formulation of Wilson's Lattice Gauge Theories}",
    reportNumber = "Print-74-1186 (CORNELL)",
    doi = "10.1103/PhysRevD.11.395",
    journal = "Phys. Rev. D",
    volume = "11",
    pages = "395--408",
    year = "1975"
}

@article{Zou:2026cfk,
    author = "Zou, Dairui and Li, Tianyin and Liang, Jian and Wang, Enke and Xing, Hongxi",
    title = "{Hadronic tensor in lattice gauge theories by quantum computing}",
    eprint = "2606.17003",
    archivePrefix = "arXiv",
    primaryClass = "hep-ph",
    reportNumber = "RIKEN-iTHEMS-Report-26",
    month = "6",
    year = "2026"
}

@article{10.1021/jp992939g,
    author = {Hahn, Susanne and Stock, Gerhard},
    title = {Quantum-Mechanical Modeling of the Femtosecond Isomerization in Rhodopsin},
    journal = {The Journal of Physical Chemistry B},
    volume = {104},
    number = {6},
    pages = {1146-1149},
    year = {2000},
    issn = {1520-6106},
    doi = {10.1021/jp992939g},
}

@article{Jordan:2011ci,
    author = "Jordan, Stephen P. and Lee, Keith S. M. and Preskill, John",
    title = "{Quantum Computation of Scattering in Scalar Quantum Field Theories}",
    eprint = "1112.4833",
    archivePrefix = "arXiv",
    primaryClass = "hep-th",
    journal = "Quant. Inf. Comput.",
    volume = "14",
    pages = "1014--1080",
    year = "2014"
}

@article{Kreshchuk:2020aiq,
    author = "Kreshchuk, Michael and Jia, Shaoyang and Kirby, William M. and Goldstein, Gary and Vary, James P. and Love, Peter J.",
    title = "{Simulating Hadronic Physics on NISQ devices using Basis Light-Front Quantization}",
    eprint = "2011.13443",
    archivePrefix = "arXiv",
    primaryClass = "quant-ph",
    doi = "10.1103/PhysRevA.103.062601",
    journal = "Phys. Rev. A",
    volume = "103",
    number = "6",
    pages = "062601",
    year = "2021"
}

@article{Chiari:2025lwq,
    author = "Chiari, Even and Makhlouf, Wafa and Pepe, Lucie and Koridon, Emiel and Klein, Johanna and Senjean, Bruno and Lasorne, Benjamin and Yalouz, Saad",
    title = "{Ab initio polaritonic chemistry on diverse quantum computing platforms: Ansatz circuit design for qubit, qudit, and hybrid qubit-qumode architectures}",
    eprint = "2506.12504",
    archivePrefix = "arXiv",
    primaryClass = "quant-ph",
    doi = "10.1103/1l5j-dfh4",
    journal = "Phys. Rev. A",
    volume = "112",
    number = "5",
    pages = "052433",
    year = "2025"
}

@article{Zohar:2015hwa,
    author = "Zohar, Erez and Cirac, J. Ignacio and Reznik, Benni",
    title = "{Quantum Simulations of Lattice Gauge Theories using Ultracold Atoms in Optical Lattices}",
    eprint = "1503.02312",
    archivePrefix = "arXiv",
    primaryClass = "quant-ph",
    doi = "10.1088/0034-4885/79/1/014401",
    journal = "Rept. Prog. Phys.",
    volume = "79",
    number = "1",
    pages = "014401",
    year = "2016"
}

@article{Banuls:2019bmf,
	author = {Ba{\~n}uls, Mari Carmen and Blatt, Rainer and Catani, Jacopo and Celi, Alessio and Cirac, Juan Ignacio and Dalmonte, Marcello and Fallani, Leonardo and Jansen, Karl and Lewenstein, Maciej and Montangero, Simone and Muschik, Christine A. and Reznik, Benni and Rico, Enrique and Tagliacozzo, Luca and Van Acoleyen, Karel and Verstraete, Frank and Wiese, Uwe-Jens and Wingate, Matthew and Zakrzewski, Jakub and Zoller, Peter},
	date = {2020/08/04},
	doi = {10.1140/epjd/e2020-100571-8},
	id = {Ba{\~n}uls2020},
	isbn = {1434-6079},
	journal = {The European Physical Journal D},
	number = {8},
	pages = {165},
	title = {Simulating lattice gauge theories within quantum technologies},
	url = {https://doi.org/10.1140/epjd/e2020-100571-8},
	volume = {74},
	year = {2020}
}

@article{Zohar:2021nyc,
    author = "Zohar, Erez",
    title = "{Quantum simulation of lattice gauge theories in more than one space dimension---requirements, challenges and methods}",
    eprint = "2106.04609",
    archivePrefix = "arXiv",
    primaryClass = "quant-ph",
    doi = "10.1098/rsta.2021.0069",
    journal = "Phil. Trans. A. Math. Phys. Eng. Sci.",
    volume = "380",
    number = "2216",
    pages = "20210069",
    year = "2022"
}

@article{Bauer:2022hpo,
  title = {Quantum Simulation for High-Energy Physics},
  author = {Bauer, Christian W. and Davoudi, Zohreh and Balantekin, A. Baha and Bhattacharya, Tanmoy and Carena, Marcela and de Jong, Wibe A. and Draper, Patrick and El-Khadra, Aida and Gemelke, Nate and Hanada, Masanori and Kharzeev, Dmitri and Lamm, Henry and Li, Ying-Ying and Liu, Junyu and Lukin, Mikhail and Meurice, Yannick and Monroe, Christopher and Nachman, Benjamin and Pagano, Guido and Preskill, John and Rinaldi, Enrico and Roggero, Alessandro and Santiago, David I. and Savage, Martin J. and Siddiqi, Irfan and Siopsis, George and Van Zanten, David and Wiebe, Nathan and Yamauchi, Yukari and Yeter-Aydeniz, K\"ubra and Zorzetti, Silvia},
  journal = {PRX Quantum},
  volume = {4},
  issue = {2},
  pages = {027001},
  numpages = {70},
  year = {2023},
  month = {May},
  publisher = {American Physical Society},
  doi = {10.1103/PRXQuantum.4.027001},
  url = {https://link.aps.org/doi/10.1103/PRXQuantum.4.027001}
}

@article{Chalermpusitarak:2025cod,
    author = "Chalermpusitarak, Teerawat and Schwennicke, Kai and Kassal, Ivan and Tan, Ting Rei",
    title = "{Programmable generation of arbitrary continuous-variable anharmonicities and nonlinear couplings}",
    eprint = "2511.22286",
    archivePrefix = "arXiv",
    primaryClass = "quant-ph",
    month = "11",
    year = "2025"
}

@article{Cahill:1969it,
    author = "Cahill, Kevin E. and Glauber, R. J.",
    title = "{Ordered expansions in boson amplitude operators}",
    doi = "10.1103/PhysRev.177.1857",
    journal = "Phys. Rev.",
    volume = "177",
    pages = "1857--1881",
    year = "1969"
}

@article{Albarelli:2018ujl,
    author = "Albarelli, Francesco and Genoni, Marco G. and Paris, Matteo G. A. and Ferraro, Alessandro",
    title = "{Resource theory of quantum non-Gaussianity and Wigner negativity}",
    eprint = "1804.05763",
    archivePrefix = "arXiv",
    primaryClass = "quant-ph",
    doi = "10.1103/PhysRevA.98.052350",
    journal = "Phys. Rev. A",
    volume = "98",
    number = "5",
    pages = "052350",
    year = "2018"
}

@misc{Hong:2025dct,
      title={Oscillator-qubit generalized quantum signal processing for vibronic models: a case study of uracil cation}, 
      author={Jungsoo Hong and Seong Ho Kim and Seung Kyu Min and Joonsuk Huh},
      year={2025},
      eprint={2510.10495},
      archivePrefix={arXiv},
      primaryClass={quant-ph},
      url={https://arxiv.org/abs/2510.10495}, 
}

@ARTICLE{Liu:2024ncw,
  author={Liu, Yuan and Martyn, John M. and Sinanan-Singh, Jasmine and Smith, Kevin C. and Girvin, Steven M. and Chuang, Isaac L.},
  journal={IEEE Transactions on Signal Processing}, 
  title={Toward Mixed Analog-Digital Quantum Signal Processing: Quantum AD/DA Conversion and the Fourier Transform}, 
  year={2025},
  volume={73},
  number={},
  pages={3641-3655},
  doi={10.1109/TSP.2025.3599462}}

@article{Sinanan-Singh:2023xzs,
    author = "Sinanan-Singh, Jasmine and Mintzer, Gabriel L. and Chuang, Isaac L. and Liu, Yuan",
    title = "{Single-shot Quantum Signal Processing Interferometry}",
    eprint = "2311.13703",
    archivePrefix = "arXiv",
    primaryClass = "quant-ph",
    doi = "10.22331/q-2024-07-30-1427",
    journal = "Quantum",
    volume = "8",
    pages = "1427",
    year = "2024"
}

@article{Jordan:1928wi,
    author = "Jordan, Pascual and Wigner, Eugene P.",
    title = "{About the Pauli exclusion principle}",
    doi = "10.1007/BF01331938",
    journal = "Z. Phys.",
    volume = "47",
    pages = "631--651",
    year = "1928"
}

@article{Blais:2020wjs,
    author = "Blais, Alexandre and Grimsmo, Arne L. and Girvin, S. M. and Wallraff, Andreas",
    title = "{Circuit quantum electrodynamics}",
    eprint = "2005.12667",
    archivePrefix = "arXiv",
    primaryClass = "quant-ph",
    doi = "10.1103/RevModPhys.93.025005",
    journal = "Rev. Mod. Phys.",
    volume = "93",
    number = "2",
    pages = "025005",
    year = "2021"
}

@inproceedings{Stavenger:2022wzz,
    author = "Stavenger, Timothy J. and Crane, Eleanor and Smith, Kevin C. and Kang, Christopher T. and Girvin, Steven M. and Wiebe, Nathan",
    title = "{C2QA - Bosonic Qiskit}",
    booktitle = "{26th IEEE High Performance Extreme Computing}",
    eprint = "2209.11153",
    archivePrefix = "arXiv",
    primaryClass = "quant-ph",
    doi = "10.1109/HPEC55821.2022.9926318",
    month = "9",
    year = "2022"
}

@article{Araz:2024dcy,
    author = "Araz, Jack Y. and Grau, Matt and Montgomery, Jake and Ringer, Felix",
    title = "{Hybrid quantum simulations with qubits and qumodes on trapped-ion platforms}",
    eprint = "2410.07346",
    archivePrefix = "arXiv",
    primaryClass = "quant-ph",
    reportNumber = "JLAB-THY-24-4200",
    doi = "10.1103/kbv4-jj51",
    journal = "Phys. Rev. A",
    volume = "112",
    number = "1",
    pages = "012620",
    year = "2025"
}

@article{Abel:2025zxb,
    author = "Abel, Steven and Spannowsky, Michael and Williams, Simon",
    title = "{Real-time scattering processes with continuous-variable quantum computers}",
    eprint = "2502.01767",
    archivePrefix = "arXiv",
    primaryClass = "quant-ph",
    reportNumber = "IPPP/24/82",
    doi = "10.1103/q36d-w649",
    journal = "Phys. Rev. A",
    volume = "112",
    number = "1",
    pages = "012614",
    year = "2025"
}

@article{Cogburn:2026aqy,
    author = "Cogburn, Cameron V. and Grieninger, Sebastian and Kharzeev, Dmitri E.",
    title = "{Quantum Simulation of Nucleon-Antinucleon Interaction in Large-$N$ QCD$_2$ on an IBM Quantum Nighthawk Processor}",
    eprint = "2606.02574",
    archivePrefix = "arXiv",
    primaryClass = "quant-ph",
    reportNumber = "IQuS@UW-21-127, NT@UW-26-11",
    month = "6",
    year = "2026"
}

@article{Banerjee:2012pg,
    author = "Banerjee, D. and Dalmonte, M. and Muller, M. and Rico, E. and Stebler, P. and Wiese, U. -J. and Zoller, P.",
    title = "{Atomic Quantum Simulation of Dynamical Gauge Fields coupled to Fermionic Matter: From String Breaking to Evolution after a Quench}",
    eprint = "1205.6366",
    archivePrefix = "arXiv",
    primaryClass = "cond-mat.quant-gas",
    doi = "10.1103/PhysRevLett.109.175302",
    journal = "Phys. Rev. Lett.",
    volume = "109",
    pages = "175302",
    year = "2012"
}

@article{Gustafson:2024kym,
    author = "Gustafson, Erik J. and Ji, Yao and Lamm, Henry and Murairi, Edison M. and Perez, Sebastian Osorio and Zhu, Shuchen",
    title = "{Primitive quantum gates for an SU(3) discrete subgroup: {\ensuremath{\Sigma}}(36{\texttimes}3)}",
    eprint = "2405.05973",
    archivePrefix = "arXiv",
    primaryClass = "hep-lat",
    reportNumber = "FERMILAB-PUB-24-0132-SQMS-T, TUM-HEP-1506/24",
    doi = "10.1103/PhysRevD.110.034515",
    journal = "Phys. Rev. D",
    volume = "110",
    number = "3",
    pages = "034515",
    year = "2024"
}

@article{Klco:2018zqz,
    author = "Klco, Natalie and Savage, Martin J.",
    title = "{Digitization of scalar fields for quantum computing}",
    eprint = "1808.10378",
    archivePrefix = "arXiv",
    primaryClass = "quant-ph",
    reportNumber = "INT-PUB-18-044",
    doi = "10.1103/PhysRevA.99.052335",
    journal = "Phys. Rev. A",
    volume = "99",
    number = "5",
    pages = "052335",
    year = "2019"
}

@article{Lloyd:1998jk,
    author = "Lloyd, Seth and Braunstein, Samuel L.",
    title = "{Quantum computation over continuous variables}",
    eprint = "quant-ph/9810082",
    archivePrefix = "arXiv",
    doi = "10.1103/PhysRevLett.82.1784",
    journal = "Phys. Rev. Lett.",
    volume = "82",
    pages = "1784--1787",
    year = "1999"
}

@article{Athanasakos:2026upp,
    author = "Athanasakos, Dimitrios and Tejedor-Garc{\'\i}a, Gloria and Araz, Jack Y. and Ram{\^o}a, Mafalda and Sambasivam, Bharath and Economou, Sophia E. and Ringer, Felix",
    title = "{Continuous-variable ADAPT-VQE for bosonic lattice models}",
    eprint = "2606.05297",
    archivePrefix = "arXiv",
    primaryClass = "quant-ph",
    month = "6",
    year = "2026"
}

@article{Liu:2024mbr,
  title = {Hybrid Oscillator-Qubit Quantum Processors: Instruction Set Architectures, Abstract Machine Models, and Applications},
  author = {Liu, Yuan and Singh, Shraddha and Smith, Kevin C. and Crane, Eleanor and Martyn, John M. and Eickbusch, Alec and Schuckert, Alexander and Li, Richard D. and Sinanan-Singh, Jasmine and Soley, Micheline B. and Tsunoda, Takahiro and Chuang, Isaac L. and Wiebe, Nathan and Girvin, Steven M.},
  journal = {PRX Quantum},
  volume = {7},
  issue = {1},
  pages = {010201},
  numpages = {166},
  year = {2026},
  month = {Jan},
  publisher = {American Physical Society},
  doi = {10.1103/4rf7-9tfx},
  url = {https://link.aps.org/doi/10.1103/4rf7-9tfx}
}

@article{Davoudi:2021ney,
    author = "Davoudi, Zohreh and Linke, Norbert M. and Pagano, Guido",
    title = "{Toward simulating quantum field theories with controlled phonon-ion dynamics: A hybrid analog-digital approach}",
    eprint = "2104.09346",
    archivePrefix = "arXiv",
    primaryClass = "quant-ph",
    reportNumber = "UMD-PP-021-02",
    doi = "10.1103/PhysRevResearch.3.043072",
    journal = "Phys. Rev. Res.",
    volume = "3",
    number = "4",
    pages = "043072",
    year = "2021"
}

@misc{Crane:2024tlj,
    author={Eleanor Crane and Kevin C. Smith and Teague Tomesh and Alec Eickbusch and John M. Martyn and Stefan Kühn and Lena Funcke and Michael Austin DeMarco and Isaac L. Chuang and Nathan Wiebe and Alexander Schuckert and Steven M. Girvin},
    title = "Hybrid Oscillator-Qubit Quantum Processors: Simulating Fermions, Bosons, and Gauge Fields",
    eprint = "2409.03747",
    archivePrefix = "arXiv",
    primaryClass = "quant-ph",
    month = "9",
    year = "2024"
}

@article{Than:2025gso,
  author = "Than, Anton T. and Kadam, Saurabh V. and Vikramaditya, Vinay and Nguyen, Nhung H. and Liu, Xingxin and Davoudi, Zohreh and Green, Alaina M. and Linke, Norbert M.",
  title = "{Observation of quantum-field-theory dynamics on a spin-phonon quantum computer}",
  eprint = "2509.11477",
  archivePrefix = "arXiv",
  primaryClass = "quant-ph",
  reportNumber = "UMD-PP-025-05, IQuS@UW-21-107",
  month = "9",
  year = "2025"
}

@article{Haase:2020kaj,
    author = "Haase, Jan F. and Dellantonio, Luca and Celi, Alessio and Paulson, Danny and Kan, Angus and Jansen, Karl and Muschik, Christine A.",
    title = "{A resource efficient approach for quantum and classical simulations of gauge theories in particle physics}",
    eprint = "2006.14160",
    archivePrefix = "arXiv",
    primaryClass = "quant-ph",
    reportNumber = "DESY-20-146",
    doi = "10.22331/q-2021-02-04-393",
    journal = "Quantum",
    volume = "5",
    pages = "393",
    year = "2021"
}

@article{Leibfried:2003zz,
  title = {Quantum dynamics of single trapped ions},
  author = {Leibfried, D. and Blatt, R. and Monroe, C. and Wineland, D.},
  journal = {Rev. Mod. Phys.},
  volume = {75},
  issue = {1},
  pages = {281--324},
  numpages = {0},
  year = {2003},
  month = {Mar},
  publisher = {American Physical Society},
  doi = {10.1103/RevModPhys.75.281},
  url = {https://link.aps.org/doi/10.1103/RevModPhys.75.281}
}

@misc{Wineland:1997mg,
      title={Experimental issues in coherent quantum-state manipulation of trapped atomic ions}, 
      author={D. J. Wineland and C. Monroe and W. M. Itano and D. Leibfried and B. E. King and D. M. Meekhof},
      year={1998},
      eprint={quant-ph/9710025},
      archivePrefix={arXiv},
      primaryClass={quant-ph},
      url={https://arxiv.org/abs/quant-ph/9710025}, 
}

@article{Chen:2022auh,
	author = {Chen, Wentao and Lu, Yao and Zhang, Shuaining and Zhang, Kuan and Huang, Guanhao and Qiao, Mu and Su, Xiaolu and Zhang, Jialiang and Zhang, Jing-Ning and Banchi, Leonardo and Kim, M. S. and Kim, Kihwan},
	date = {2023/06/01},
	doi = {10.1038/s41567-023-01952-5},
	id = {Chen2023},
	isbn = {1745-2481},
	journal = {Nature Physics},
	number = {6},
	pages = {877--883},
	title = {Scalable and programmable phononic network with trapped ions},
	url = {https://doi.org/10.1038/s41567-023-01952-5},
	volume = {19},
	year = {2023}
}

@article{Katz:2022gra,
	archiveprefix = {arXiv},
	author = {Katz, Or and Monroe, Christopher},
	doi = {10.1103/PhysRevLett.131.033604},
	eprint = {2207.13653},
	journal = {Phys. Rev. Lett.},
	number = {3},
	pages = {033604},
	primaryclass = {quant-ph},
	title = {{Programmable Quantum Simulations of Bosonic Systems with Trapped Ions}},
	volume = {131},
	year = {2023}}

@article{Chen:2021nbo,
	author = {Chen, Wentao and Gan, Jaren and Zhang, Jing-Ning and Matuskevich, Dzmitry and Kim, Kihwan},
	doi = {10.1088/1674-1056/ac01e3},
	journal = {Chinese Physics B},
	month = {jun},
	number = {6},
	pages = {060311},
	publisher = {{IOP} Publishing},
	title = {Quantum computation and simulation with vibrational modes of trapped ions},
	url = {https://doi.org/10.1088%2F1674-1056%2Fac01e3},
	volume = {30},
	year = 2021}

@article{Heris:2026dmx,
    author = "Heris, Masoud Hakimi and Liu, Yuan and Mueller, Frank",
    title = "{HyPulse: A Pulse Synthesis Framework for Hybrid Qubit-Oscillator Gates on Trapped-Ion Platform}",
    eprint = "2604.26804",
    archivePrefix = "arXiv",
    primaryClass = "quant-ph",
    month = "4",
    year = "2026"
}

@article{Saner:2025nrq,
    author = "Saner, S. and B{\u{a}}z{\u{a}}van, O. and Webb, D. J. and Araneda, G. and Ballance, C. J. and Srinivas, R. and Lucas, D. M. and Berm{\'u}dez, A.",
    title = "{Real-Time Observation of Aharonov-Bohm Interference in a $\mathbb{Z}_2$ Lattice Gauge Theory on a Hybrid Qubit-Oscillator Quantum Computer}",
    eprint = "2507.19588",
    archivePrefix = "arXiv",
    primaryClass = "quant-ph",
    month = "7",
    year = "2025"
}

@article{Mohapatra:2026tyk,
    author = "Mohapatra, Shubdeep and Liu, Yuan and Zhang, Eddy Z. and Zhou, Huiyang",
    title = "{HyQBench: A Benchmark Suite for Hybrid CV-DV Quantum Computing}",
    eprint = "2603.04398",
    archivePrefix = "arXiv",
    primaryClass = "quant-ph",
    month = "3",
    year = "2026"
}

@article{Briceno:2023xcm,
    author = "Brice{\~n}o, Ra{\'u}l A. and Edwards, Robert G. and Eaton, Miller and Gonz{\'a}lez-Arciniegas, Carlos and Pfister, Olivier and Siopsis, George",
    title = "{Toward coherent quantum computation of scattering amplitudes with a measurement-based photonic quantum processor}",
    eprint = "2312.12613",
    archivePrefix = "arXiv",
    primaryClass = "quant-ph",
    reportNumber = "JLAB-THY-24-3948",
    doi = "10.1103/PhysRevResearch.6.043065",
    journal = "Phys. Rev. Res.",
    volume = "6",
    number = "4",
    pages = "043065",
    year = "2024"
}

@misc{Kemper:2025ldr,
      title={Hybrid continuous-discrete-variable quantum computing: a guide to utility}, 
      author={A. F. Kemper and Antonios Alvertis and Muhammad Asaduzzaman and Bojko N. Bakalov and Dror Baron and Joel Bierman and Blake Burgstahler and Srikar Chundury and Elin Ranjan Das and Jim Furches and Fucheng Guo and Raghav G. Jha and Katherine Klymko and Arvin Kushwaha and Ang Li and Aishwarya Majumdar and Carlos Ortiz Marrero and Shubdeep Mohapatra and Christopher Mori and Frank Mueller and Doru Thom Popovici and Tim Stavenger and Mastawal Tirfe and Norm M. Tubman and Muqing Zheng and Huiyang Zhou and Yuan Liu},
      year={2025},
      eprint={2511.13882},
      archivePrefix={arXiv},
      primaryClass={quant-ph},
      url={https://arxiv.org/abs/2511.13882}, 
}

@article{Wiebe:2008cbb,
    author = "Wiebe, Nathan and Berry, Dominic W. and H{\o}yer, Peter and Sanders, Barry C.",
    title = "{Higher order decompositions of ordered operator exponentials}",
    eprint = "0812.0562",
    archivePrefix = "arXiv",
    primaryClass = "math-ph",
    doi = "10.1088/1751-8113/43/6/065203",
    journal = "J. Phys. A",
    volume = "43",
    number = "6",
    pages = "065203",
    year = "2010"
}

@article{Suzuki:2002zuq,
title = {Fractal decomposition of exponential operators with applications to many-body theories and Monte Carlo simulations},
journal = {Physics Letters A},
volume = {146},
number = {6},
pages = {319-323},
year = {1990},
issn = {0375-9601},
doi = {https://doi.org/10.1016/0375-9601(90)90962-N},
url = {https://www.sciencedirect.com/science/article/pii/037596019090962N},
author = {Masuo Suzuki}
}

@article{Ale:2024uxf,
    author = "Ale, Victor and Bauer, Nora M. and Jha, Raghav G. and Ringer, Felix and Siopsis, George",
    title = "{Quantum computation of SU(2) lattice gauge theory with continuous variables}",
    eprint = "2410.14580",
    archivePrefix = "arXiv",
    primaryClass = "hep-lat",
    reportNumber = "JLAB-THY-24-4217",
    doi = "10.1007/JHEP06(2025)084",
    journal = "JHEP",
    volume = "06",
    pages = "084",
    year = "2025"
}

@article{Matsos:2024qer,
    author = "Matsos, V. G. and Valahu, C. H. and Millican, M. J. and Navickas, T. and Kolesnikow, X. C. and Biercuk, M. J. and Tan, T. R.",
    title = "{Universal quantum gate set for Gottesman{\textendash}Kitaev{\textendash}Preskill logical qubits}",
    eprint = "2409.05455",
    archivePrefix = "arXiv",
    primaryClass = "quant-ph",
    doi = "10.1038/s41567-025-03002-8",
    journal = "Nature Phys.",
    volume = "21",
    number = "10",
    pages = "1664--1669",
    year = "2025"
}

@article{Mercurio:2025alt,
  title = {Quantum{T}oolbox.jl: {A}n efficient {J}ulia framework for simulating open quantum systems},
  author = {Mercurio, Alberto and Huang, Yi-Te and Cai, Li-Xun and Chen, Yueh-Nan and Savona, Vincenzo and Nori, Franco},
  journal = {{Quantum}},
  issn = {2521-327X},
  publisher = {{Verein zur F{\"{o}}rderung des Open Access Publizierens in den Quantenwissenschaften}},
  volume = {9},
  pages = {1866},
  month = sep,
  year = {2025},
  doi = {10.22331/q-2025-09-29-1866},
  url = {https://doi.org/10.22331/q-2025-09-29-1866}
}

@article{Bezanson:2014pyv,
    title={Julia: A fresh approach to numerical computing},
    author={Bezanson, Jeff and Edelman, Alan and Karpinski, Stefan and Shah, Viral B},
    journal={SIAM {R}eview},
    volume={59},
    number={1},
    pages={65--98},
    year={2017},
    publisher={SIAM},
    doi={10.1137/141000671},
    url={https://epubs.siam.org/doi/10.1137/141000671}
}

@misc{McGarry:2026mkk,
      title={Programmable quantum simulation of anharmonic dynamics}, 
      author={Cameron McGarry and Teerawat Chalermpusitarak and Kai Schwennicke and Frank Scuccimarra and Maverick J. Millican and Vassili G. Matsos and Christophe H. Valahu and Prachi Nagpal and Hon-Kwan Chan and Henry L. Nourse and Ivan Kassal and Ting Rei Tan},
      year={2026},
      eprint={2603.04744},
      archivePrefix={arXiv},
      primaryClass={quant-ph},
      url={https://arxiv.org/abs/2603.04744}, 
}

@article{Gottesman:2000di,
    author = "Gottesman, Daniel and Kitaev, Alexei and Preskill, John",
    title = "{Encoding a qubit in an oscillator}",
    eprint = "quant-ph/0008040",
    archivePrefix = "arXiv",
    reportNumber = "CALT-68-2273, CALT-68-2273",
    doi = "10.1103/PhysRevA.64.012310",
    journal = "Phys. Rev. A",
    volume = "64",
    pages = "012310",
    year = "2001"
}

@article{Brock:2024vkc,
    author = "Brock, Benjamin L. and Singh, Shraddha and Eickbusch, Alec and Sivak, Volodymyr V. and Ding, Andy Z. and Frunzio, Luigi and Girvin, Steven M. and Devoret, Michel H.",
    title = "{Quantum error correction of qudits beyond break-even}",
    eprint = "2409.15065",
    archivePrefix = "arXiv",
    primaryClass = "quant-ph",
    doi = "10.1038/s41586-025-08899-y",
    journal = "Nature",
    volume = "641",
    number = "8063",
    pages = "612--618",
    year = "2025"
}

@article{Fluhmann:2018rvj,
    author = {Fl{\"u}hmann, Christa and Nguyen, Thanh Long and Marinelli, Matteo and Negnevitsky, Vlad and Mehta, Karan and Home, Jonathan},
    title = "{Encoding a qubit in a trapped-ion mechanical oscillator}",
    eprint = "1807.01033",
    archivePrefix = "arXiv",
    primaryClass = "quant-ph",
    doi = "10.1038/s41586-019-0960-6",
    journal = "Nature",
    volume = "566",
    number = "7745",
    pages = "513--517",
    year = "2019"
}

@article{deNeeve:2020ugg,
	author = {de Neeve, Brennan and Nguyen, Thanh-Long and Behrle, Tanja and Home, Jonathan P.},
	date = {2022/03/01},
	doi = {10.1038/s41567-021-01487-7},
	id = {de Neeve2022},
	isbn = {1745-2481},
	journal = {Nature Physics},
	number = {3},
	pages = {296--300},
	title = {Error correction of a logical grid state qubit by dissipative pumping},
	url = {https://doi.org/10.1038/s41567-021-01487-7},
	volume = {18},
	year = {2022}
}

@article{Dutta:2024cso,
	author = {Dutta, Rishab and Cabral, Delmar G. A. and Lyu, Ningyi and Vu, Nam P. and Wang, Yuchen and Allen, Brandon and Dan, Xiaohan and Corti{\~n}as, Rodrigo G. and Khazaei, Pouya and Sch{\"a}fer, Max and Albornoz, Alejandro C. C. d. and Smart, Scott E. and Nie, Scott and Devoret, Michel H. and Mazziotti, David A. and Narang, Prineha and Wang, Chen and Whitfield, James D. and Wilson, Angela K. and Hendrickson, Heidi P. and Lidar, Daniel A. and P{\'e}rez-Bernal, Francisco and Santos, Lea F. and Kais, Sabre and Geva, Eitan and Batista, Victor S.},
	date = {2024/08/13},
	doi = {10.1021/acs.jctc.4c00544},
	isbn = {1549-9618},
	journal = {Journal of Chemical Theory and Computation},
	journal1 = {Journal of Chemical Theory and Computation},
	journal2 = {J. Chem. Theory Comput.},
	month = {08},
	number = {15},
	pages = {6426--6441},
	publisher = {American Chemical Society},
	title = {Simulating Chemistry on Bosonic Quantum Devices},
	type = {doi: 10.1021/acs.jctc.4c00544},
	url = {https://doi.org/10.1021/acs.jctc.4c00544},
	volume = {20},
	year = {2024},
	year1 = {2024}
}

@article{Hu:2018mmi,
	author = {Hu, L. and Ma, Y. and Cai, W. and Mu, X. and Xu, Y. and Wang, W. and Wu, Y. and Wang, H. and Song, Y. P. and Zou, C. -L. and Girvin, S. M. and Duan, L-M. and Sun, L.},
	date = {2019/05/01},
	doi = {10.1038/s41567-018-0414-3},
	id = {Hu2019},
	isbn = {1745-2481},
	journal = {Nature Physics},
	number = {5},
	pages = {503--508},
	title = {Quantum error correction and universal gate set operation on a binomial bosonic logical qubit},
	url = {https://doi.org/10.1038/s41567-018-0414-3},
	volume = {15},
	year = {2019}}

@article{Campagne-Ibarcq:2019nmy,
	author = {Campagne-Ibarcq, P. and Eickbusch, A. and Touzard, S. and Zalys-Geller, E. and Frattini, N. E. and Sivak, V. V. and Reinhold, P. and Puri, S. and Shankar, S. and Schoelkopf, R. J. and Frunzio, L. and Mirrahimi, M. and Devoret, M. H.},
	date = {2020/08/01},
	doi = {10.1038/s41586-020-2603-3},
	id = {Campagne-Ibarcq2020},
	isbn = {1476-4687},
	journal = {Nature},
	number = {7821},
	pages = {368--372},
	title = {Quantum error correction of a qubit encoded in grid states of an oscillator},
	url = {https://doi.org/10.1038/s41586-020-2603-3},
	volume = {584},
	year = {2020}}

@article{Ale:2025sxz,
    author = "Ale, Victor and Rainaldi, Tommaso and Rico, Enrique and Ringer, Felix and Siopsis, George",
    title = "{Simulating quantum electrodynamics in 2+1 dimensions with qubits and qumodes}",
    eprint = "2511.14506",
    archivePrefix = "arXiv",
    primaryClass = "quant-ph",
    doi = "10.1007/JHEP04(2026)122",
    journal = "JHEP",
    volume = "04",
    pages = "122",
    year = "2026"
}

@ARTICLE{Clark:2021mxw,
  author={Clark, Susan M. and Lobser, Daniel and Revelle, Melissa C. and Yale, Christopher G. and Bossert, David and Burch, Ashlyn D. and Chow, Matthew N. and Hogle, Craig W. and Ivory, Megan and Pehr, Jessica and Salzbrenner, Bradley and Stick, Daniel and Sweatt, William and Wilson, Joshua M. and Winrow, Edward and Maunz, Peter},
  journal={IEEE Transactions on Quantum Engineering}, 
  title={Engineering the Quantum Scientific Computing Open User Testbed}, 
  year={2021},
  volume={2},
  number={},
  pages={1-32},
  doi={10.1109/TQE.2021.3096480}
}

@article{Fluhmann:2019slh,
  title = {Direct Characteristic-Function Tomography of Quantum States of the Trapped-Ion Motional Oscillator},
  author = {Fl\"uhmann, C. and Home, J. P.},
  journal = {Phys. Rev. Lett.},
  volume = {125},
  issue = {4},
  pages = {043602},
  numpages = {6},
  year = {2020},
  month = {Jul},
  publisher = {American Physical Society},
  doi = {10.1103/PhysRevLett.125.043602},
  url = {https://link.aps.org/doi/10.1103/PhysRevLett.125.043602}
}

@article{Bazavan:2024yya,
	author = {B{\u a}z{\u a}van, O. and Saner, S. and Webb, D. J. and Ainley, E. M. and Drmota, P. and Nadlinger, D. P. and Araneda, G. and Lucas, D. M. and Ballance, C. J. and Srinivas, R.},
	date = {2026/05/01},
	doi = {10.1038/s41567-026-03222-6},
	id = {B{\u a}z{\u a}van2026},
	isbn = {1745-2481},
	journal = {Nature Physics},
	number = {5},
	pages = {757--762},
	title = {Squeezing, trisqueezing and quadsqueezing in a hybrid oscillator--spin system},
	url = {https://doi.org/10.1038/s41567-026-03222-6},
	volume = {22},
	year = {2026}
}

@article{Braunstein:2005zz,
  title = {Quantum information with continuous variables},
  author = {Braunstein, Samuel L. and van Loock, Peter},
  journal = {Rev. Mod. Phys.},
  volume = {77},
  issue = {2},
  pages = {513--577},
  numpages = {0},
  year = {2005},
  month = {Jun},
  publisher = {American Physical Society},
  doi = {10.1103/RevModPhys.77.513},
  url = {https://link.aps.org/doi/10.1103/RevModPhys.77.513}
}

@article{Andersen:2014xqr,
	author = {Andersen, Ulrik L. and Neergaard-Nielsen, Jonas S. and van Loock, Peter and Furusawa, Akira},
	date = {2015/09/01},
	doi = {10.1038/nphys3410},
	id = {Andersen2015},
	isbn = {1745-2481},
	journal = {Nature Physics},
	number = {9},
	pages = {713--719},
	title = {Hybrid discrete- and continuous-variable quantum information},
	url = {https://doi.org/10.1038/nphys3410},
	volume = {11},
	year = {2015}
}

@article{Shen:2017txh,
	author = {Shen, Yangchao and Lu, Yao and Zhang, Kuan and Zhang, Junhua and Zhang, Shuaining and Huh, Joonsuk and Kim, Kihwan},
	doi = {10.1039/C7SC04602B},
	issue = {4},
	journal = {Chem. Sci.},
	pages = {836-840},
	publisher = {The Royal Society of Chemistry},
	title = {Quantum optical emulation of molecular vibronic spectroscopy using a trapped-ion device},
	url = {http://dx.doi.org/10.1039/C7SC04602B},
	volume = {9},
	year = {2018}
}

@article{Valahu:2022kfg,
	author = {Valahu, C. H. and Olaya-Agudelo, V. C. and MacDonell, R. J. and Navickas, T. and Rao, A. D. and Millican, M. J. and P{\'e}rez-S{\'a}nchez, J. B. and Yuen-Zhou, J. and Biercuk, M. J. and Hempel, C. and Tan, T. R. and Kassal, I.},
	date = {2023/11/01},
	doi = {10.1038/s41557-023-01300-3},
	id = {Valahu2023},
	isbn = {1755-4349},
	journal = {Nature Chemistry},
	number = {11},
	pages = {1503--1508},
	title = {Direct observation of geometric-phase interference in dynamics around a conical intersection},
	url = {https://doi.org/10.1038/s41557-023-01300-3},
	volume = {15},
	year = {2023}}

@article{Meekhof:1996zz,
  title = {Generation of Nonclassical Motional States of a Trapped Atom},
  author = {Meekhof, D. M. and Monroe, C. and King, B. E. and Itano, W. M. and Wineland, D. J.},
  journal = {Phys. Rev. Lett.},
  volume = {76},
  issue = {11},
  pages = {1796--1799},
  numpages = {0},
  year = {1996},
  month = {Mar},
  publisher = {American Physical Society},
  doi = {10.1103/PhysRevLett.76.1796},
  url = {https://link.aps.org/doi/10.1103/PhysRevLett.76.1796}
}

@phdthesis{Fluhmann:2019thesis,
    title = {Encoding a qubit in the motion of a trapped ion using superpositions of displaced squeezed states},
    author = {Flühmann, Christa},
    school = {ETH Zürich},
    year = {2019},
    url = {https://doi.org/10.3929/ethz-b-000355836}
}

@misc{Revelle:2020xpg,
      title={Phoenix and Peregrine Ion Traps}, 
      author={Melissa C. Revelle},
      year={2020},
      eprint={2009.02398},
      archivePrefix={arXiv},
      primaryClass={physics.app-ph},
      url={https://arxiv.org/abs/2009.02398}, 
}

@article{Haljan:2004nbn,
  title = {Spin-Dependent Forces on Trapped Ions for Phase-Stable Quantum Gates and Entangled States of Spin and Motion},
  author = {Haljan, P. C. and Brickman, K.-A. and Deslauriers, L. and Lee, P. J. and Monroe, C.},
  journal = {Phys. Rev. Lett.},
  volume = {94},
  issue = {15},
  pages = {153602},
  numpages = {4},
  year = {2005},
  month = {Apr},
  publisher = {American Physical Society},
  doi = {10.1103/PhysRevLett.94.153602},
  url = {https://link.aps.org/doi/10.1103/PhysRevLett.94.153602}
}

@article{Um:2015vwj,
	author = {Um, Mark and Zhang, Junhua and Lv, Dingshun and Lu, Yao and An, Shuoming and Zhang, Jing-Ning and Nha, Hyunchul and Kim, M. S. and Kim, Kihwan},
	date = {2016/04/21},
	doi = {10.1038/ncomms11410},
	id = {Um2016},
	isbn = {2041-1723},
	journal = {Nature Communications},
	number = {1},
	pages = {11410},
	title = {Phonon arithmetic in a trapped ion system},
	url = {https://doi.org/10.1038/ncomms11410},
	volume = {7},
	year = {2016}
}

@article{Matsos:2023iqx,
  title = {Robust and Deterministic Preparation of Bosonic Logical States in a Trapped Ion},
  author = {Matsos, V. G. and Valahu, C. H. and Navickas, T. and Rao, A. D. and Millican, M. J. and Kolesnikow, X. C. and Biercuk, M. J. and Tan, T. R.},
  journal = {Phys. Rev. Lett.},
  volume = {133},
  issue = {5},
  pages = {050602},
  numpages = {7},
  year = {2024},
  month = {Jul},
  publisher = {American Physical Society},
  doi = {10.1103/PhysRevLett.133.050602},
  url = {https://link.aps.org/doi/10.1103/PhysRevLett.133.050602}
}

@article{Lv:2017tmn,
  title = {Quantum Simulation of the Quantum Rabi Model in a Trapped Ion},
  author = {Lv, Dingshun and An, Shuoming and Liu, Zhenyu and Zhang, Jing-Ning and Pedernales, Julen S. and Lamata, Lucas and Solano, Enrique and Kim, Kihwan},
  journal = {Phys. Rev. X},
  volume = {8},
  issue = {2},
  pages = {021027},
  numpages = {11},
  year = {2018},
  month = {Apr},
  publisher = {American Physical Society},
  doi = {10.1103/PhysRevX.8.021027},
  url = {https://link.aps.org/doi/10.1103/PhysRevX.8.021027}
}

@article{Leibfried:1996zz,
  title = {Experimental Determination of the Motional Quantum State of a Trapped Atom},
  author = {Leibfried, D. and Meekhof, D. M. and King, B. E. and Monroe, C. and Itano, W. M. and Wineland, D. J.},
  journal = {Phys. Rev. Lett.},
  volume = {77},
  issue = {21},
  pages = {4281--4285},
  numpages = {0},
  year = {1996},
  month = {Nov},
  publisher = {American Physical Society},
  doi = {10.1103/PhysRevLett.77.4281},
  url = {https://link.aps.org/doi/10.1103/PhysRevLett.77.4281}
}

@article{Lv:2016obh,
  title = {Reconstruction of the Jaynes-Cummings field state of ionic motion in a harmonic trap},
  author = {Lv, Dingshun and An, Shuoming and Um, Mark and Zhang, Junhua and Zhang, Jing-Ning and Kim, M. S. and Kim, Kihwan},
  journal = {Phys. Rev. A},
  volume = {95},
  issue = {4},
  pages = {043813},
  numpages = {7},
  year = {2017},
  month = {Apr},
  publisher = {American Physical Society},
  doi = {10.1103/PhysRevA.95.043813},
  url = {https://link.aps.org/doi/10.1103/PhysRevA.95.043813}
}

\addtocontents{toc}{\protect\appendixtocsectionsonly}
\appendix

\section{Weak-angle Wigner negativity: tail nucleation and linear response}
\label{app:weak_theta_negativity}
In this Appendix we analyze the weak-angle limit of the Wigner negativity generated by
\begin{equation}
    U(\theta,c)=e^{-i\theta\cos(c\hat x)}
\end{equation}
acting on the vacuum. Particular care is required, because the small-$\theta$ expansion of the Wigner function is nonuniform in momentum, and because the nature of that nonuniformity depends on the spatial frequency $c$. We show that the weak-angle analysis yields two physically distinct regimes separated by $|\theta|\sim\zeta_0(c)e^{-c^2/4}$, derive a uniform formula valid across both, and obtain the corresponding asymptotics.
\subsection{Tail resummation and the function $\Hc$}
\label{app:weak_setup}
Setting $n=0$ in Eq.~\eqref{eq:Wigner_cos_n}, the exact Wigner function after application of the trigonometric gate is
\begin{equation}
W_0(x,p;\theta,c)
=
\frac{e^{-x^2}}{\pi}
\sum_{k=-\infty}^{\infty}
e^{-(p-kc/2)^2}
J_k\!\left(2\theta\sin(cx)\right).
\label{eq:W_exact_appendix}
\end{equation}
Extracting the overall Gaussian, $e^{-(p-kc/2)^2}=e^{-p^2}e^{kcp-k^2c^2/4}$,
\begin{align}
&W_0
=
\Wvac\,G,\nonumber\\
&G(x,p;\theta,c)
=
\sum_{k\in\mathbb Z}
J_k\!\left(2\theta\sin(cx)\right)
e^{kcp-k^2c^2/4},
\label{eq:G_appendix}
\end{align}
so that $\operatorname{sgn}W_0=\operatorname{sgn}G$ exactly. The mana is
\begin{equation}
    \mathcal M
    =
    \ln\!\left[
    \int dx\,dp\,
    |W_0|
    \right]
    =
    \ln\left(1+2\mathcal N_W\right),
\label{eq:mana_def_appendix}
\end{equation}
with the negative volume
\begin{equation}
    \mathcal N_W
    =
    \frac{1}{\pi}
    \int dx\,e^{-x^2}
    \int_{G<0}dp\,e^{-p^2}\left(-G\right).
\label{eq:NW_appendix}
\end{equation}
At any fixed $(x,p)$ one may expand
\begin{equation}
    W_0
    =
    \Wvac+\theta W^{(1)}+O(\theta^2),
    \quad
    \Wvac=\frac{1}{\pi}e^{-x^2-p^2}>0,
\label{eq:W_expansion_appendix}
\end{equation}
with, from $J_{\pm1}(z)=\pm z/2+O(z^3)$,
\begin{equation}
W^{(1)}(x,p)
=
\frac{e^{-x^2}}{\pi}\sin(cx)
\left[
e^{-(p-c/2)^2}-e^{-(p+c/2)^2}
\right].
\label{eq:W1_appendix}
\end{equation}
On any fixed compact set the Wigner function is positive for sufficiently small $\theta$, and since $\int dx\,dp\,W^{(1)}=0$ there is no ordinary $O(\theta)$ contribution to the $L_1$ norm from any fixed region of phase space. The negativity is instead controlled by where $\theta W^{(1)}$ overwhelms $\Wvac$, i.e.\ by the ratio
\begin{equation}
\frac{\theta W^{(1)}}{\Wvac}
=
\theta\sin(cx)
\left[
e^{cp-c^2/4}-e^{-cp-c^2/4}
\right],
\label{eq:ratio_appendix}
\end{equation}
which grows exponentially in the momentum tails. Let $s(x)=\sin(cx)$. For definiteness take $\theta s<0$ and consider the $p>0$ tail. The case $\theta s>0$ follows by $p\rightarrow-p$. With $\varepsilon=|\theta s|\ll1$ and nonnegative integer $k$,
\begin{equation}
    J_k(-2\varepsilon)
    =
    (-1)^k\frac{\varepsilon^k}{k!}
    \left[1+O(\varepsilon^2)\right],
\label{eq:Bessel_small_appendix}
\end{equation}
so that, using $\varepsilon^k e^{kcp-k^2c^2/4}=\zeta^k\exp[-\tfrac{c^2}{4}k(k-1)]$, the combined tail variable is
\begin{equation}
    \zeta
    =
    \varepsilon\,e^{cp-c^2/4}
    =
    |\theta\sin(cx)|\,e^{cp-c^2/4}.
\label{eq:zeta_tail}
\end{equation}
Taking $\varepsilon\rightarrow0$ and $p\rightarrow\infty$ with $\zeta$ fixed, the nonnegative-$k$ terms resum to
\begin{equation}
G
\sim
\Hc(\zeta),
\quad
\Hc(\zeta)
\equiv
\sum_{k=0}^{\infty}
\frac{(-\zeta)^{k}}{k!}
\exp\!\left[-\frac{c^2}{4}k(k-1)\right].
\label{eq:Hc_def}
\end{equation}
Terms with $k<0$ contribute $(\varepsilon^2/\zeta)^{|k|}\exp[-\tfrac{c^2}{4}|k|(|k|+1)]$ and are suppressed by additional powers of $\varepsilon^2$. The $O(\varepsilon^2)$ corrections in Eq.~\eqref{eq:Bessel_small_appendix} shift $\zeta_0$ by $O(\theta^2)$ and hence $\mathcal M$ by a relative $O(\theta^2\ln(1/|\theta|))$, which is negligible throughout. Equation~\eqref{eq:Hc_def} contains an infinite resummation of Bessel orders even though $\theta$ is arbitrarily small. Keeping only $J_0$ and $J_{\pm1}$ is sufficient only in the linear-response regime identified below, and never sufficient in the tail-nucleation regime. Because its coefficients decay as $e^{-c^2k^2/4}$, $\Hc$ is an entire function of order zero, and by Hadamard's theorem it possesses infinitely many zeros. We denote by $\zeta_0(c)$ its smallest positive zero, $\Hc(\zeta_0)=0$, and define
\begin{equation}
    A_c
    =
    -c\,\zeta_0\Hc'(\zeta_0) .
\label{eq:Ac_def}
\end{equation}
Numerical values are collected in Table~\ref{tab:mana_constants}. For all $c$ studied here the zeros of smallest modulus are real and positive. The series obeys a functional-differential equation that fixes most of the properties we need. Differentiating Eq.~\eqref{eq:Hc_def} term by term and shifting the summation index, using $(j+1)j=j(j-1)+2j$, gives the pantograph equation
\begin{equation}
\Hc'(\zeta)=-\Hc\!\left(e^{-c^2/2}\zeta\right),
\quad
\Hc(0)=1 ,
\label{eq:Hc_pantograph}
\end{equation}
equivalently $\Hc=\exp[-\tfrac{c^2}{4}(\hat D^2-\hat D)]e^{-\zeta}$ with $\hat D=\zeta\partial_\zeta$. Equations of this type are classical. For a contraction factor $\exp(-c^2/2)\in(0,1)$ the asymptotic behavior is oscillatory, with a modulation periodic in $\ln\zeta$, see Ref.~\cite{KatoMcLeod1971}. This is the origin of the zeros of $\Hc$. Evaluating Eq.~\eqref{eq:Hc_pantograph} at $\zeta_0$,
\begin{equation}
A_c=c\,\zeta_0\,\Hc\!\left(e^{-c^2/2}\zeta_0\right)>0 ,
\label{eq:Ac_pantograph}
\end{equation}
where positivity holds because $e^{-c^2/2}\zeta_0<\zeta_0$ lies below the smallest positive zero, so that $\Hc'(\zeta_0)<0$ follows. Equation~\eqref{eq:Hc_pantograph} is moreover form invariant under $\zeta\rightarrow e^{c^2/2}\zeta$, so successive zeros are asymptotically geometric with ratio $e^{c^2/2}$. In the lobe variable of Sec.~\ref{app:uniform} this is a spacing $\Delta u=c/2$, which quantifies the suppression of successive negative lobes and justifies retaining only the first. The two limiting behaviors of $\zeta_0(c)$ follow from the operator form. Writing $\hat D=\partial_{\ln\zeta}$ exhibits $\Hc$ as a Gaussian smoothing of $e^{-Z}$ in $\ln Z$, with $Z=\zeta e^{c^2/4}$, or explicitly
\begin{align}
\Hc(\zeta)
&=
\frac{1}{c\sqrt\pi}
\int_{-\infty}^{\infty}\!\!du\,
e^{-u^2/c^2}
e^{-Z\cos u}
\cos\!\left(Z\sin u\right).
\label{eq:Hc_integral_rep}
\end{align}
Expanding the exponent to quadratic order in $u$ shows that the integrand is sign-definite while $Z\lesssim c^{-2}$, so a sign change requires $Z=O(c^{-2})$ and hence $\zeta_0(c)=O(c^{-2})$ as $c\rightarrow0$, with numerically $\zeta_0(c)\,c^2\simeq0.66,\,0.83,\,0.90$ at $c=0.1,\,0.2,\,0.3$. The first zero migrates to large argument as the generator becomes effectively quadratic, which is one of the two mechanisms suppressing the mana in the Gaussian limit. The other, and the faster, is the theta-function factor in Eq.~\eqref{eq:Cc_def}, $\vartheta_2(0,e^{-\pi^2/c^2})\simeq2e^{-\pi^2/4c^2}\rightarrow0$. At large $c$, by contrast, only the first two terms of Eq.~\eqref{eq:Hc_def} survive near $\zeta\sim1$, giving
\begin{align}
\Hc(\zeta)&=1-\zeta+\frac{e^{-c^2/2}}{2}\zeta^2-\cdots,
\nonumber\\
\zeta_0(c)&=1+\frac{e^{-c^2/2}}{2}+O(e^{-c^2}),\,
A_c\rightarrow c .
\label{eq:Hc_largec}
\end{align}
From Eq.~\eqref{eq:zeta_tail}, the momentum at which the first negative lobe begins is
\begin{align}
p_0(x)
&=
\frac{1}{c}
\left[
\ln\!\left(
\frac{\zeta_0(c)}{|\theta\sin(cx)|}
\right)
+
\frac{c^2}{4}
\right]\nonumber
\\
&=
\mathcal P-\frac{1}{c}\ln|\sin(cx)| ,
\label{eq:p0_x}
\end{align}
with
\begin{equation}
\mathcal P(\theta,c)
=
\frac{1}{c}
\left[
\ln\!\left(\frac{\zeta_0(c)}{|\theta|}\right)
+\frac{c^2}{4}
\right]
\ge
\sqrt{\ln\frac{\zeta_0(c)}{|\theta|}} ,
\label{eq:P_def}
\end{equation}
the inequality following from the arithmetic--geometric mean and being saturated at $c=2\sqrt{\ln(\zeta_0/|\theta|)}$. The negative tail lies at $p>p_0(x)$ when $\theta\sin(cx)<0$ and at $p<-p_0(x)$ when $\theta\sin(cx)>0$. In either case there is exactly one lobe per $x$ at leading order, and no factor of two arises from the two signs of $p$. At fixed $c$,
\begin{equation}
    |p_0(x)|
    \sim
    \frac{1}{c}\ln\frac{1}{|\theta|}
    \quad
    (\theta\rightarrow0).
\label{eq:p0_scaling}
\end{equation}
\subsection{Uniform evaluation of the negative volume}
\label{app:uniform}
Write $|p|=p_0(x)+u$ with $u\ge0$, so that $\zeta=\zeta_0e^{cu}$. Near the zero, $\Hc(\zeta)=c\zeta_0\Hc'(\zeta_0)u+O(u^2)=-A_cu+O(u^2)$, while for larger $u$ the leading behavior of $-\Hc$ is $\zeta_0e^{cu}$. Both limits are captured by the interpolation
\begin{equation}
    -\Hc\!\left(\zeta_0e^{cu}\right)
    \simeq
    \frac{A_c}{c}\left(e^{cu}-1\right),
\label{eq:Hc_interp}
\end{equation}
which is exact to first order at $u=0$ and has the correct exponential growth. The momentum integral then evaluates in closed form,
\begin{align}
&\int_0^\infty\! du\,\left(e^{cu}-1\right)e^{-(p_0+u)^2}
=
\mathcal F(p_0,c),
\nonumber\\
&\mathcal F(p_0,c)
=
\frac{\sqrt\pi}{2}
\left[
e^{c^2/4-cp_0}\erfc\!\left(p_0-\frac{c}{2}\right)
-
\erfc(p_0)
\right],
\label{eq:Ffun_def}
\end{align}
and therefore
\begin{align}
e^{\mathcal M}
&=
1+
\frac{2A_c}{\pi c}
\int_{-\infty}^{\infty}dx\,
e^{-x^2}\,
\mathcal F\!\left[p_0(x),c\right]
\nonumber\\
&\times
\left[1+O(\mathcal P^{-1})\right].
\label{eq:mana_uniform_asymptotic}
\end{align}
Equation~\eqref{eq:mana_uniform_asymptotic} is a single quadrature, uniformly valid throughout the weak-angle region. The structure of $\mathcal F$ makes the regime dependence explicit. The integrand of Eq.~\eqref{eq:Ffun_def} has a saddle at
\begin{equation}
u_*=\frac{c}{2}-p_0 ,
\label{eq:saddle_appendix}
\end{equation}
so that for $p_0>c/2$ the saddle lies outside the physical range and the integral is dominated by the edge $u=0$, i.e.\ by the immediate neighborhood of the first zero of $\Hc$, whereas for $p_0<c/2$ the saddle lies inside the negative lobe and the integral is dominated by a bulk region far from that zero. Since $p_0(x)\ge\mathcal P$ with equality at the maxima of $|\sin(cx)|$, the crossover occurs at
\begin{equation}
c=2\mathcal P
\Longleftrightarrow
c^2=4\ln\frac{\zeta_0(c)}{|\theta|}
\Longleftrightarrow
|\theta|=\zeta_0(c)\,e^{-c^2/4} .
\label{eq:crossover_appendix}
\end{equation}
This is the condition under which the truncation of Eq.~\eqref{eq:Hc_interp} to its linear term, and hence the entire ``first-lobe edge'' picture, is legitimate.
\subsection{Tail-nucleation regime: $c<2\mathcal P$}
\label{app:regime_I}
For $p_0\gg c/2$, using the asymptotics of the complementary error function,
\begin{equation}
\mathcal F(p_0,c)
=
\frac{c}{4}\,
\frac{e^{-p_0^2}}{p_0^2}
\left[1+O(p_0^{-1})\right],
\label{eq:F_regimeI}
\end{equation}
so that
\begin{equation}
\mathcal N_W
\sim
\frac{A_c}{4\pi}
\int_{-\infty}^{\infty}dx\,
e^{-x^2}
\frac{e^{-p_0(x)^2}}{p_0(x)^2}.
\label{eq:N_intermediate}
\end{equation}
As $\theta\rightarrow0$, $\mathcal P\rightarrow\infty$ and the integral is dominated by the maxima of $|\sin(cx)|$,
\begin{equation}
    x_m
    =
    \frac{(m+\tfrac12)\pi}{c},
    \quad
    m\in\mathbb Z .
\end{equation}
Writing $x=x_m+\delta$ and using $-\ln|\sin(cx)|=\tfrac{c^2}{2}\delta^2+O(\delta^4)$, one has $p_0=\mathcal P+\tfrac{c}{2}\delta^2+O(\delta^4)$ and $p_0^2=\mathcal P^2+\mathcal Pc\,\delta^2+O(\delta^4)$. Retaining the Gaussian factor $e^{-x^2}=e^{-x_m^2}e^{-2x_m\delta-\delta^2}$ in full, Laplace's method gives
\begin{align}
\int dx\,
e^{-x^2}
\frac{e^{-p_0(x)^2}}{p_0(x)^2}
&\simeq
\frac{e^{-\mathcal P^2}}{\mathcal P^2}
\sqrt{\frac{\pi}{1+\mathcal Pc}}\,
\vartheta_2\!\left(0,e^{-\pi^2\lambda/c^2}\right),
\nonumber\\
\lambda&=\frac{\mathcal Pc}{1+\mathcal Pc},
\label{eq:Laplace_result}
\end{align}
where $\vartheta_2$ is the second Jacobi theta function and we used $\sum_m\exp[-x_m^2\lambda]=\vartheta_2(0,e^{-\pi^2\lambda/c^2})$. Hence the refined result
\begin{align}
&e^{\mathcal M}
=\nonumber\\
&1+
\frac{A_c}{2\pi}
\frac{e^{-\mathcal P^2}}{\mathcal P^2}
\sqrt{\frac{\pi}{1+\mathcal Pc}}\,
\vartheta_2\!\left(0,e^{-\pi^2\lambda/c^2}\right)
\left[1+O(\mathcal P^{-1})\right].
\label{eq:mana_refined_app}
\end{align}
Dropping the subleading $1/(1+\mathcal Pc)\rightarrow1/(\mathcal Pc)$ corrections gives the leading closed form
\begin{align}
\mathcal M
&\sim
C(c)\,
\mathcal P^{-5/2}e^{-\mathcal P^{2}},
\nonumber\\
C(c)
&=
-\frac{\sqrt{c}}{2\sqrt{\pi}}\,
\zeta_0(c)\Hc'\!\left[\zeta_0(c)\right]
\vartheta_2\!\left(0,e^{-\pi^2/c^2}\right),
\label{eq:Cc_def}
\end{align}
where we used $\mathcal M=\ln(1+2\mathcal N_W)\simeq2\mathcal N_W$. Three distinct effects contribute at relative order $\mathcal P^{-1}$: the $\erfc$ expansion in Eq.~\eqref{eq:F_regimeI}, which enters at $O(\mathcal P^{-2})$, the curvature of $\Hc$ and of $\zeta=\zeta_0e^{cu}$, entering at relative order $\tfrac{c}{2p_0}[1+\zeta_0\Hc''/\Hc']$, and the linear term $e^{-2x_m\delta}$ dropped in the naive Laplace evaluation, entering at relative order $(x_m^2-\tfrac12)/(\mathcal Pc)$. The last of these is resummed in Eq.~\eqref{eq:mana_refined_app}. Since $\mathcal P^{-1}\simeq c/\ln(1/|\theta|)$ is only logarithmically small, these corrections are of order tens of percent at practical gate strengths, and Eq.~\eqref{eq:Cc_def} should be regarded as fixing the exponent, not the prefactor. A convenient way to display the asymptotics is to plot $\ln\mathcal M+\mathcal P^2+\tfrac52\ln\mathcal P\rightarrow\ln C(c)$ instead of $\mathcal M$ itself. The leading exponential in Eq.~\eqref{eq:Cc_def} is
\begin{equation}
e^{-\mathcal P^2}
=
\exp\!\left\{
-\frac{1}{c^2}
\left[
\ln\!\left(\frac{\zeta_0(c)}{|\theta|}\right)
+\frac{c^2}{4}
\right]^2
\right\},
\label{eq:log_gaussian}
\end{equation}
so that at fixed $c$
\begin{equation}
\mathcal M=o(|\theta|^n)
\quad
\text{for every fixed }n>0 .
\label{eq:beyond_all_orders}
\end{equation}
The weak-angle Wigner negativity is therefore not linear, quadratic, or described by any finite algebraic order in $\theta$: it is a beyond-all-orders contribution associated with a phase-space region whose position itself diverges as $\theta\rightarrow0$. For $c\neq0$ and $\theta\neq0$ the state $\ket{\psi_\theta}=e^{-i\theta\cos(c\hat x)}\ket{0}$ is pure and non-Gaussian, so Hudson's theorem \cite{HUDSON1974249} requires a negative region. There is no contradiction with pointwise positivity because the negative region escapes to $|p|\sim c^{-1}\ln(1/|\theta|)$, i.e.\ the limits $\theta\rightarrow0$ and $|p|\rightarrow\infty$ do not commute. Equation~\eqref{eq:beyond_all_orders} is, however, a statement at fixed $c$. Equation~\eqref{eq:log_gaussian} appears to say that increasing $c$ monotonically weakens the log-Gaussian suppression, but $\mathcal P=c^{-1}\ln(\zeta_0/|\theta|)+c/4$ is itself non-monotonic in $c$ and, by Eq.~\eqref{eq:P_def}, attains its minimum exactly on the crossover line Eq.~\eqref{eq:crossover_appendix}. Past that line Eq.~\eqref{eq:Cc_def} ceases to apply, and the correct description is the one derived next.
\subsection{Linear response and a first-order bound}
\label{app:regime_II}
For $p_0\ll c/2$ the saddle Eq.~\eqref{eq:saddle_appendix} lies well inside the negative lobe, $\erfc(p_0-c/2)\rightarrow2$, and
\begin{equation}
\mathcal F(p_0,c)
\rightarrow
\sqrt\pi\,e^{c^2/4-cp_0}
=
\sqrt\pi\,\frac{|\theta\sin(cx)|}{\zeta_0(c)} .
\label{eq:F_regimeII}
\end{equation}
Substituting into Eq.~\eqref{eq:mana_uniform_asymptotic} and using Eq.~\eqref{eq:Ac_def},
\begin{align}
\mathcal N_W
&\rightarrow
\frac{-\Hc'(\zeta_0)}{\sqrt\pi}\,
|\theta|\int dx\,e^{-x^2}|\sin(cx)|
\nonumber\\
&\xrightarrow[c\rightarrow\infty]{}
\frac{2|\theta|}{\pi}.
\label{eq:NW_regimeII}
\end{align}
This is ordinary linear response: when the negative set of $W_0$ coincides at leading order with the negative set of $W^{(1)}$, the constraint $\int W^{(1)}=0$ gives $\mathcal N_W=\tfrac12|\theta|\,\|W^{(1)}\|_1$, and
\begin{equation}
\mathcal M
\simeq
\kappa(c)\,|\theta| ,
\quad
\kappa(c)
\equiv
\|W^{(1)}\|_1 .
\label{eq:mana_linear_app}
\end{equation}
The coefficient is available in closed form. From Eq.~\eqref{eq:W1_appendix},
\begin{equation}
\int dp\left|e^{-(p-c/2)^2}-e^{-(p+c/2)^2}\right|
=
2\sqrt\pi\,\erf\!\left(\frac{c}{2}\right),
\end{equation}
while the Fourier series $|\sin u|=\tfrac{2}{\pi}-\tfrac{4}{\pi}\sum_{n\ge1}\cos(2nu)/(4n^2-1)$ gives
\begin{align}
B(c)
&\equiv
\int dx\,e^{-x^2}|\sin(cx)|
\nonumber\\&=
\sqrt\pi
\left[
\frac{2}{\pi}
-\frac{4}{\pi}
\sum_{n=1}^{\infty}
\frac{e^{-n^2c^2}}{4n^2-1}
\right],
\end{align}
and therefore
\begin{align}
\kappa(c)
&=
\frac{2}{\sqrt\pi}B(c)\erf\!\left(\frac{c}{2}\right)
\nonumber\\
&=
\frac{4}{\pi}
\erf\!\left(\frac{c}{2}\right)
\left[
1-2\sum_{n=1}^{\infty}
\frac{e^{-n^2c^2}}{4n^2-1}
\right]
\nonumber\\
&\xrightarrow[c\rightarrow\infty]{}
\frac{4}{\pi}.
\label{eq:kappa_def}
\end{align}
The linear-response result is also the maximum attainable at first order. Since $\Wvac>0$ everywhere, on the set where $W_0<0$ one has $W_0-\Wvac<W_0<0$ and hence $|W_0|\le|W_0-\Wvac|$, and moreover $\{W_0<0\}\subset\{W_0-\Wvac<0\}$. Because $\int(W_0-\Wvac)=0$, the negative part of $W_0-\Wvac$ carries exactly half of its $L_1$ norm, so
\begin{equation}
\mathcal N_W
\le
\frac{1}{2}\left\|W_0-\Wvac\right\|_1
=
\frac{|\theta|}{2}\kappa(c)+O(\theta^2),
\end{equation}
and since $\mathcal M=\ln(1+2\mathcal N_W)$ increases with $\mathcal N_W$,
\begin{equation}
\mathcal M(\theta,c)
\le
\ln\!\left[1+\kappa(c)\,|\theta|\right]
=
\kappa(c)\,|\theta|+O(\theta^2)
\quad
\text{for all }c .
\label{eq:mana_bound_app}
\end{equation}
Equation~\eqref{eq:mana_linear_app} shows that the bound is saturated for $c>2\mathcal P$, so $\kappa(c)|\theta|$ is attained, not only bounded. At fixed weak $\theta$, the maximum resource extractable by tuning $c$ is therefore $4|\theta|/\pi$, approached at $c\simeq c_\times=2\sqrt{\ln(\zeta_0(c)/|\theta|)}$ and constant thereafter. In particular the two limits do not commute,
\begin{equation}
\lim_{c\rightarrow\infty}
\lim_{\theta\rightarrow0}
\frac{\mathcal M}{|\theta|}
=
0,
\quad
\lim_{\theta\rightarrow0}
\lim_{c\rightarrow\infty}
\frac{\mathcal M}{|\theta|}
=
\frac{4}{\pi}.
\label{eq:noncommuting_app}
\end{equation}
\subsection{Numerical checks}
\label{app:weak_numerics}
For the representative value $c=2$ used in the main text, Eq.~\eqref{eq:Hc_def} becomes
\begin{equation}
\mathcal H_2(\zeta)
=
1-\zeta+\frac{e^{-2}}{2}\zeta^2
-\frac{e^{-6}}{6}\zeta^3+\cdots ,
\end{equation}
with
\begin{equation}
    \zeta_0(2)
    =
    1.078\ldots,
    \quad
    \mathcal H_2'\!\left[\zeta_0(2)\right]
    =
    -0.856\ldots ,
\end{equation}
and $\vartheta_2(0,e^{-\pi^2/4})=1.087\ldots$, giving $C(2)=0.4\ldots$ Hence
\begin{equation}
\mathcal M
\sim
0.4\,
\mathcal P^{-5/2}e^{-\mathcal P^{2}},
\quad
\mathcal P=
\frac{
\ln(1.078/|\theta|)+1
}{2},
\label{eq:c2_asymptotic}
\end{equation}
valid for $|\theta|\ll\zeta_0(2)e^{-1}=0.397$, i.e.\ for $\mathcal P>c/2=1$. For $|\theta|$ approaching this value the linear-response bound $\mathcal M\le1.06|\theta|$ takes over. Three independent numerical schemes were used. The first evaluates Eq.~\eqref{eq:NW_appendix} directly, locating the zeros of $G$ and factoring $e^{-\mathcal P^2}$ analytically so that double precision suffices even when $\mathcal M\sim10^{-130}$. The second evaluates
\begin{align}
&W_0(x,p;\theta,c)
=\nonumber\\
&
\frac{e^{-x^2}}{2\pi\sqrt\pi}
\int dy\,
e^{ipy}e^{-y^2/4}
e^{-2i\theta\sin(cx)\sin(cy/2)}
\label{eq:W_fourier_appendix}
\end{align}
by fast Fourier transform, making no use of the Bessel expansion or of any tail approximation, and integrates $|W_0|$ over phase space. These two agree to four or five significant digits for $|\theta|\le0.2$, and both reproduce $\|W_0\|_1=1$ at $\theta=0$ to thirteen digits. The third, used for Fig.~\ref{fig:W0_Mana}, exploits the fact that in Eq.~\eqref{eq:W_exact_weak} the momentum dependence enters only through $z=2\theta\sin(cx)$, so that
\begin{align}
\left\|W_0\right\|_1
&=
\frac{1}{\pi}\int dx\,e^{-x^2} I_c(z),
\nonumber\\
I_c(z)
&=
\int dp\left|\sum_k e^{-(p-kc/2)^2}J_k(z)\right| ,
\label{eq:app_Iz_reduction}
\end{align}
and a single tabulation of $I_c$ serves every gate strength. The negative part is accumulated directly rather than as $[I_c(z)-\sqrt\pi]/2$, which would be a cancellation of nearly equal quantities at small $z$. Equation~\eqref{eq:Hc_def} itself suffers cancellation when evaluated in double precision below $c\simeq0.3$. Extended precision was used there, and all entries of Table~\ref{tab:mana_constants} were checked against a $60$-digit evaluation. The comparison with the analytic results is shown in Fig.~\ref{fig:W0_Mana}. At $c=10$ and $|\theta|=10^{-3}$, for instance, the exact mana is $1.26\times10^{-3}$, in agreement with Eq.~\eqref{eq:mana_linear_app} to $0.3\%$, whereas Eq.~\eqref{eq:Cc_def} would give $1.0\times10^{-5}$.
\subsection{Generality of the mechanism}
\label{app:generality}
The derivation above used the Jacobi--Anger structure of the cosine gate, but the mechanism behind it does not. Let $U=e^{-i\theta f(\hat x)}$ act on a pure Gaussian state of Wigner function $\Wvac>0$, and write
\begin{equation}
W_\theta
=
\Wvac+\theta W^{(1)}+O(\theta^2),
\quad
R
\equiv
\frac{W^{(1)}}{\Wvac} .
\label{eq:R_general}
\end{equation}
Three properties hold for any such $f$. First, $\Wvac$ is strictly positive, so on any fixed compact set $W_\theta>0$ once $\theta$ is small enough. Second, $\int dx\,dp\,W^{(1)}=0$ by trace preservation, so no fixed region contributes to $\|W_\theta\|_1$ at order $\theta$. Third, if $f$ is not quadratic the evolved state is pure and non-Gaussian, so Hudson's theorem \cite{HUDSON1974249} requires a negative region somewhere. It can only lie where $\theta|W^{(1)}|\gtrsim\Wvac$, and recedes as $\theta\rightarrow0$. Because $\Wvac$ is Gaussian,
\begin{equation}
\mathcal N_W
\lesssim
\sup_{|R|\ge1/|\theta|}\Wvac
\sim
e^{-\mathcal P(\theta)^2}
\label{eq:NW_general}
\end{equation}
up to algebraic prefactors, where $\mathcal P(\theta)$ is the phase-space distance at which $|R|$ first reaches $1/|\theta|$, generalizing Eq.~\eqref{eq:P_def}. The generator enters only through $R$, and whenever $\mathcal P\rightarrow\infty$ the negative volume is beyond all algebraic orders in $\theta$.

Only the rate depends on $f$. For the cosine gate $R$ grows exponentially in the momentum tail, giving $\mathcal P\sim c^{-1}\ln(1/|\theta|)$ and the log-Gaussian law of Eq.~\eqref{eq:Cc_def}. A polynomial generator grows algebraically instead, and the suppression is stronger. For the cubic phase gate $U=e^{-i\theta\hat x^3/3}$ acting on the vacuum, Eq.~\eqref{eq:R_general} is elementary,
\begin{equation}
R(x,p)=\frac{2}{3}p^3-2x^2p-p ,
\label{eq:R_cubic}
\end{equation}
so $\mathcal P\sim|\theta|^{-1/3}$ and $\mathcal N_W$ is suppressed as $\exp[-O(|\theta|^{-2/3})]$. This is the quantitative content of the remark in Ref.~\cite{Moore:2025qzc} that the negativity generated by a weak cubic gate is too small to resolve numerically.
\section{Large-angle stationary-phase asymptotics}
\label{app:large_theta}
In this Appendix we analyze the large-angle behavior of the Wigner logarithmic negativity generated by a single-coordinate phase gate
\begin{equation}
    U(\theta)=e^{-i\theta f(\hat x)}
\label{eq:A_U}
\end{equation}
acting on an arbitrary pure one-mode Gaussian state. We show that the scaling
\begin{equation}
    \mathcal M(\theta)
    =
    \frac{1}{2}\ln|\theta|+O(1)
\label{eq:A_generic_scaling}
\end{equation}
arises generically when the Wigner integral is dominated by regular nondegenerate stationary points. This result should not, however, be interpreted as a universal bound following from unitarity alone. Gaussian phase gates provide an immediate counterexample, and more degenerate stationary-point structures require a separate asymptotic analysis.
\subsection{Wigner representation and stationary-phase scaling}
The most general one-mode pure Gaussian state may be written in the position representation as
\begin{equation}
    \psi_G(x)
    =
    \left(\frac{2a}{\pi}\right)^{1/4}
    \exp\left[
       -(a+ib)(x-x_0)^2+ip_0x
    \right],
\label{eq:A_gaussian}
\end{equation}
with $a>0$. After the application of Eq.~\eqref{eq:A_U}, its Wigner function is
\begin{align}
W(x,p;\theta)
&=
K(x)
\int_{-\infty}^{\infty}dy\,
e^{-ay^2/2}
e^{iy[p-p_0+2b(x-x_0)]}
\nonumber\\
&\times
\exp\left\{
i\theta
\left[
f\left(x+\frac{y}{2}\right)
-
f\left(x-\frac{y}{2}\right)
\right]
\right\},
\label{eq:A_W}
\\
K(x)
&=
\frac{1}{2\pi}
\sqrt{\frac{2a}{\pi}}\,
e^{-2a(x-x_0)^2}.
\label{eq:A_K}
\end{align}
It is convenient to remove the momentum displacement and shear of the initial Gaussian by introducing
\begin{equation}
p
=
p_0-2b(x-x_0)+\theta v .
\label{eq:A_vdef}
\end{equation}
The Jacobian is $dx\,dp = |\theta|\,dx\,dv$. For definiteness we take $\theta>0$ below. Equation~\eqref{eq:A_W} becomes
\begin{equation}
    W(x,v;\theta)
    =
    K(x)
    \int_{-\infty}^{\infty}dy\,
    e^{-ay^2/2}
    e^{i\theta\Phi(y;x,v)},
\label{eq:A_osc}
\end{equation}
where
\begin{equation}
\Phi(y;x,v)
=
vy+
f\left(x+\frac{y}{2}\right)
-
f\left(x-\frac{y}{2}\right).
\label{eq:A_phase}
\end{equation}
The Wigner $L_1$ norm is therefore
\begin{equation}
e^{\mathcal M}
=
\theta
\int dx\,dv\,|W(x,v;\theta)|.
\label{eq:A_L1}
\end{equation}
The stationary points of the $y$ integral satisfy
\begin{equation}
0
=
\Phi_y
=
v+
\frac{1}{2}
\left[
f'\left(x+\frac{y}{2}\right)
+
f'\left(x-\frac{y}{2}\right)
\right].
\label{eq:A_stationary}
\end{equation}
Since $\Phi(y;x,v)$ is odd in $y$, every nonzero stationary point $y_*>0$ is accompanied by a stationary point $-y_*$. Moreover,
\begin{equation}
\Phi_{yy}(y;x,v)
=
\frac{1}{4}
\left[
f''\left(x+\frac{y}{2}\right)
-
f''\left(x-\frac{y}{2}\right)
\right]
\label{eq:A_second}
\end{equation}
is odd in $y$. Consider first a regular phase-space region in which the relevant stationary points are isolated and nondegenerate, $\Phi_{yy}(y_*;x,v)\neq0$. For a single pair $\pm y_*$, the standard stationary-phase expansion gives
\begin{align}
W(x,v;\theta)
&=
2K(x)e^{-ay_*^2/2}
\sqrt{
\frac{2\pi}
{\theta|\Phi_{yy}(y_*;x,v)|}
}
\nonumber\\
&\times
\cos\left[
\theta\Phi(y_*;x,v)
+
\frac{\pi}{4}
\operatorname{sgn}\Phi_{yy}(y_*;x,v)
\right]\nonumber\\
&+
O(\theta^{-3/2}).
\label{eq:A_SP}
\end{align}
Thus the Wigner function in a regular stationary-phase region has amplitude $O(\theta^{-1/2})$. The factor of $\theta$ in the phase-space measure in Eq.~\eqref{eq:A_L1} then implies a contribution of order $\sqrt{\theta}$ to the Wigner $L_1$ norm. More explicitly, on a regular region ${\cal R}$ containing one stationary pair and bounded away from caustics, rapid phase oscillations give $\langle|\cos\varphi|\rangle_\varphi=2/\pi$, and therefore
\begin{align}
&\theta
\int_{\cal R}dx\,dv\,|W|
=
C_{\cal R}\sqrt{\theta}
+
o(\sqrt{\theta}),
\label{eq:A_regular_L1}
\\
&C_{\cal R}
=
\frac{4\sqrt{2\pi}}{\pi}
\int_{\cal R}
dx\,dv\,
K(x)
\frac{e^{-ay_*^2/2}}
{\sqrt{|\Phi_{yy}(y_*;x,v)|}} .
\label{eq:A_CR}
\end{align}
The phase averaging in Eq.~\eqref{eq:A_regular_L1} follows, for example, whenever the stationary phase varies continuously across ${\cal R}$. Indeed, evaluated on the stationary manifold, $\partial_v\Phi(y_*;x,v)=y_*$, so the phase is nonconstant on any regular region with $y_*\neq0$. Using the stationary-point condition,
\begin{equation}
v
=
-\frac{1}{2}
\left[
f'\left(x+\frac{y_*}{2}\right)
+
f'\left(x-\frac{y_*}{2}\right)
\right],
\end{equation}
one also has $|\partial v/\partial y_*|=|\Phi_{yy}(y_*;x,v)|$. When the stationary branch is one-to-one, Eq.~\eqref{eq:A_CR} may therefore equivalently be written as
\begin{equation}
C_{\cal R}
=
\frac{4\sqrt{2\pi}}{\pi}
\int_{\widetilde{\cal R}}
dx\,dy_*\,
K(x)e^{-ay_*^2/2}
\sqrt{
|\Phi_{yy}(y_*;x,v)|
}.
\label{eq:A_CR_y}
\end{equation}
If several stationary-point pairs contribute, the leading expression is instead
\begin{align}
W(x,v;\theta)
&\sim
2K(x)\sqrt{\frac{2\pi}{\theta}}
\sum_j
\frac{e^{-ay_j^2/2}}
{\sqrt{|\Phi_{yy}(y_j;x,v)|}}
\nonumber\\
&\times
\cos\left[
\theta\Phi(y_j;x,v)
+
\frac{\pi}{4}\operatorname{sgn}\Phi_{yy}(y_j;x,v)
\right].
\label{eq:A_multi}
\end{align}
The overall $\theta^{-1/2}$ scaling remains unchanged, but the coefficient multiplying $\sqrt{\theta}$ in the integrated $L_1$ norm depends on interference between the different stationary branches. Consequently, stationary-phase analysis fixes the generic exponent but not a universal prefactor.
\subsection{Caustics and the uniform scaling near $y=0$}
Special care is required near degenerate stationary points. In particular, $y=0$ becomes stationary along
\begin{equation}
    v_c(x)=-f'(x),
\label{eq:A_caustic_curve}
\end{equation}
where $\Phi_{yy}(0;x,v_c)=0$. The fact that this caustic is lower-dimensional is not by itself sufficient to conclude that it is asymptotically negligible. Its neighborhood must be treated with a uniform approximation. Expanding Eq.~\eqref{eq:A_phase} about $y=0$ gives
\begin{equation}
\Phi(y;x,v)
=
\left[v+f'(x)\right]y
+
\frac{f'''(x)}{24}y^3
+
O(y^5).
\label{eq:A_fold_phase}
\end{equation}
When $f'''(x)\neq0$ the degeneracy is a generic fold caustic. With $\delta v=v+f'(x)$, the corresponding uniform integral has Airy scaling,
\begin{equation}
W_{\rm fold}
=
O(\theta^{-1/3})
\,
{\cal A}
\left(
\theta^{2/3}\delta v
\right),
\label{eq:A_Airy_scaling}
\end{equation}
where ${\cal A}$ is an Airy-type scaling function whose precise normalization depends on $f'''(x)$ and on the Gaussian amplitude. The width of the caustic layer in the rescaled momentum variable is then $\Delta v=O(\theta^{-2/3})$, and its contribution to the Wigner $L_1$ norm scales as
\begin{equation}
\theta\,
\Delta v\,
|W_{\rm fold}|
=
\theta
\times
O(\theta^{-2/3})
\times
O(\theta^{-1/3})
=
O(1).
\label{eq:A_caustic_L1}
\end{equation}
Thus generic fold caustics are subleading compared with the $O(\sqrt{\theta})$ contribution from regular stationary-phase regions. If $f'''(x)=0$ at a degenerate stationary point, the Airy approximation does not apply. Higher-order catastrophes must then be analyzed separately, and no universal large-$\theta$ exponent can be inferred without specifying the corresponding degeneracy.
\subsection{Conditions for square-root growth}
The preceding analysis gives a controlled statement of the large-$\theta$ behavior. Suppose that the following conditions hold.
\begin{enumerate}
\item $f$ is sufficiently smooth and independent of $\theta$.
\item The stationary points contributing within the Gaussian envelope
are isolated regular branches except for generic fold caustics.
\item The sum over stationary branches is finite or absolutely
convergent after inclusion of the Gaussian factor $e^{-ay^2/2}$.
\item The phase-averaged leading stationary contribution is nonzero on
a phase-space region of nonzero measure.
\end{enumerate}
Then the regular stationary regions provide a nonzero contribution proportional to $\sqrt{\theta}$, while generic fold-caustic and nonstationary regions are subleading. Consequently,
\begin{equation}
    e^{\mathcal M(\theta)}
    =
    \Theta(\sqrt{\theta}),
    \quad
    \theta\rightarrow\infty ,
\label{eq:A_theta_half}
\end{equation}
or equivalently
\begin{equation}
    \mathcal M(\theta)
    =
    \frac{1}{2}\ln\theta+O(1).
\label{eq:A_M_half}
\end{equation}
If only the regularity assumptions required for the stationary-phase upper estimate are imposed, one obtains the weaker statement
\begin{equation}
    e^{\mathcal M(\theta)}
    =
    O(\sqrt{\theta}),
\quad
    \mathcal M(\theta)
    \le
    \frac{1}{2}\ln\theta+O(1),
\label{eq:A_upper}
\end{equation}
within this class of phase functions. Equation~\eqref{eq:A_upper} is a statement about the specified stationary-phase class and is not a general consequence of unitarity. The qualification above is essential. If
\begin{equation}
    f(x)=\alpha x^2+\beta x+\gamma,
\label{eq:A_quadratic}
\end{equation}
then $U(\theta)$ is a Gaussian unitary for every $\theta$. Acting on a Gaussian input, it produces another Gaussian state with an everywhere nonnegative Wigner function. Hence
\begin{equation}
    e^{\mathcal M}=1,
    \quad
    \mathcal M=0
\label{eq:A_gaussian_exception}
\end{equation}
for all $\theta$. Thus neither unitarity nor dependence on a single phase-space coordinate alone implies $\mathcal M\sim\frac12\ln\theta$. The square-root law instead characterizes a broad class of genuinely nonlinear single-coordinate phase gates whose semiclassical Wigner integrals possess regular stationary manifolds of nonzero measure. The stationary-phase estimates above are controlled by a large phase, not simply by $\theta>1$. Decomposing the generator,
\begin{equation}
\theta f(x)
=
\theta\left[f(0)+f'(0)x+\tfrac12f''(0)x^2\right]
+
\theta\,\delta f(x),
\end{equation}
the bracketed part generates a Gaussian unitary and contributes nothing to the Wigner negativity of a Gaussian input, by Eq.~\eqref{eq:A_gaussian_exception}. The relevant large parameter is therefore the size of $\theta\,\delta f$ over the support of the input state. For $f(x)=\cos(cx)$ the leading non-Gaussian term is $\theta c^4x^4/24$, and with $\langle x^4\rangle=3/4$ for the vacuum this gives the validity condition
\begin{equation}
|\theta|\gg\frac{32}{c^4}.
\label{eq:strong_validity_app}
\end{equation}
This is confirmed numerically: at $c=3,5,10$ the offset $\mathcal M-\tfrac12\ln|\theta|$ has already settled by $|\theta|\sim10$, whereas at $c=1$ it is still drifting at $|\theta|=10^2$ and at $c=0.5$ convergence is not reached until $|\theta|\gtrsim10^3$, in each case consistent with Eq.~\eqref{eq:strong_validity_app}. The exponent $\tfrac12$ is reproduced for every $c$ examined. Only the onset of the asymptotic regime, and the constant $I(c)$, depend on the spatial frequency.
\subsection{Application to the cosine gate}
For the gate considered in the main text, $f(x)=\cos(cx)$, the phase in Eq.~\eqref{eq:A_phase} simplifies to
\begin{equation}
\Phi(y;x,v)
=
vy
-
2\sin(cx)\sin\left(\frac{cy}{2}\right).
\label{eq:A_cos_phase}
\end{equation}
The stationary-point equation is
\begin{equation}
    v
    =
    c\sin(cx)
    \cos\left(\frac{cy}{2}\right),
\label{eq:A_cos_stationary}
\end{equation}
and
\begin{equation}
\Phi_{yy}
=
\frac{c^2}{2}
\sin(cx)
\sin\left(\frac{cy}{2}\right).
\label{eq:A_cos_second}
\end{equation}
Thus regular stationary regions with $\sin(cx)\sin(cy/2)\neq0$ are present for any $c\neq0$. The corresponding caustics occur at
\begin{equation}
    y_n=\frac{2\pi n}{c},
    \quad
    v_n=(-1)^n c\sin(cx),
\label{eq:A_cos_caustics}
\end{equation}
for integer $n$. At these points,
\begin{equation}
\Phi_{yyy}(y_n)
=
\frac{c^3}{4}
(-1)^n\sin(cx).
\label{eq:A_cos_third}
\end{equation}
Except at the discrete values $\sin(cx)=0$, these are generic fold caustics and obey the Airy scaling discussed above. Furthermore, the Gaussian factor
\begin{equation}
    e^{-ay_n^2/2}
    =
    \exp\left(
       -\frac{2a\pi^2n^2}{c^2}
    \right)
\end{equation}
exponentially suppresses distant stationary branches. The stationary-phase analysis therefore explains the square-root large-angle behavior observed for the cosine gate,
\begin{equation}
    e^{\mathcal M}
    \sim
    I(c)\sqrt{\theta},
\label{eq:A_cos_scaling}
\end{equation}
when the regular stationary contribution dominates and Eq.~\eqref{eq:strong_validity_app} is satisfied. The coefficient $I(c)$ is not universal: it depends on $c$, the Gaussian input, and the phase-averaged interference among the contributing stationary branches. The momentum dependence in Eq.~\eqref{eq:W_exact_appendix} enters only through $z=2\theta\sin(cx)$, so the reduction Eq.~\eqref{eq:app_Iz_reduction} together with the large-$z$ scaling $I_c(z)\rightarrow G_c\sqrt{|z|}$ gives
\begin{equation}
I(c)
=
\frac{\sqrt2}{\pi}\,G_c
\int_{-\infty}^{\infty}dx\,e^{-x^2}\sqrt{|\sin(cx)|} ,
\label{eq:Ic_reduction}
\end{equation}
with $G_c=\lim_{z\rightarrow\infty}I_c(z)/\sqrt{|z|}$ obtained from the Debye form of $J_k(z)$. Table~\ref{tab:mana_constants} lists the result. At large $c$ the Gaussians in Eq.~\eqref{eq:app_Iz_reduction} decouple and Eq.~\eqref{eq:Ic_reduction} can be evaluated in closed form,
\begin{equation}
I(c)
\;\xrightarrow[c\rightarrow\infty]{}\;
\frac{128\sqrt\pi}{\Gamma(1/4)^4}
=
1.3130 ,
\label{eq:Iinf}
\end{equation}
already attained by the $c=10$ entry. Consequently,
\begin{equation}
    \mathcal M
    =
    \frac{1}{2}\ln\theta
    +
    \ln I(c)
    +
    o(1) .
\label{eq:A_cos_mana}
\end{equation}
Equations~\eqref{eq:A_cos_scaling} and \eqref{eq:A_cos_mana} describe the large-angle semiclassical regime of the nonlinear cosine phase gate. They do not constitute a unitarity bound on arbitrary one-mode operations. In particular, Gaussian phase gates have identically zero Wigner logarithmic negativity, while nongeneric higher-order stationary-point degeneracies can require different asymptotic treatments.
\section{Finite-step convergence of the hybrid gate construction}\label{app:2}
We consider the operator $\hat{\mathcal{C}}_1(\theta, c) = \hat{G}_1 \otimes \mathbb{I} + \hat{G}_2 \otimes Z + \hat{B} \otimes X$ as in Eq.~\eqref{eq:op_Trotter1_1mode_new}. Given that all the bosonic operators commute with each other, the $N$-th power is found to have the same structure $\hat{\mathcal{C}}_1^N = A_N \otimes \mathbb{I} + Z_N \otimes Z + X_N \otimes X$. The coefficients follow the recurrence relation
\begin{equation}
    \begin{pmatrix} A_{N+1} \\ Z_{N+1} \\ X_{N+1} \end{pmatrix} = \begin{pmatrix} \hat{G}_1 & \hat{G}_2 & \hat{B} \\ \hat{G}_2 & \hat{G}_1 & 0 \\ \hat{B} & 0 & \hat{G}_1 \end{pmatrix} \begin{pmatrix} A_N \\ Z_N \\ X_N \end{pmatrix}.
\end{equation}
Since $\hat{G}_2$ and $\hat{B}$ are anti-Hermitian, their squares are negative semi-definite. We define the positive semi-definite Hermitian operator $\hat{R} = \sqrt{-(\hat{G}_2^2 + \hat{B}^2)}$. Diagonalizing the coefficient matrix yields the closed-form coefficients
\begin{align}
    A_N &= \frac{1}{2} \left[ (\hat{G}_1 + i\hat{R})^N + (\hat{G}_1 - i\hat{R})^N \right], \\
    Z_N &= \frac{\hat{G}_2}{2i\hat{R}} \left[ (\hat{G}_1 + i\hat{R})^N - (\hat{G}_1 - i\hat{R})^N \right], \\
    X_N &= \frac{\hat{B}}{2i\hat{R}} \left[ (\hat{G}_1 + i\hat{R})^N - (\hat{G}_1 - i\hat{R})^N \right].
\end{align}
To verify the limit $\theta \mapsto \theta/N$ as $N \to \infty$, we note $\hat{G}_1 \to \mathbb{I}$, $\hat{G}_2 \to -i\frac{\theta}{N}\cos(c\hat{x})$, $\hat{B} \to 0$, and thus $\hat{R} \to \frac{\theta}{N}\cos(c\hat{x})$. It follows that
\begin{align}
    A_N &\to \frac{1}{2} \left[ \left(1 + i\frac{\theta}{N}\cos(c\hat{x})\right)^N + \left(1 - i\frac{\theta}{N}\cos(c\hat{x})\right)^N \right]\nonumber\\
    &\to \cos(\theta \cos(c\hat{x})), \nonumber\\
    Z_N &\to \frac{-1}{2} \left[ \left(1 + i\frac{\theta}{N}\cos(c\hat{x})\right)^N - \left(1 - i\frac{\theta}{N}\cos(c\hat{x})\right)^N \right] \nonumber\\
    &\to -i\sin(\theta \cos(c\hat{x})),\nonumber\\
    X_N &\to 0.
\end{align}
Consequently, the operator converges to the trigonometric gate
\begin{align}
    \hat{\mathcal{C}}_1^N &\to \cos(\theta \cos(c\hat{x})) \otimes \mathbb{I} - i\sin(\theta \cos(c\hat{x})) \otimes Z  \nonumber\\
    &= \exp(-i \theta \cos(c\hat{x}) \otimes Z).
\end{align}
\section{Equivalence of the ancilla constructions}\label{app:3}
Here, we briefly show how the two-ancilla gate qubit construction of the trigonometric gates presented in Ref.~\cite{Rainaldi:2025ymn} is related to the single-ancilla construction of Ref.~\cite{Chalermpusitarak:2025cod}. The two-ancilla construction is based on the following structure of hybrid qubit-qumode unitaries
\begin{equation}
\begin{split}
    \Sigma_{ij}&\equiv \cos(\hat{A})\otimes \sigma_i + \sin(\hat{A})\otimes \sigma_j,\\
    \overline{\Sigma}_{ij}&\equiv \cos(\hat{A})\otimes \sigma_i - \sin(\hat{A})\otimes \sigma_j,
\end{split}
\end{equation}
that can be exponentiated by exploiting a qubit-ancilla based method to approximate continuous-variable trigonometric gates of the form
\begin{equation}
    \begin{split}
        e^{-i\frac{\theta}{2}\Sigma_{ij}}e^{-i\frac{\theta}{2}\overline{\Sigma}_{ij}} = e^{-i\theta\cos(\hat{A})\otimes \sigma_i} + \mathcal{O}(\theta^2),\\
        e^{-i\frac{\theta}{2}\Sigma_{ij}}e^{i\frac{\theta}{2}\overline{\Sigma}_{ij}} = e^{-i\theta\sin(\hat{A})\otimes \sigma_j} + \mathcal{O}(\theta^2),
    \end{split}
\end{equation}
with $\hat{A}$ any Hermitian operator. The method described in Ref.~\cite{Rainaldi:2025ymn} exploits two ancilla gate qubits for the trigonometric gates construction. However, the final circuit can be unitarily reduced to requiring solely a single ancilla gate qubit \cite{Chalermpusitarak:2025cod}. Without loss of generality, we work with $\Sigma_{zy}\equiv \Sigma$, which corresponds to the same choice of both Refs.~\cite{Rainaldi:2025ymn,Chalermpusitarak:2025cod}. The ancilla-based exponentiation of Ref.~\cite{Rainaldi:2025ymn} exactly provides the hybrid gate $e^{-i \theta \Sigma\otimes Z_b}$, as opposed to simply $e^{-i \theta \Sigma\otimes \mathbb{I}_b}$. The latter being created using the circuit in Ref.~\cite{Chalermpusitarak:2025cod}. It is clear that there exists a unitary transformation $V$ such that
\begin{equation}
    Ve^{-i \theta \Sigma\otimes Z_b}V^\dagger = e^{-i \theta \Sigma\otimes \mathbb{I}_b}\,.
\end{equation}
For our choice, it is $V = \text{CNOT}_{b\to a}$, where $a$ labels the ancilla gate qubit in $\Sigma$. Note that this is an operator identity. The secondary gate qubit $b$ decouples exactly, and it is not projected onto its eigenstate.
\section{Experimental readout and modeling details}
\label{app:experimental_methods}
In this Appendix we provide details of the motional-state reconstruction, postselection procedure, and open-system modeling used in the experimental analysis. We distinguish throughout between two physically separate dephasing processes. The parameter $\Gamma$ denotes decoherence during the blue-sideband (BSB) spectroscopy used to reconstruct Fock-state populations, whereas $\gamma$ denotes motional dephasing during implementation of the trigonometric-gate circuit itself. The former enters only the measurement model, while the latter enters the simulated quantum channel describing the gate sequence.
\subsection{Blue-sideband reconstruction of Fock-state populations}
\label{app:bsb_readout}
The motional state is not measured directly. Instead, the Fock-state populations are mapped onto the dynamics of an ancillary probe qubit through a blue-sideband interaction. Denoting the motional state after the trigonometric-gate circuit by $\rho_{\rm m}(\theta)$, the probe qubit is initialized in $\ket{\uparrow}$ and coupled to the selected motional mode through the blue-sideband Hamiltonian
\begin{equation}
    H_{\rm BSB}
    =
    \frac{\eta\Omega}{2}
    \left(
       \hat a\,\sigma_{-}
       +
       \hat a^{\dagger}\sigma_{+}
    \right),
\label{eq:app_bsb_H}
\end{equation}
where $\eta$ is the Lamb--Dicke factor of the probe ion for the addressed mode and $\Omega$ is the carrier Rabi frequency. For a motional state diagonal in the Fock basis,
\begin{equation}
    \rho_{\rm m}
    =
    \sum_{n=0}^{\infty}
    P_n \ket{n}\bra{n},
\label{eq:app_rho_diag}
\end{equation}
ideal BSB evolution produces a superposition of oscillations with frequencies proportional to $\sqrt{n+1}$. The resulting probe-qubit population is
\begin{equation}
P_{\downarrow}^{(0)}(t)
=
\frac{1}{2}
\left[
1-
\sum_{n=0}^{\infty}
P_n
\cos\!\left(
\eta\Omega\sqrt{n+1}\,t
\right)
\right].
\label{eq:app_bsb_ideal}
\end{equation}
Thus the amplitudes of the distinct BSB frequency components encode the Fock-state populations $P_n$. To account for decoherence during the BSB interrogation, we model the readout dynamics using a dephased sideband oscillation. With the convention used here, the exact single-frequency contribution is
\begin{equation}
P_{\downarrow}(t)
=
\frac{1}{2}
\left[
1-
\sum_{n=0}^{\Lambda}
P_n e^{-\Gamma t}
\left(
\cos(\widetilde{\Omega}_n t)
+
\frac{\Gamma}{\widetilde{\Omega}_n}
\sin(\widetilde{\Omega}_n t)
\right)
\right],
\label{eq:app_bsb_exact}
\end{equation}
where
\begin{equation}
    \widetilde{\Omega}_n
    =
    \sqrt{
       \eta^2\Omega^2(n+1)-\Gamma^2
    }.
\label{eq:app_bsb_frequency}
\end{equation}
In the experimentally relevant weak-readout-dephasing regime, $\Gamma\ll\eta\Omega$, this reduces to
\begin{align}
P_{\downarrow}(t)
&=
\frac{1}{2}
\left[
1-
\sum_{n=0}^{\Lambda}
P_n e^{-\Gamma t}
\cos\!\left(
\eta\Omega\sqrt{n+1}\,t
\right)
\right]\nonumber\\
&+
O\!\left(
e^{-\Gamma t}
\frac{\Gamma}{\eta\Omega}
\right).
\label{eq:app_bsb_approx}
\end{align}
We use Eq.~\eqref{eq:app_bsb_approx} to reconstruct the motional populations. For the single-mode measurements, the BSB evolution time is sampled at 101 uniformly spaced points between $0$ and $300\,\mu{\rm s}$. For the two-mode experiments, each addressed mode is sampled independently over a window containing approximately eight BSB oscillations, again using 101 points. The effective sideband coupling is close to $\eta\Omega/2\pi\simeq 16.7$--$17.5\,{\rm kHz}$ for the measurements reported here. The fitted readout-dephasing rates remain small compared with $\eta\Omega$, validating the use of Eq.~\eqref{eq:app_bsb_approx}.
\subsection{Population reconstruction and statistical uncertainties}
\label{app:population_fit}
For each value of the gate parameter $\theta$, the measured BSB trace is fit to Eq.~\eqref{eq:app_bsb_approx}. We retain reconstructed populations $\{P_n\}_{n=0}^{n_{\rm rec}}$ with $n_{\rm rec}=20$, and impose the physical constraints
\begin{equation}
    P_n\geq0,
    \quad
    \sum_{n=0}^{n_{\rm rec}}P_n=1.
\label{eq:app_probability_constraints}
\end{equation}
The readout-dephasing rate $\Gamma$ and effective BSB coupling $\eta\Omega$ are included as nuisance parameters in the reconstruction. For measured BSB times $t_\ell$, we minimize the weighted least-squares objective
\begin{align}
{\cal L}_{\rm BSB}
&=
\frac{
\displaystyle
\sum_{\ell}
w_{\ell}
\left[
P_{\downarrow,{\rm data}}(t_\ell)
-
P_{\downarrow,{\rm model}}(t_\ell)
\right]^2
}{
\displaystyle
\sum_{\ell} w_{\ell}
},
\nonumber\\
w_{\ell}
&=
\frac{1}{\sigma_{\ell}^2}.
\label{eq:app_bsb_loss}
\end{align}
For binomial qubit readout, the statistical uncertainty is estimated from
\begin{equation}
\sigma_{\ell}
=
\sqrt{
\frac{
P_{\downarrow,{\rm data}}(t_\ell)
\left[
1-P_{\downarrow,{\rm data}}(t_\ell)
\right]
}{
N_{\uparrow}
}
},
\label{eq:app_binomial_sigma}
\end{equation}
where $N_{\uparrow}$ is the number of experimental shots surviving postselection on the gate qubit. The reconstructed populations and their uncertainties are obtained from the local covariance of the constrained fit. We additionally verify that the inferred low-lying populations are stable against increasing the Fock cutoff used in the forward model. Unless otherwise stated, a Hilbert-space cutoff $\Lambda=40$ is used for single-mode classical simulations. The two-mode simulations use $\Lambda=15$ per mode for computational efficiency. Convergence of the reported low-lying marginal populations is checked against changes of this cutoff. For the $c=1$ single-mode data, $1200$ shots per BSB time and value of $\theta$ were collected for the one-, two-, and four-step circuits, and $1000$ shots for the eight-step circuit. The $c=2$ and $c=3$ single-step data and the two-mode measurements use $200$ experimental shots per BSB time and value of $\theta$. In all cases, the quoted experimental Fock-state probabilities refer to the postselected ensemble.
\subsection{Postselection}
\label{app:postselection}
The gate qubit provides a natural herald for finite-step errors in the hybrid implementation. Let $\rho_{\rm out}$ denote the joint state of the motional register and gate qubit after the circuit. The probability for successful postselection is
\begin{equation}
p_{\rm succ}
=
\tr
\left[
\left(
\mathbb I_{\rm m}\otimes
\ket{\uparrow}\bra{\uparrow}
\right)
\rho_{\rm out}
\right],
\label{eq:app_psucc}
\end{equation}
and the normalized postselected motional state is
\begin{equation}
\rho_{\rm m}^{(\uparrow)}
=
\frac{
\bra{\uparrow}
\rho_{\rm out}
\ket{\uparrow}
}{
p_{\rm succ}
}.
\label{eq:app_postselected_state}
\end{equation}
Experimentally, the postselection probability is typically $0.8$--$0.9$ over the parameter range studied. All experimental Fock-state distributions shown in the main text are conditioned on this successful outcome. The same projection and renormalization are applied to the numerical simulations, so that experiment and theory are compared on the same conditional ensemble. The success probability is nevertheless a separate performance metric and is not absorbed into the normalized Fock-state distributions.
\subsection{Thermal initialization}
\label{app:thermal_initialization}
Residual motional excitation after sideband cooling is modeled by a thermal state,
\begin{align}
\rho_{\rm th}(\bar n)
&=
\sum_{n=0}^{\infty}
\frac{\bar n^n}{(1+\bar n)^{n+1}}
\ket{n}\bra{n}
\nonumber\\
&=
\left(1-e^{-\beta\omega}\right)
\sum_{n=0}^{\infty}
e^{-n\beta\omega}
\ket{n}\bra{n},
\label{eq:app_thermal_state}
\end{align}
with $\bar n=(e^{\beta\omega}-1)^{-1}$ and purity $\tr[\rho_{\rm th}^2]=(2\bar n+1)^{-1}$. The non-center-of-mass radial modes used in the experiment are cooled close to the vacuum, with typical residual occupations of order $\bar n\sim0.1$. Instead of assuming perfect vacuum initialization, the open-system comparisons retain $\bar n$ explicitly. For the single-mode simulation the initial qubit--qumode state is
\begin{equation}
\rho_{\rm in}
=
\rho_{\rm th}(\bar n)
\otimes
\ket{\uparrow}\bra{\uparrow},
\label{eq:app_single_initial}
\end{equation}
and for the two-mode experiment,
\begin{equation}
\rho_{\rm in}^{(2)}
=
\rho_{\rm th}(\bar n_1)
\otimes
\rho_{\rm th}(\bar n_2)
\otimes
\ket{\uparrow}\bra{\uparrow}.
\label{eq:app_two_initial}
\end{equation}
\subsection{Open-system model for motional dephasing}
\label{app:dephasing_model}
The dominant dynamical imperfection included in the circuit model is motional dephasing. The independently measured motional coherence times of the addressed radial modes are approximately $500$--$700\,\mu{\rm s}$. Heating rates are substantially slower: representative values are of order $10~{\rm quanta/s}$ for the zig-zag mode and $100~{\rm quanta/s}$ for the drum mode. Over the circuit durations used here, the corresponding increase in mean occupation is small compared with the effects of motional phase noise, and heating is therefore neglected in the minimal model. The evolution during each motional gate is described by the Lindblad master equation
\begin{equation}
\frac{d\rho}{dt}
=
-i[H(t),\rho]
+
\gamma
\left[
\hat n\rho\hat n
-
\frac{1}{2}
\left\{
\hat n^2,\rho
\right\}
\right],
\quad
\hat n=\hat a^\dagger\hat a .
\label{eq:app_lindblad_single}
\end{equation}
This channel suppresses Fock-basis coherences according to
\begin{equation}
\rho_{nm}(t)
\propto
\exp\!\left[
-\frac{\gamma t}{2}(n-m)^2
\right]
\rho_{nm}(0)
\label{eq:app_dephasing_elements}
\end{equation}
in the absence of the driven Hamiltonian. Physically, the model represents phase diffusion caused by fluctuations of the motional-mode frequency. Such fluctuations generate stochastic relative phases between Fock states and are consistent with the measured finite motional coherence times. The Lindblad operator $\hat n$ is, however, an effective description: agreement with this model establishes consistency with motional dephasing as an important error mechanism, but does not by itself uniquely identify the microscopic noise spectrum. The conditional-displacement portions of the circuit are simulated by integrating Eq.~\eqref{eq:app_lindblad_single} with the corresponding spin-dependent-force Hamiltonian. In dimensionless units, the two directions of the displacement may be represented by
\begin{equation}
    H_{+}=-\hat x\otimes X,
    \quad
    H_{-}=+\hat x\otimes X,
\label{eq:app_cd_hamiltonians}
\end{equation}
with the integration times chosen to reproduce the required displacement amplitudes. Qubit-only rotations are much faster than the motional gates and their contribution to the total dephasing time is neglected. We denote the resulting noisy quantum channel corresponding to an $N$-step implementation by ${\cal G}^{(\gamma)}_N$, so that for the input state in Eq.~\eqref{eq:app_single_initial},
\begin{equation}
\rho_{\rm out}^{[N]}
=
{\cal G}^{(\gamma)}_N
\left[
\rho_{\rm in}
\right].
\label{eq:app_noisy_output}
\end{equation}
The conditional Fock-state distribution compared with experiment is then
\begin{equation}
P_n^{[N]}
(\theta,c;\bar n,\gamma\,|\,\uparrow)
=
\frac{
\bra{n,\uparrow}
\rho_{\rm out}^{[N]}
\ket{n,\uparrow}
}{
\displaystyle
\sum_{n'=0}^{\Lambda}
\bra{n',\uparrow}
\rho_{\rm out}^{[N]}
\ket{n',\uparrow}
}.
\label{eq:app_single_post_prob}
\end{equation}
The denominator is precisely the simulated postselection probability.
\subsection{Two-mode open-system simulation}
\label{app:two_mode_noise}
For the two-mode circuit, independent thermal occupations and effective dephasing rates are retained for the two addressed motional modes. The master equation becomes
\begin{align}
\frac{d\rho}{dt}
&=
-i[H(t),\rho]
+
\sum_{j=1}^{2}
\gamma_j
\left[
\hat n_j\rho\hat n_j
-
\frac{1}{2}
\left\{
\hat n_j^2,\rho
\right\}
\right].
\label{eq:app_lindblad_two}
\end{align}
Starting from Eq.~\eqref{eq:app_two_initial}, the final state is denoted
\begin{equation}
\rho_{\rm out}^{[N],2}
=
{\cal G}^{(\gamma_1,\gamma_2)}_N
\left[
\rho_{\rm in}^{(2)}
\right].
\label{eq:app_two_output}
\end{equation}
Because the experimental BSB measurements interrogate the two modes separately, the reconstructed quantities are the mode-resolved marginals, not the full joint Fock distribution. For mode 1,
\begin{equation}
P_{n_1}^{[N]}
=
\frac{
\tr
\left[
\left(
\ket{n_1}\bra{n_1}
\otimes\mathbb I_2
\otimes\ket{\uparrow}\bra{\uparrow}
\right)
\rho_{\rm out}^{[N],2}
\right]
}{
\tr
\left[
\left(
\mathbb I_{12}
\otimes\ket{\uparrow}\bra{\uparrow}
\right)
\rho_{\rm out}^{[N],2}
\right]
},
\label{eq:app_two_marginal}
\end{equation}
with the expression for mode 2 obtained by interchanging the two modes. Consequently, these measurements characterize population transfer within each mode but do not by themselves determine intermode correlations or the joint distribution $P_{n_1,n_2}$. The ordering of the conditional-displacement operations in the physical two-mode sequence can lead to different effective exposure of the two modes to dephasing. The corresponding asymmetry between their marginal distributions is included automatically in the time-resolved open-system simulation and disappears in the noiseless limit.
\subsection{Inference of circuit-noise parameters}
\label{app:noise_fit}
The comparison between the reconstructed experimental populations and the open-system simulation constitutes a second fitting stage, distinct from the BSB reconstruction described in Sec.~\ref{app:population_fit}. For a set of experimental probabilities $P_{{\rm data},\ell}$ with standard errors $\sigma_\ell$, we minimize
\begin{equation}
{\cal L}_{\rm circ}(\boldsymbol{\lambda})
=
\frac{
\displaystyle
\sum_{\ell}
\frac{
\left[
P_{{\rm data},\ell}
-
P_{{\rm sim},\ell}(\boldsymbol{\lambda})
\right]^2
}{
\sigma_\ell^2
}
}{
\displaystyle
\sum_{\ell}\frac{1}{\sigma_\ell^2}
},
\label{eq:app_circuit_loss}
\end{equation}
where $\boldsymbol{\lambda}=(\bar n,\gamma)$ for the single-mode data and $\boldsymbol{\lambda}=(\bar n_1,\bar n_2,\gamma_1,\gamma_2)$ for the two-mode model. Where several circuits are described by a common experimental configuration, shared parameters are fit globally. In particular, the $c=1$ single-mode measurements at Trotter depths $N=1,2,4,8$ are described using common thermal and dephasing parameters. For data sets analyzed with independent fits, the fitted parameters provide an effective phenomenological description of that particular experimental sequence. They do not independently measure the underlying noise rate. At each optimum we verify local stability by evaluating the Hessian of ${\cal L}_{\rm circ}$ and checking that the fitted solution is a local minimum within numerical precision. Uncertainty bands shown for simulated populations are obtained by propagating the covariance of the fitted parameters through the open-system calculation. Because the noise parameters are optimized against the same populations they are compared with, the agreement shows that a minimal thermal-plus-dephasing model suffices to account for the principal deviations from the ideal circuit, not that no other noise mechanisms are present.
\subsection{Displacement-axis calibration errors}
\label{app:phase_calibration}
A small error in the motional-mode frequency produces a phase-space rotation of the implemented conditional displacement. If the displacement axis is miscalibrated by an angle $\varphi$, then
\begin{equation}
    {\rm CD}_{x}(\alpha)
    \longrightarrow
    {\rm CD}_{x}(\alpha e^{i\varphi}).
\label{eq:app_cd_rotation}
\end{equation}
For a constant phase rotation common to the sequence, this is equivalent to conjugating the ideal oscillator circuit by $R(\varphi)=e^{i\varphi\hat n}$, so that
\begin{equation}
{\cal C}_N(\varphi)
=
R^\dagger(\varphi)
{\cal C}_N
R(\varphi).
\label{eq:app_rotated_circuit}
\end{equation}
Since $R(\varphi)\ket{n}=e^{in\varphi}\ket{n}$, the corresponding Fock-basis matrix elements obey
\begin{equation}
\bra{m}
{\cal C}_N(\varphi)
\ket{n}
=
e^{i(n-m)\varphi}
\bra{m}
{\cal C}_N
\ket{n},
\label{eq:app_rotated_matrix}
\end{equation}
and therefore
\begin{equation}
\left|
\bra{m}
{\cal C}_N(\varphi)
\ket{n}
\right|^2
=
\left|
\bra{m}
{\cal C}_N
\ket{n}
\right|^2 ,
\label{eq:app_fock_phase_invariance}
\end{equation}
so that a rigid phase-space rotation cannot be inferred from Fock-state transition probabilities alone. Phase-sensitive observables do respond to this calibration error. The Wigner and characteristic functions transform as
\begin{equation}
    W(\alpha)
    \longrightarrow
    W(\alpha e^{i\varphi}),
    \quad
    \chi(\beta)
    \longrightarrow
    \chi(\beta e^{i\varphi}).
\label{eq:app_phase_space_rotation}
\end{equation}
This effect is included when comparing the experimentally reconstructed characteristic function with the theoretical prediction. For the characteristic-function measurement discussed in the main text, the expected displacement-angle error follows from
\begin{equation}
    \varphi\simeq-\delta t_{\rm CD},
\label{eq:app_phi_detuning}
\end{equation}
where $\delta$ is the residual sideband detuning and $t_{\rm CD}$ is the physical displacement-gate duration. The experimentally relevant detunings and gate durations correspond to phase rotations of several degrees, consistent with the rotation required to align the measured and simulated phase-space structures.
\subsection{Characteristic-function measurement}
\label{app:chi_measurement}
In addition to BSB spectroscopy, we characterize one representative output state through its characteristic function
\begin{equation}
    \chi(\beta)
    =
    \tr
    \left[
       \rho_{\rm m}^{(\uparrow)}
       {\rm D}(\beta)
    \right].
\label{eq:app_chi_def}
\end{equation}
The characteristic function is mapped onto an ancillary probe qubit using conditional displacements. By choosing the initial probe-qubit state and the phase of the analysis pulse, the real and imaginary components of $\chi(\beta)$ can be accessed from the measured qubit polarization. The measurement is repeated over a grid of complex displacements $\beta$ to reconstruct the phase-space structure of the postselected motional state. Unlike Fock-state populations, $\chi(\beta)$ retains phase information and so is sensitive to a global rotation of the displacement axis. For this reason, the comparison with open-system simulation includes the rotation in Eq.~\eqref{eq:app_phase_space_rotation}, in addition to thermal initialization and motional dephasing. The resulting comparison is a phase-sensitive consistency check of the error model, not an independent determination of its parameters.
\subsection{Numerical implementation and convergence}
\label{app:numerical_details}
All classical calculations reported in this work are performed using \texttt{QuantumToolbox.jl} \cite{Mercurio:2025alt} in \texttt{Julia} \cite{Bezanson:2014pyv}. The open-system dynamics are evaluated in a truncated Fock basis. The single-mode calculations use a default cutoff $\Lambda=40$, while the two-mode calculations use a reduced cutoff $\Lambda=15$ per mode. For all reported observables, the truncation is chosen sufficiently above the experimentally occupied Fock levels that increasing the cutoff does not change the displayed low-energy populations within their quoted uncertainties. The finite-step circuits are simulated using the same ordering and physical durations of the conditional-displacement operations as in the experiment. In the presence of dephasing, two circuits that are algebraically equivalent in the noiseless limit need not experience identical open-system evolution when their gate ordering or duration differs. Postselection is performed only after propagation of the full qubit--qumode density matrix, according to Eq.~\eqref{eq:app_postselected_state}.
\end{document}